\makeatletter
\@ifundefined{@parse@version@dash}{%
\def\@parse@version#1{\@parse@version@0#1}
\def\@parse@version@#1/#2/#3#4#5\@nil{%
\@parse@version@dash#1-#2-#3#4\@nil}
\def\@parse@version@dash#1-#2-#3#4#5\@nil{%
  \if\relax#2\relax\else#1\fi#2#3#4 }
}{}
\makeatother
\documentclass[aps,prd,reprint,superscriptaddress,nofootinbib,floatfix]{revtex4-2}
\usepackage{graphicx,color}
\usepackage{amsmath}
\usepackage[utf8]{inputenc}
\usepackage{ulem}
\def\be{\begin{eqnarray}&&} 
\def\nonu{\nonumber \\ &&} 
\def\ee{\end{eqnarray}} 

\def\bwt{\begin{widetext}}
\def\ewt{\end{widetext}}
\newcommand{\bma}[1] {\mbox{\boldmath{$#1$}}}
\newcommand{\blf}[1]{\tilde {\bf #1}}

\def\sumint{\int \! \!\ \! \! \! \! \!\ \! \! \!\! \!\sum}

\def\Tr{\rm Tr} 
\def\bq{\begin{eqnarray}}
\def\eq{\end{eqnarray}}

\usepackage{color}

\begin{document}
\title{ 
{ Light-Front Transverse Momentum Distributions for 
{ ${\cal J}$=1/2 Hadronic  Systems in  Valence Approximation. }
 }}
\author{Rocco Alessandro}
\affiliation{Universit\`a di Roma ``Tor Vergata'', 
Via della Ricerca Scientifica 1, 00133 Rome, Italy}
\author{ Alessio Del Dotto}
\affiliation{Istituto  Nazionale di Fisica Nucleare, Laboratori Nazionali di Frascati, Via Enrico Fermi 40, 00044 Frascati (Roma), Italy\\
} 
\author{Emanuele Pace}
\affiliation{Universit\`a di Roma ``Tor Vergata'', 
Via della Ricerca Scientifica 1, 00133 Rome, Italy}
 \author{Gabriele Perna}
 \affiliation{Dipartimento di Fisica e Geologia, Universit\`a di Perugia, Via Alessandro Pascoli,
06123 Perugia, Italy}
  \author{Giovanni Salm\`e}
  \affiliation{Istituto  Nazionale di Fisica Nucleare, Sezione di Roma, P.le A. Moro 2,
00185 Rome, Italy}
\author{Sergio Scopetta}
\affiliation{Dipartimento di Fisica e Geologia, Universit\`a di Perugia, Via Alessandro Pascoli,
06123 Perugia, Italy}
\affiliation{Istituto  Nazionale di Fisica Nucleare, Sezione di Perugia, Via Alessandro Pascoli,
06123 Perugia, Italy}
\begin{abstract}
 {The semi-inclusive correlator for a  ${\cal J}$=1/2 bound-system,
composed by  $A$ spin-1/2 fermions,  is linearly expressed 
in terms of 
 the light-front Poincar\'e covariant spin-dependent spectral function, in valence approximation.
 {The light-front spin-dependent spectral function is fully determined by six scalar functions that} allow for a complete description of the six T-even
 transverse-momentum distributions, suitable for a detailed investigation of the dynamics inside the bound system. The application of the
 developed formalism to a case with a sophisticated dynamical content, like 
 $^3$He, reaches two goals: (i) to illustrate a prototype 
 of an investigation path 
 for gathering a rich wealth of
 information on the dynamics and also finding valuable constraints to be exploited from the phenomenological standpoint; (ii) to support 
 for the three-nucleon system a dedicated  experimental effort for obtaining a detailed 3D picture 
 {in momentum space}.
  In particular, the
 orbital-angular momentum decomposition of the bound state can be studied through the assessment of relations among the transverse-momentum
 distributions, as well as the relevance of the relativistic effect generated by the implementation of 
  macroscopic locality. A fresh evaluation
 of the longitudinal and transverse polarizations of the neutron and proton is also provided, confirming}  
 essentially the values
 used in the standard {procedure for extracting the neutron structure functions from both deep-inelastic scattering and   semi-inclusive reactions, in the same
 kinematical regime.} 
 \end{abstract}
 
\date{\today}
 \maketitle
 \section{Introduction}
A fully relativistic treatment is needed to describe 
{hadronic bound
systems}, when high energy  processes are studied and/or a high degree 
of accuracy is required. {Among  the phenomenological efforts to implement
 a Poincar\'e-covariant description of a bound system, we recall the proposal in 
Ref. \cite{DelDotto:2016vkh},  where the   light-front  Hamiltonian dynamics (LFHD)  \cite{Dirac:1949cp,KP,Coester:1992cg,Lev:1993pfz,Polyzou:2012ut,Polyzou:2018eis}
was adopted in order 
to obtain   the light-front (LF) 
 spin-dependent spectral  function,  whose diagonal elements  yield the distribution probability to
 find a constituent   with given spin,  LF momentum  and (off-shell) energy, 
 inside the bound system. } 
 The spectral function {is} {primarily} used to study
  the nucleon momentum distributions {in nuclei, but the formalism can be notably extended to a 
  hadronic bound system 
 and, as it will be illustrated in detail,  to eventually  obtain}  the six T-even
 transverse momentum distributions (TMDs) \cite{Barone:2001sp}.
 {Through the latter quantities,  one can achieve}  a detailed description of {{the system}}, much richer than the one given 
 by the usual  distribution in terms of the  
 {constituent}
  momentum $|{\bf p}|$ 
 in the lab  frame. {As a matter of fact, one can address the
 correlations  between spin and momentum, substantially deepening our
 understanding of the inner dynamics.} 
 {{For the nucleon, TMDs}}
(see, e.g.,  \cite{Sivers:1989cc,Sivers:1990fh,Collins:1992kk,Mulders:1995dh,JMR,Boer:1997nt,Bacchetta:2004jz,Collins:2012uy,Angeles-Martinez:2015sea,Collins:2016hqq,Avakian:2019drf}) are the object of 
 impressive theoretical and experimental efforts, both in {semi-inclusive deep inelastic scattering (SIDIS)} and in Drell-Yan processes (see, e.g., \cite{Anselmino:2005ea,Kang:2015msa,Anselmino:2020nrk,JLAB-E12-11-007,JLAB-E12-09-018,JLAB-E12-10-006,JLAB-E06-011,JeffersonLabHallA:2013qky,HERMES:2004mhh}  and   \cite{COMPASS:2010hbb,Martin:2014wua,COMPASS:2017jbv},
 respectively). In particular light-cone models {and phenomenological approaches for the TMDs} have been used, {e.g., (i) }to study the 
 three-dimensional nucleon 
 structure \cite{Bacchetta:2008af,Pasquini:2008ax, Boffi:2009sh,Lorce:2011zta,Burkardt:2015qoa,Bacchetta:2016ccz,Bacchetta:2020gko}, 
  (ii) to 
 {address} the nucleon-spin puzzle \cite{Lorce:2017wkb}  
 and (iii) {hence} to disentangle the contributions  of
 different  angular-momentum components to the spin of the nucleon \cite{Ji:2002xn,Ji:2012sj}.
  Let us notice that the nucleon, {namely} a spin $1/2$ system,
    is composed by three 
  quarks, in valence approximation.
  {Therefore}, in this approximation, 
  { the approach we have elaborated in LFHD} 
  can be applied {both } to 
  the nucleon, 
  as a 
  system of three quarks, and to $^3$He or $^3$H, as systems of three nucleons. 
 
  {The  application of our formalism to the three-nucleon system has a twofold benefit. On one side, it allows one  to  illustrate a 
  realistic example,  with specific features of TMDs 
  and constraints among them that can be traced back to the inner dynamics, e.g. the impact of the orbital-momentum content  generated by the interaction. On
  the other side, it   
  establishes a first theoretical basis for supporting future experimental efforts aiming to investigate the TMDs of $^3$He, and eventually
  construct a  3D
  tomography of the nucleus {in momentum space}.
  In a  nucleus 
  with total momentum $P$ in the lab frame, TMDs
  describe the nucleon distribution as a function of $x=p^+/P^+$
  \footnote{For the definitions of the kinematical variables of the LFHD used in this paper, the reader can refer to Ref. \cite{DelDotto:2016vkh}. Let us only recall here that
the light-front components of a four vector $v$ are $(v^-,{\bf \tilde  v})$, where
${\bf \tilde  v}=(v^+,{\bf v}_ \perp)$ with $v^\pm=v^0 \pm {\bf {\hat {n}}} \cdot {\bf v}$ and 
${\bf v}_ \perp= {\bf v} - {\bf {\hat {n}}} ({\bf {\hat {n}}} \cdot {\bf v})$. The vector 
${\bf {\hat {n}}}$ is a generic unit vector. Then
the scalar product of two four vectors ${ a}$ and ${ b}$ 
is ${ a}\cdot { b} = {a^-b^+ +a^+b^-\over 2}-{\bf  a}_ \perp \cdot {\bf   b}_ \perp$. 
In this paper we choose ${\bf {\hat {n}}} \equiv {{\hat {z}}}$.}
    and of the transverse momentum 
  $p_\perp$, for any possible orientation of the spin of the nucleus and of the spin of the
   nucleon. 
   Hence, it is quite natural to look for a possible interplay  with the spin-dependent spectral function.}
 In Ref. \cite{DelDotto:2016vkh} the LF spectral  function, ${\cal P}^{\tau}_{{\cal M},\sigma'\sigma}({ \tilde {\bma \kappa}},\epsilon,S)$,
is   defined  starting from the LF wave function for a 
 three-nucleon system with polarization vector $\bf S$, spin $1/2$  and third component 
 $\cal M$, { using non symmetric intrinsic variables}. The energy $\epsilon$ is the energy  of a fully interacting {two-particle}  (23) subsystem and
 the variable $\tilde{\bma \kappa} = (\kappa^+, \bma \kappa_\perp)$ is the LF momentum of 
  particle 1 {in the intrinsic reference frame of the cluster [1,(23)]}.
  {From the
  spectator energy $\epsilon$ one can reconstruct the  component $\kappa^-$, leading to the off-shell
  energy of the constituent.}
   The spectral function is defined through the overlaps between the LF wave function of the system  and 
   tensor products of a plane wave of 
  momentum $\tilde{\bma \kappa}$ and the intrinsic state 
  of  the {two-particle spectator} subsystem.
The  mentioned tensor 
   product allows one to take care of macroscopic locality, i.e. cluster separability, \cite{KP} and  to introduce a new effect of binding in the spectral function \cite{DelDotto:2016vkh}. 
   With the help of the Bakamjan-Thomas construction of the Poincar\'e  generators \cite{Bakamjian:1953kh}, 
   the LF wave function can be obtained from the usual non relativistic wave function of the system
   with a realistic interaction between the nucleons. 
    Then the LF spectral function allows one to embed the successful phenomenology 
    for few-nucleon systems  in a Poincar\'e-covariant framework and 
    to satisfy at the same time both  normalization of the three-body system bound state 
    (i.e. the baryon number sum rule) and  momentum sum rule.
    {{Interestingly, in Ref. \cite{DelDotto:2016vkh} 
   the definition of the  LF spin-dependent spectral  function was 
   {plainly} 
   generalized to a generic  system of $A$ spin 1/2 fermions. 
   }}
  {As a first test of our approach, the calculation of the EMC effect for $^3$He is under way 
  \cite{Rina}. Preliminary results which consider only the contribution of the two-body bound-state
   channel show encouraging improvements \cite{Pace:2020ikl,Pace:2020ned,Pace:2016eiq,Pace:2017dwa} with respect to a  
   convolution approach with a momentum distribution \cite{Oelfke:1990uy}.
    
  In this work the {\it most general expressions} for the spin-dependent spectral functions and for 
  the spin-dependent momentum  distribution in terms of six scalar functions, ${\cal B}_i$ and $b_i$, 
  respectively (i=0,...,5), {{valid for any {{system of spin 1/2 fermions}}}} 
   are presented. 
  In valence approximation, we demonstrate that a linear relation 
  between the LF spectral function and the semi-inclusive fermion correlator 
  {occurs} (for preliminary results see  Ref. \cite{Scopetta:2011yak,Pace:2013wbl,Pace:2013bq,Pace:2015haa, Rinaldi:2016erw}). In turn, 
 {{for a spin 1/2 system}}
  it is straightforward to
  {relate} 
  the six T-even twist-two TMDs to the LF spectral function 
  and eventually to the system wave-function. The results for the {{$^3$He}} TMDs corresponding to 
  a realistic nuclear interaction are {also} presented. {In particular}, the LF  longitudinal and transverse effective polarizations,
  {{quantities relevant for the extraction
  of the neutron information from data
  collected with polarized nuclear targets,}} are evaluated for the proton and the neutron in $^3$He and compared with the corresponding non relativistic results {{currently used by experimental collaborations.}}
  {Moreover  we have assessed  the  approximated   
  relations 
   {between}
   the TMDs, investigated in Refs. \cite{JMR,Lorce:2011zta} with the aim to offer a guide for the extraction 
   of TMDs from experimental data.} 

It is important to stress that the validity of these
 relations  could be
 {tested} using $^3$He as a playground, since to perform a similar test 
for the proton target {is much more challenging, at the present stage}, due to quark confinement
and 
{hadron} fragmentation. {Indeed} 
 the measurement of
 $^3$He TMDs appear feasible at high luminosity facilities, such as Jefferson Lab and the future 
 electron-ion collider
\cite{AbdulKhalek:2021gbh},
through {{$^3{\overrightarrow{\rm He}}({\overrightarrow {e}},e'p)X$}}  experiments with proper polarization set-ups of beams and targets \cite{exclus}.

It should be pointed out that  the  relations investigated in Refs. \cite{JMR,Lorce:2011zta} could be applied and experimentally tested for  the nucleon, once one considers the nucleon formed by three constituents, i.e. when the valence regime of the dressed quarks is acting.

 The paper is organized as follows. In Section II,  the most general expressions  
  of the LF spin-dependent spectral function  and of the LF spin-dependent momentum distribution 
  are presented  {for any
 bound-system, composed by   $A$ fermions of spin 1/2} 
  in terms of  six suitable scalar functions.  
  {Explicit expressions for}
  {{ the six scalar
   functions 
   { defining} the momentum distribution are given in Appendix \ref{mdistr} {for a three-nucleon system
    of spin 1/2}}}.
     In Section III, the  linear relation in valence approximation between 
    the semi-inclusive correlator  and the LF spectral function is derived. In Section IV,  {the relation 
    between the LF spectral function and} 
 the T-even twist-two TMDs is illustrated. 
 {In Section V, the numerical results for the $^3$He nucleus are
 presented, ranging from the TMDs to the new calculations of 
 LF  longitudinal 
 and transverse effective polarizations} for the proton and the neutron in $^3$He. Also  
  the approximate relations between the T-even twist-two TMDs are discussed. 
In Section VI, our conclusions are drawn.

 \section{The LF spin-dependent spectral function and the LF spin-dependent {momentum distribution}}  
\label{sf}

 {Following Ref. \cite{DelDotto:2016vkh},  within the LFHD}
 a Poincar\'e covariant definition of the spin-dependent spectral function 
for an $A$-particle {bound } system polarized along
$\bf{S}$, {can be  simply obtained by} replacing the non relativistic overlaps 
$\langle{\vec{p}_1,\sigma \tau;\psi}_{f_{(A-1)}} |{\psi }_{\cal{J}\cal{M}};S, T T_z \rangle$,
which define the non relativistic spectral function,
with their LF counterparts 
$_{LF}\langle \tau_{S},T_{S};\alpha,\epsilon ;J_{z}J;\tau,\sigma,\tilde{\bma \kappa}|
{\psi}_{\cal{J}\cal{M}}; S, TT_z\rangle$. 

Hence, 
{the LF spin-dependent spectral function reads}
\bwt\be
{\cal P}^{\tau}_{{\cal M}, \sigma'\sigma}(\tilde{\bma \kappa},\epsilon,S) = \rho(\epsilon)  
\sum_{J J_{z}\alpha}\sum_{T_{S}\tau_{S} } ~
_{LF}\langle  \tau_{S},T_{S} ; 
\alpha,\epsilon ;J J_{z}; \tau\sigma',\tilde{\bma \kappa}|{\psi}_{\cal{J}\cal{M}}; S, T T_z\rangle
~ \langle S, T T_z;
{\psi}_{\cal{J}\cal{M}}|\tilde{\bma \kappa},\sigma\tau; J J_{z}; 
\epsilon, \alpha; T_{S}, \tau_{S}\rangle_{LF} ~ , 
\nonu\label{LFspf}
\ee
\ewt
where $|{\psi}_{\cal{J}\cal{M}};S, T T_z \rangle$ is the $A$-fermion   
  ground state, with total angular momentum ${\cal J}$ (third component ${\cal M}$), isospin $TT_z$ and rest-frame polarization 
  $S\equiv\{0, {\bf S}\}$.
 {The state
$|\tilde{\bma \kappa},\sigma, \tau; T_{S},\tau_{S};\alpha,\epsilon ;JJ_{z} \rangle_{LF}$ is the tensor
 product of (i) a fully interacting intrinsic {state  of the $(A-1)$-particle spectator} system, 
with intrinsic  energy $\epsilon$ (negative for bound states
and positive for the continuum spectrum ones), spin $J$ (third component $J_{z}$), isospin $T_{S}$ (third component
$ \tau_{S}$),
  and
$\alpha$  the set of quantum numbers needed to completely specify the 
   state,
  and  
(ii) a plane wave for  the 
{acting} particle with  LF-momentum 
$\tilde{\bma \kappa}$    in the intrinsic reference frame of the cluster    $[1,(A-1)]$. The total LF-momentum of the cluster is
${\blf P}_{intr}[1,(A-1)] \equiv \{{\cal M}_0, {\bf 0}_\perp\}$ with ${\cal M}_0$  its free mass. }
{In Eq. \eqref{LFspf},  $\rho(\epsilon)$ is the energy density of the $(A-1)$-particle states. Notice that
 for $(A-1)$ = 2, one has 
 $\rho(\epsilon) = 1$ for the bound states, e.g. for the deuteron, and 
 $\rho(\epsilon)= m\sqrt{m \epsilon}/2$ for the continuum{, with $m$ the constituent mass.}}
 
Denoting with ${\bf p}_i$ (i=1,...,A) and $\bf P$ the momenta of the particles and 
the {whole} system in the lab frame, respectively, {one gets the following expression for
the intrinsic} momentum 
$\tilde{\bma \kappa}$ 
\be
\kappa^+= \xi_1~{\cal{M}}_0[1,(A-1)]~, \quad 
{\bma \kappa}_{\perp }=
{\bf p}_{1 \perp }-\xi_1 {\bf P}_{\perp}~,
\label{kappa}
\ee
where $\xi_1={p^+_1 / P^+}$ and ${\cal{M}}_0[1,(A-1)]$,  the {previously mentioned} free mass of the $[1,(A-1)]$ 
cluster \cite{DelDotto:2016vkh},  {is} given by
\be
{\cal{M}}^2_0[1,(A-1)] = 
{ m^2 + {\bma \kappa}_{\perp }^2 \over \xi_1 } + {M_S^2 + {\bma \kappa}_{\perp }^2 \over (1 - \xi_1) }
~,
\label{calM}
\ee
with $M_S$ the mass of the interacting (A-1) system. 
Let us assume that the {\it system is at rest in the laboratory}.  {Then it follows that ${\bma \kappa}_{\perp }=
{\bf p}_{1 \perp }$ (recall that the LF-momentum $\tilde{\bma \kappa}$ is relative to the cluster frame). }
 {For completeness let us introduce the LF momentum $\tilde {\bf k} \equiv (k^+, {\bf k}_\perp)$ of the acting particle in the intrinsic $A$-particle system, with   transverse component  
 ${\bf k}_\perp = {\bf p}_{1 \perp } - \xi_1 {\bf P}_{\perp} = {\bf p}_{1 \perp }$, and  plus component 
 $k^+=\xi_1 ~M_0 \ne p^+_1$ { ($M_0$ is the free mass of
 $A$ particles)}. } In what follows the subscript $1$ will be dropped out and the notations 
$\xi_1= x$ and ${\bf p}_{1 \perp }= {\bf p}_{ \perp }$ will be adopted.

 It is worth 
 {reminding} that the states 
 $|\tilde{\bma \kappa},\sigma, \tau; T_{S},\tau_{S};\alpha,\epsilon ;JJ_{z} \rangle_{LF}$,
  to be used for the definition of the spectral function  in the LF overlaps
  $_{LF}\langle \tau_{S},T_{S};\alpha,\epsilon ;J_{z}J;\tau\sigma,\tilde{\bma \kappa}|{\psi}_{\cal{J}\cal{M}}; S, TT_z\rangle$,
  fulfill the macroscopic locality. The property of macroscopic locality means that the unitary representation of the Poincar\'e group for a system composed of two separated subsystems can be expressed as the tensor product of the unitary representations of the two subsystems. Hence, subsystem observables, associated with different space-time regions, must commute for large enough space-time separation (see Refs. [1, 3]). This is the mathematical formulation of the physical insight that when a system is separated in disjoint subsystems, these subsystems must behave as independent subsystems. Obviously the notion of disjoint subsystems does not apply to systems of quarks, where asymptotic states do not exist due to the confinement.

  Furthermore the use of the momentum  $\tilde{\bma \kappa}$,  instead of the  momentum { ${\bf{p}}$} of the fermion in 
  the lab frame,  introduces a new effect of binding in the spectral function. 



The LF overlaps
where the ground state 
has a generic polarization vector $\bf S$, i.e. $|{\psi}_{\cal{J}\cal{M}}; S, TT_z\rangle$,  
can be obtained from the overlaps where the ground state 
{is polarized}
along the $z$ axis, i.e. $|{{\cal{J}}{m}}; \epsilon^A, \Pi; TT_z \rangle_z$, by using
the Wigner rotation 
matrixes, $D^{\cal J}_{m, \cal M}(\alpha, \beta, \gamma)$, viz

\be
\hspace{-4mm}|{\psi}_{\cal{J}\cal{M}}; S, TT_z\rangle =\hspace{-1mm}{{\sum}_{m}} \hspace{-1mm}|{{\cal{J}}{m}};\epsilon^A,\Pi;TT_z\rangle_z  
D^{\cal J}_{m, \cal M}(\alpha, \beta, \gamma),
\label{rot}
\ee
where $\alpha, \beta$ and $\gamma$ are the Euler angles describing the proper rotation from the $z$-axis 
to the polarization vector $\bf S$ {and $\epsilon^A$, $\Pi$ are the energy and the parity of the state, respectively}. 
{Let us recall that the rotations involved act on the  bound
system as a whole,
and therefore they are interaction-free.}
{ Through Eq. (\ref{rot}) one can relate the spin-dependent spectral function with a given polarization to the one with polarization $\bf S = {\hat z}$ (see Appendix \ref{Sdep}).}

As explained in Ref. \cite{DelDotto:2016vkh}, in the {three-nucleon} case 
the overlaps 
$_{LF}\langle \tau_{S},T_{S};\alpha,\epsilon ;J_{z}J;\tau\sigma,\tilde{{\bma \kappa}}|{\psi}_{{\cal{J}}{m}}; S, T T_z \rangle_z$ 
 can be evaluated  
in terms of canonical {(or instant-form)} two- and three-body wave functions, replaced by the non relativistic ones,
{after applying the Melosh rotation
 matrixes \cite{Melosh:1974cu,Terentev:1976jk}(needed  for obtaining the LF spin states from the canonical ones).}
 {This follows, once}  the  Bakamjian-Thomas  construction of the Poincar\'e  generators \cite{Bakamjian:1953kh}
 {is adopted. Then, it turns out  that} 
 the two- and three-body non relativistic wave functions  have all the needed properties with  respect 
to rotations and translations  of the corresponding canonical wave functions.


 In conclusion the LF  spin-dependent spectral function is a $2 \times 2$ matrix,
${\bma{\widehat{\cal{P}}}}^{\tau}_{\cal{M}}({\tilde{\bma \kappa}},\epsilon,S)$, which depends on the direction 
of the polarization
vector $\bf{S}$, {and it is  usually normalized   
 for each isospin channel ${{\tau=p(n)}}$, i.e.,
\be
 \sumint {d\epsilon
  }
  \int {d {\bma \kappa} \over 2E({\bf \kappa}) 
  (2\pi)^3}~ 
Tr \left [ {\bma {\widehat {\cal P}}}^{\tau}_{\cal M}(\tilde{{\bma \kappa}},\epsilon,S) \right ]~=~1 
~,
\label{normFSLF1}
\ee 
with $E(\kappa) = \sqrt {m^2 + |{\bma \kappa}|^2}$.}
 
A general expression of ${\bma{\widehat{\cal{P}}}}^{\tau}_{\cal{M}}({\tilde{\bma \kappa}},\epsilon,S)$
can be obtained in terms
  of the vectors at our disposal in the {\it rest frame of the system},
   i.e. {(i)} the unit vector ${\hat {\bf n}}$, which defines the LF components of a four vector,
    {(ii)} the  polarization vector $\bf S$, and {(iii)}
the transverse (with respect to the ${\hat {\bf n}}$ axis) momentum component ${\bf k}_\perp={\bf p}_\perp=\bma \kappa_\perp$.
  Let us recall that we 
  {adopt} ${\hat {\bf n}}\equiv \hat z $.
   Then the LF  spin-dependent spectral function    
   {reads}
\be
{{\cal P}}^{\tau}_{{\cal M}, \sigma'\sigma}({\tilde{\bma \kappa}},\epsilon,S)=
{1 \over 2}  \left[{\cal {B}}_{0,{\cal M}}^{\tau}+
{\bma \sigma} \cdot {\bma {\cal F}}^{\tau}_{{\cal M}}({\tilde{\bma \kappa}},\epsilon, \bf S) 
\right]_{\sigma'\sigma} ,
\label{speclf1}
\ee
where the function ${\cal{B}}_{0,{\cal M}}^{\tau}$ is the trace of 
$
{{
{\widehat{
{\bma{\cal P}}}}^{\tau}_{\cal M} ({\tilde{\bma \kappa}},\epsilon,S)
}}
$
and yields the unpolarized spectral function, while
\be
{\bma {\cal F}}^{\tau}_{{\cal M}}({\tilde{\bma \kappa}},\epsilon, {\bf S}) ~=~
Tr \left [ {\bma {\widehat {\cal P}}}^{\tau}_{\cal M} ({\tilde{\bma \kappa}},\epsilon,S) 
~ \bma \sigma
\right ]
 ~.
\ee
The quantity 
${\bma {\cal F}}^{\tau}_{\cal{M}}({\tilde{\bma \kappa}},\epsilon, \bf S)$ is a pseudovector and depends on the direction of the polarization
vector $\bf{S}$. Therefore it can
be written as a linear combination of the independent pseudovectors at our disposal,
viz. $\bf{S}$, $~\hat {\bf k}_{\perp} ({\bf  S} \cdot \hat {\bf k}_{\perp})$, 
$~\hat {\bf k}_{\perp} ({\bf  S} \cdot \hat z)$, 
 $~\hat z ~({\bf  S} \cdot \hat {\bf k}_{\perp})$, and $~ \hat z ~({\bf  S} \cdot \hat {z})$.
 Furthermore ${\bma {\cal F}}^{\tau}_{\cal{M}}$ depends on $x$
  \be
 x =
 {\kappa^+ \over {\cal{M}}_0[1,(A-1)]} ~,
 \label{csi}
 \ee
where  ${{\cal{M}}_0[1,(A-1)]}$ {(cf. Eq. \eqref{calM})}, 
 is given in terms of 
${\tilde{\bma \kappa}}$ by \cite{DelDotto:2016vkh}
 \be
 {\cal{M}}_0[1,(A-1)] = E( \kappa) + E_S = {(\kappa^+)^2 + (m^2 + |{\bf k}_\perp|^2) \over 2 ~ \kappa^+ }
 \nonu
 +\left \{ \left [(\kappa^+)^2 + (m^2 + |{\bf k}_\perp|^2) \over 2 ~ \kappa^+ \right ]^2 ~ +
 ~ M_S^2 ~ - ~ m^2 \right \}^{1/2} ~ ,
 \label{CallM0}
 \ee
 where $E_S = \sqrt {M_S^2 + |{\bma \kappa}|^2}$, with   $M_S^2 = 4m^2 + 4m \epsilon$  for $(A-1)=2$, {and ${\bf k}_\perp={\bma
 \kappa}_\perp={\bf p}_\perp$}.
 
 Hence ${\bma {\cal F}}^{\tau}_{\cal{M}}$ can be expressed as a sum of the five available independent pseudovectors  multiplied   by
five scalar quantities, ${\cal B}^{\tau}_{i,{\cal M}}$ ($i= 1, . . .,5)$, viz.
 \be
 {\bma {\cal F}}^{\tau}_{{\cal M}}(x,{\bf k}_{\perp};\epsilon,{\bf S}) = 
{\bf  S} 
 {\cal B}^{\tau}_{1,{\cal M}}
 +
\hat {\bf k}_{\perp} ({\bf  S} \cdot \hat {\bf k}_{\perp})
{\cal B}^{\tau}_{2,{\cal M}}  
 \nonu+  \hat {\bf k}_{\perp} ({\bf  S} \cdot \hat z) 
{\cal B}^{\tau}_{3,{\cal M}} 
+ 
 \hat z ({\bf  S} \cdot \hat {\bf k}_{\perp})
  {\cal B}^{\tau}_{4,{\cal M}} 
  + 
 \hat z ({\bf  S} \cdot \hat {z}) 
 {\cal B}^{\tau}_{5,{\cal M}}
\label{dv1}
\ee
 where the dependence upon $\hat n\equiv \hat z$ is understood for making light the notation.
{ It should be pointed out that to  fully address  the issue of TMDs one has to distinguish  between the transverse dof's  and the one associate to $\hat n$. Therefore one cannot anymore use the parametrization with  only three scalar functions considered in Ref. \cite{CiofidegliAtti:1994cm}.


The six   scalar quantities ${\cal B}^{\tau}_{i,{\cal M}}$
 in general can depend on the possible scalars at our disposal, i.e.,
$ x, |{\bf k}_{\perp}|,\epsilon,({\bf  S}\cdot \hat {\bf k}_{\perp})^2,({\bf  S} \cdot {\hat z})^2, {{(  \hat {\bf k}_{\perp} \times {\hat z}  ) \cdot  {\bf S}}} $. 
However, as shown in Appendix \ref{Sdep}, for a system of total angular momentum ${\cal J}=1/2$, as $^3$He or 
$^3$H, the quantities ${\cal B}^{\tau}_{i,{\cal M}}$ can depend only on $x, |{\bf k}_{\perp}|$ {and }$\epsilon$.

 Through the trace of the spectral function 
$
{{
{\widehat{
{\bma{\cal P}}}}^{\tau}_{\cal M} ({\tilde{\bma \kappa}},\epsilon,S)
}}
$
 one can define  the LF {\it spin-independent} nucleon momentum distribution, averaged on the spin directions,
 as follows (see Ref. \cite{DelDotto:2016vkh})
\bwt\be
 n^\tau(x,{\bf k}_{\perp})  =  \sumint {d\epsilon ~
 }
 {1 \over 2 \kappa^+~ (2\pi)^3}~ 
 ~ {\partial \kappa^+ \over \partial x} 
 ~
 Tr {\cal P}^{\tau}(\tilde{\bma \kappa},\epsilon,S) 
 =
\sumint {d\epsilon ~
} 
{1 \over 2 ~ (2\pi)^3}~
{ E_S \over (1- {x })  ~ \kappa^+} ~ \rho (\epsilon)
 \nonu
 \times~ 
  \sum _{\sigma}  
\sum_{J J_{z}\alpha}\sum_{T_{S}\tau_{S} } 
 ~_{LF}\langle  \tau_{S},T_{S} ; 
\alpha,\epsilon ;J_{z}J; \tau\sigma,\tilde{\bma \kappa}|{\psi}_{\cal{J}\cal{M}}; S, TT_z\rangle
~ \langle TT_z, S;
{\psi}_{\cal{J}\cal{M}}|\tilde{\bma \kappa},\sigma\tau; J J_{z}; 
\epsilon, \alpha; T_{S}, \tau_{S}\rangle_{LF}
~ .
\label{momdisLF}
\ee  
\ewt
The completeness relation of the nonsymmetric basis for three-interacting-particle systems (see Eq. (51) of Ref. \cite{DelDotto:2016vkh}), 
 immediately leads to the normalization of the  nucleon momentum distribution, i.e. the baryon number sum rule,
\be
\int d x ~ \int d{\bf k}_{\perp} ~ n^\tau(x,{\bf k}_{\perp}) ~ =~1 ~ ,
\label{momnorm}
\ee  
and to
 the momentum sum rule 
\be
\int x ~ d x
~ \int d{\bf k}_{\perp} ~ n^\tau(x,{\bf k}_{\perp}) ~ =~{ 1 \over 3} ~.
\label{momsumrule}
\ee

{From the LF spin-dependent spectral function, after performing the integration shown in  Eq. \eqref{momdisLF}, 
one can obtain the LF {\it spin-dependent} momentum distribution,  
a $ 2 \times 2$ matrix}
 defined by 
(see also  Appendix \ref{mdistr}) 
\be
 \left [ {\bma {\cal N}}_{\cal M}^\tau(x,{\bf k}_{\perp};{\bf S}) \right ]_{\sigma ' \sigma } =
    \sumint  d\epsilon ~{1 \over 2 ~ {(2\pi)^3}} ~ {1 \over 1 - x}  ~ {E_S \over \kappa^+}  ~
  \nonu\times {{\cal P}}^{\tau}_{{\cal M}, \sigma'\sigma}({\tilde{\bma \kappa}},\epsilon,S)  
 = 
  {\pi \over 4 m}~ \int \frac{d{ p^+} d{ p^-}}{(2\pi)^4} ~ \delta(p^+ - \xi P^+) ~ P^+  
  \nonu \times~ {E_S \over \kappa^+} ~
   {{\cal P}}^{\tau}_{{\cal M}, \sigma'\sigma}({\tilde{\bma \kappa}},\epsilon,S)
 ~ .
  \label{N}
  \ee
As shown in Appendix \ref{mdistr}, 
one can obtain the LF spin-dependent momentum distribution 
from the three-body wave function, using Eq. (\ref{rot}) and the expression for the LF spin-dependent spectral 
function given by Eq. (72) of Ref. \cite{DelDotto:2016vkh}. 

As it occurs  for the spectral function, the momentum distribution can be expressed through the three 
independent vectors available in the rest frame of the system, i.e. $\bf k_\perp$, $\bf S$, and 
  $\hat n \equiv \hat z$, and 
   six scalar functions $b^{\tau}_{i,{\cal M}}$
  ($i= 0,1, . . .,5)$, viz  
  \be
  {\bma {\cal N}}_{\cal M}^\tau(x,{\bf k}_{\perp};{\bf S}) = 
  {1 \over 2} \left\{ b_{0,{\cal M}} +
 {\bma \sigma} \cdot {\bma {f}^{\tau}_{{\cal M}}(x,{\bf k}_{\perp};\bf S)}\right\} ~ , 
  \label{distr1}
  \ee
 where ${{\bma f}}^{\tau}_{{\cal M}}(x,{\bf k}_{\perp};\bf S)$ is a pseudovector  (recall that the dependence upon $\hat n\equiv
 \hat z$ has been dropped out for simplicity) 
 that can be decomposed as follows
  \be
 {\bma {f}}^{\tau}_{{\cal M}}(x,{\bf k}_{\perp};{\bf S}) = 
{\bf  S} 
 b^{\tau}_{1,{\cal M}}
 + 
\hat {\bf k}_{\perp} ({\bf  S} \cdot \hat {\bf k}_{\perp})
b^{\tau}_{2,{\cal M}}  
\nonu
 +  \hat {\bf k}_{\perp} 
 ({\bf  S} \cdot \hat z) 
b^{\tau}_{3,{\cal M}} + 
 \hat z ({\bf  S} \cdot \hat {\bf k}_{\perp})
  b^{\tau}_{4,{\cal M}} 
  + 
 \hat z ({\bf  S} \cdot \hat {z}) 
 b^{\tau}_{5,{\cal M}}  ~.
\label{dv2}
\ee
  The functions $b^{\tau}_{i,{\cal M}}$, that depend upon $ x$, $|{\bf k}_{\perp}|$,  $({\bf  S}\cdot \hat {\bf k}_{\perp})^2$,
   $({\bf  S} \cdot {\hat z})^2$ and  ${{
  (  \hat {\bf k}_{\perp} \times {\hat z}  ) \cdot  {\bf S} 
  }}$ ,  are   integrals over the energy $\epsilon$ of the  functions 
 ${\cal B}^{\tau}_{i,{\cal M}}$
  (see Eq. (\ref{N})), viz.
 \be
\hspace{-5mm} b^{\tau}_{i,{\cal M}}\left [x, |{\bf k}_{\perp}|,({\bf  S}\cdot \hat {\bf k}_{\perp})^2,({\bf  S} \cdot {\hat z})^2, {{
  (  \hat {\bf k}_{\perp} \times {\hat z}  ) \cdot  {\bf S} 
  }}  \right]  
 \nonu
 \hspace{-5mm} =   {\pi \over 4 m} \int {{d p^+ d p^-} \over {(2\pi)^4}} ~ \delta[p^+ - x P^+] ~ P^+ ~ {E_S \over \kappa^+} 
\nonu 
\hspace{-5mm} \times ~{\cal B}^{\tau}_{i,{\cal M}}\left [x, |{\bf k}_{\perp}|,\epsilon,({\bf  S}\cdot \hat {\bf k}_{\perp})^2,({\bf  S} \cdot {\hat z})^2, {{
  (  \hat {\bf k}_{\perp} \times {\hat z}  ) \cdot  {\bf S} 
  }}  \right] 
.
\label{dv12}
\ee

  In {Appendix \ref{mdistr}}, for a three-nucleon system of total angular momentum ${\cal J} = 1/2$
    {explicit expressions for}
  the quantities $b^{\tau}_{i,{\cal M}}$ ($i= 0,1, . . .,5$) are 
  obtained in terms of the wave function of the three-nucleon system, according to the BT procedure.
   It is also shown that
   these  
functions do not depend on $\bf S$, while they do depend on $|{\bf k}_{\perp}|$ and $x$.  
{ Moreover,} 
the quantity $b_{0}$ is independent of 
 $\cal M$, while for $i = 1, . . . ,5$ the dependence on $\cal M$ is through the factor $(-1)^{{\cal M} +1/2}$.
 
 From the {actual} expressions of the quantities $b^{\tau}_{i,{\cal M}}$, 
 the essential role  of the Melosh matrixes to generate the six different quantities $b^{\tau}_{i,{\cal M}}$ clearly emerges. 
 Their effect is parametrized by the angle $\varphi$ present in the expression of {the Melosh rotations, $ {\cal R}_M ({\blf k}) $, given in Appendix \ref{Melosh}, }
 viz.
 \be
D^{{1 \over 2}} [{\cal R}_M ({\blf k})]_{\sigma\sigma'}
= \left [ \cos {\varphi \over 2} ~ + ~i ~ \sin {\varphi \over 2} ~ \hat {\bf n} \cdot  {\bma  \sigma} 
 \right ]_{\sigma\sigma'}~,
\label{defmelo}
\ee
where 
\be
\varphi = 2 ~ arctg  ~{ |{\bf k}_\perp| \over k^+ + m}~.
\label{phidef}
\ee 
It has to be pointed  out that $\varphi$ is small if the relevant values of $|{\bf k}_\perp|/m$ do too, making   
 the effect of the Melosh rotations  small.
 
Indeed, in absence of the Melosh matrixes, the quantities $b^{\tau}_{2,{\cal M}}$, 
 $b^{\tau}_{3,{\cal M}}$, 
 $b^{\tau}_{4,{\cal M}}$, and $b^{\tau}_{5,{\cal M}}$ are related to each other by factors as $\cos^2 \theta$,
 $\sin^2 \theta$, and $\cos \theta \sin \theta$, with $\theta$ the angle between the momentum $\bf k$ and the $z$ axis. 
 Then in this case the spin-dependent momentum distribution can be expressed in terms  of only three independent scalar quantities, $b^{\tau}_{0}$, $b^{\tau}_{1,{\cal M}}$, and $b^{\tau}_{2,{\cal M}} / \sin^2 \theta = b^{\tau}_{3,{\cal M}} / \cos \theta \sin \theta = b^{\tau}_{4,{\cal M}} / \cos \theta \sin \theta = b^{\tau}_{5,{\cal M}} / \cos^2 \theta $, as in the non relativistic approximation.
 
 Our aim is 
 to obtain an expression of TMDs 
 from the
 functions $b^{\tau}_{i, {\cal M}}$, just introduced. In order to  accomplish this task,  another ingredient,
the fermion correlator for a
semi-inclusive process, {has to be added, along with} 
its relation to the LF spectral function
 This is detailed in the following section.

 \section{ The Semi-Inclusive Correlator  for a ${\cal J}=1/2$ bound system and the LF spectral function}

Let  $p$ be  the momentum  in the lab frame of an off-mass-shell
spin 1/2 {particle}, with isospin ${\tau}$,
  inside a bound system of $A$ {{spin 1/2}} {particles} with total momentum $P$ and spin $S$.
   The  {semi-inclusive} fermion correlator in terms of the LF coordinates is \cite{Barone:2001sp}
   
\bwt\be
\Phi^{\tau}_{\alpha,\beta}(p,P,S) ={ 1\over 2}\int {d\zeta^- d\zeta^+ 
d{\bma \zeta}_\perp}~e^{ip^-\zeta^+/ 2}~e^{ip^+ \zeta^-/ 2}~
e^{-i{\bf p}_\perp\cdot {\bma \zeta}_\perp}
 \langle P,S,A|\bar{\psi}^{\tau}_{\beta}(0)~{{\cal W}(\hat n \cdot {\cal A})}~\psi^{\tau}_{\alpha}(\zeta)|A,S,P  \rangle ~ ,
\label{corr1}
\ee
\ewt
where $|A,S,P  \rangle$ is the $A$-particle state (e.g. a nucleus
or a nucleon), $\psi^{\tau}_{\alpha}(\zeta)$ the particle field (e.g. a nucleon of isospin ${\tau}$
if the system is a nucleus,
or a quark if the system is a nucleon) and {${\cal W}(\hat n \cdot {\cal A})$} is  a  link operator which makes 
$\Phi^{\tau}_{\alpha,\beta}(p,P,S)$ gauge invariant. By working in the ${\cal A}^+=0$ gauge, ${{\cal W}}$ can be reduced to unity.
 Let us notice that the ${\cal A}^+$ gauge condition is preserved under Lorentz transformation in the LF dynamics \cite{Chiu:2017ycx,Brodsky:1997de}. 
 Hereafter, we shall assume that the link operator is
the unity operator. 
Using the translation invariance relation
\be
\psi^{\tau}_{\alpha}(\zeta)=e^{i\widehat{P}\cdot \zeta}~\psi^{\tau}_{\alpha}(0)~e^{-i\widehat{P}\cdot \zeta} \, ,
\ee
one can rewrite the correlator as 
\be
\Phi^{\tau}_{\alpha,\beta}(p,P,S) = { 1\over 2}~\int {d\zeta^- d\zeta^+ d{\bma \zeta}_\perp}~e^{i(p-P)\cdot \zeta}
\nonu \times~ 
\langle P,S,A | \bar{\psi}^{\tau}_{\beta}(0)~e^{i\zeta\cdot \widehat{P}}~\psi^{\tau}_{\alpha}(0)
|A,S,P  \rangle ~,
\label{corr2}
\ee
where the explicit expression for the fermion field in { $\zeta^\mu=0$} is \cite{Brodsky:1997de}
\be
\psi^{\tau}_{\alpha}(0) =\int {d\tilde{\bf p} \over 2p^+(2 \pi)^{3}}
\nonu \times ~\sum_{\sigma}
 \left [ b^{\tau}_{\sigma}({\tilde {\bf  p}})~u_{\alpha}( \tilde {\bf p},\sigma)+	     		     		     
  d^{{\tau}\dagger}_{\sigma}( \tilde {\bf p})~v_{\alpha}( \tilde {\bf p},\sigma)\right ]
  \, ,
\label{fields}
\ee
with  $\tilde{\bf p}=(p^+,{\bf p}_\perp)$ the LF particle momentum in the lab frame and $u(\tilde {\bf p},\sigma)$,
$v(\tilde {\bf  p},\sigma)$ the particle and anti-particle LF 
 spinors \cite{Brodsky:1997de,Lev:2000vm}. 
  The following spinor normalization is adopted
 \be 
\bar u( {\bf \tilde p},\sigma )   u( {\bf \tilde p},\sigma ') = 2~m ~ \delta_{\sigma\sigma '} ~.
\label{pcc}
\ee
The Fock operators satisfy the canonical
 anti-commutation relations,
\be
\left \{ b^{\tau}_\sigma(\tilde{{\bf p}}) ,b_{\sigma'}^{{\tau'}\dag}(\tilde{{\bf p'}}) \right
\}=\left \{ d^{\tau}_\sigma(\tilde{{\bf p}}) ,d_{\sigma'}^{{\tau}'\dag}(\tilde{{\bf p'}}) \right
\}
\nonu
=2p^+(2\pi)^3\delta_{\sigma\sigma'}\delta_{\tau\tau'}\delta(\tilde{{\bf p}}-\tilde{{\bf p'}}).
\label{canon}
\ee



By inserting in Eq. (\ref{corr2}) for the correlator 
$\Phi^{\tau}_{\alpha,\beta}(p,P,S)$
{the fermionic field} given in  
Eq. (\ref{fields}),
 the {particle} correlator reads
\bwt \be
\Phi^{\tau }_{\alpha,\beta}(p,P,S)
=
~\int { {d\zeta^+}\int d {\bar {\bma{\zeta}}}~e^{{i(p^--P^-) \zeta^+}/ 2} ~
e^{i({\tilde {\bf p}}-{\tilde {\bf P}})\cdot {\bar {\bma \zeta}}}  
~\int {d\tilde{\bf p}' \over 2p^{\prime +}(2 \pi)^{3}}}
\nonu
\times ~
 \int {d\tilde {\bf p}'' \over 2p^{\prime \prime+}(2 \pi)^{3}}
 \langle P,S,A |
  ~\sum _{\sigma ''}
 \left [ b^{\tau \dagger}_{\sigma ''}({\bf \tilde p''})~
 { \bar u}_\beta({\bf \tilde  p''}, \sigma '') 
 +	     		     		     
  d^{\tau}_{\sigma ''}({\bf \tilde p''})~{ \bar v}_\beta({\bf \tilde p''},\sigma '')\right]~ 
  ~e^{i\widehat{P}^-\zeta^+ /2 }~
   \nonu
\times  \sum_{\sigma'}
 \left [ e^{i(\bf{\tilde P}-\bf{\tilde p'})\cdot {\bma{\bar \zeta}}} ~
 b^{\tau}_{\sigma'}({\bf \tilde  p'})~u_{\alpha}({\bf \tilde  p'},\sigma') +	    
  e^{i(\bf{\tilde P}+\bf{\tilde p'})\cdot {\bma{\bar \zeta}}} 	~	    
 d^{\tau \dagger}_{\sigma'}({\bf \tilde  p'})~v_{\alpha}({\bf \tilde p'},\sigma')\right ]
|A,S,P \rangle ~ ,
\label{corr3}
\ee
with ${\bma {\bar \zeta}}=(\zeta^-/2,{\bma \zeta}_\perp)$. {Recall that  
 $ b^{\tau}_{\sigma'}({\bf \tilde  p'})|A,S,P \rangle$ and  $d^{\tau \dagger}_{\sigma'}({\bf \tilde  p'})|A,S,P \rangle$ are  eigenvectors of the LF-momentum operator (it is a
 kinematical one) with eigenvalues
$\tilde P-\tilde p'$ and  
$\tilde P+\tilde p'$, respectively.  } 
 
By performing the integration on 
 $\{\zeta^+/2,{\bar {\bma \zeta}}\}$,  the term in (\ref{corr3}) with $v_{\alpha}({\bf \tilde p'},\sigma')$ vanishes, since $p'^+$ cannot be negative. Then one has
\be
\Phi_{\alpha,\beta}^{\tau }(p,P,S)
= 
\sum_{\sigma'} u_{\alpha}({\bf \tilde p},\sigma')
{2 \pi \over p^+}
 \int {d\tilde{\bf p}'' \over 2p^{\prime \prime +}(2 \pi)^{3}}
~
   \langle P,S,A |
  \sum _{\sigma ''}
 \left [ b^{\tau \dagger}_{\sigma ''}(\tilde{\bf  p}'')
 { \bar u}_\beta(\tilde{\bf   p}'',\sigma '') 
 +d^\tau_{\sigma ''}({\bf \tilde p''}){ \bar v}_\beta({\bf \tilde p''}, 
  \sigma '')\right] 
  \nonu \times  ~\delta(\hat{P}^-+p^--P^-) 
 b^\tau_{\sigma '}({\bf \tilde  p}) 	     		     		     
|A,S,P \rangle
\nonu
={2 \pi \over p^+}~
\sum_\sigma \sum_{\sigma'}~ u_{\alpha}({\bf \tilde p},\sigma') ~
~
 \int {\tilde{\bf  p}'' \over 2p^{\prime\prime +}(2 \pi)^3}
  \langle P,S,A |~
  b^{\tau \dagger}_{\sigma}({\bf \tilde p''})~\delta(\hat{P}^- + p^- - P^-) ~
 b^\tau_{\sigma '}({\bf \tilde  p})~ 	     		     		     
|A,S,P \rangle
~\bar{u}_{{\beta}}( {\bf \tilde p''},\sigma) ~,
\ee
\ewt
{where in the last step  the anti-particle contribution has been eliminated, 
since the LF-momentum has to be conserved and $p''^+$ cannot be negative.
}

{In the previous equation, let us introduce  the
completeness  for the  states of $(A-1)$-particles  in valence approximation
(see also Ref.  \cite{DelDotto:2016vkh})}
\be
\hspace{-5mm} \sum_{J J_{z} \alpha} 
  \sum_{T_{S} \tau_{S} } \sumint { \rho (\epsilon) d\epsilon}
  \int {d{\blf P}_{S} \over (2 \pi)^3 2 P^+_{S}}
  \nonu 
  \hspace{-5mm} \times 
  |{\blf P}_{S};J J_{z} \epsilon, \alpha; T_{S} \tau_{S}
  \rangle_{LF} 
  ~_{LF}
  \langle 
 \tau_{S} T_{S}; 
\alpha,\epsilon J_{z}J;{\blf P}_S |  = 1.
\label{compl}
\ee
In Eq. (\ref{compl})
  ${\blf P}_S$ is the total LF momentum  of the fully interacting $(A-1)$-particle system.
 The symbol 
 ~$
 {\LARGE{ \int }} \!\!\ \! \! \! \! \!\ \! \!\!   \sum
 $
means a sum over the bound states
of the $(A-1)$ system and
an integration over the continuum.

 {For the bound states $ |{\blf P}_{S};J J_{z} \epsilon, \alpha; T_S \tau_S
  \rangle_{LF} $, the normalization adopted is }
  \be
    _{LF}\langle T' \tau'; J' J'_{z} \epsilon' \alpha', {\blf P}'_{S}
  |{\blf P}_{S};J J_{z} \epsilon \alpha; T \tau  \rangle_{LF} 
  =
   2~P_{S}^+~(2\pi)^3
   \nonu \times~\delta ^3({\blf P}_S'  -{\blf P}_S)
   ~\delta_{T',T} 
\delta_{\tau',\tau} \delta_{\alpha',\alpha} \delta_{J',J}
\delta_{J'_z,J_z} \delta_{\epsilon',\epsilon}
 ~ ,
  \label{norm2bound}
  \ee
{while   for the LF continuum states the orthogonality  reads}  (see Appendix A of Ref. \cite{DelDotto:2016vkh})
\be
_{LF}\langle T'\tau';\alpha' \epsilon' J'_z J';{\blf P}'_{S}|{\blf P}_{S};J J_z
 \epsilon \alpha; T  \tau \rangle_{LF}
 =
 2P_{S}^+(2\pi)^3\nonu \times~\delta ^3({\blf P}'_{S}  -{\blf P}_{S})
 \delta_{T',T} 
\delta_{\tau',\tau} \delta_{\alpha',\alpha} \delta_{J',J}
\delta_{J'_z,J_z}
{\delta (\epsilon'-\epsilon)\over \rho(\epsilon)} ~  .
\label{freeStateOrt}
\ee


 Then, {one can define as follows the
 {valence} contribution to the particle correlator }  
\bwt\be
\left[\Phi^{\tau}_V(p,P,S)\right]_{\alpha,\beta} = {2\pi \over p^+}\int {d{\bf \tilde{p}}'' \over 
(2 \pi)^{3}{2 {p}}^{\prime \prime+}}
~\sum_{J J_{z} \alpha} 
  \sum_{T_{S} \tau_{S} }\sum_{J' J'_{z} \alpha'}\sum_{T'_{S} \tau'_{S} }\sumint {\rho(\epsilon)  
  d\epsilon~ 
 }
\sumint {  \rho(\epsilon')  d\epsilon'~}  
 \int {d{\blf P}_{S} \over (2 \pi)^3 2 P^+_{S}}~ 
\nonu
 \times ~
 \int {d{\blf P}'_{S} \over (2 \pi)^3 2 P^{\prime +}_{S}}~
 \sum_{\sigma\sigma '}
 \left[~u_{\alpha}({\bf \tilde p},\sigma ')
 \langle P,S,A |  {\blf  p}'' \sigma\tau;
{\blf P}'_{S};J' J'_{z} \epsilon', \alpha'; T'_{S} \tau'_{S}
\rangle_{LF}~ _{LF}\langle 
 \tau'_{S}T'_{S} ; 
\alpha',\epsilon' J'_{z}J';{\blf P}'_S |
\right. 
\nonu
 \left. \times
 ~\delta(\hat{P}^-+p^--P^-)~ 
 |{\blf P}_{S};J J_{z} \epsilon, \alpha; T_{S} \tau_{S}\rangle_{LF} 
 ~ _{LF} \langle 
 \tau_{S}T_{S} ;
\alpha,\epsilon J_{z}J;{\blf P}_S ; {\blf p} \sigma ' \tau
|A,S,P\rangle ~
{ \bar u}_\beta({\bf \tilde p},\sigma)~ \right ]
\nonu
=
{2\pi \over p^+}\int {d{\bf \tilde{p}}'' \over (2 \pi)^{3}{2 {p}}^{\prime \prime+}}
\sum_{J J_{z} \alpha} 
  \sum_{T_{S} \tau_{S} }\sumint {  \rho(\epsilon)  d\epsilon~
  }  
 \int {d{\blf P}_{S} \over (2 \pi)^3 2 P^+_{S}} 
 \sum_{\sigma\sigma '}
 \Bigl[u_{\alpha}({\bf \tilde p},\sigma ')
 \langle P,S,A | {\blf  p}''\sigma \tau;
{\blf P}_{S};J J_{z} \epsilon, \alpha; T_{S} \tau_{S}
\rangle_{LF}
\nonu
\times 
 ~ \delta(P^-_S +p^--P^-)~ 
 _{LF} \langle 
 \tau_{S}T_{S}; 
\alpha,\epsilon J_{z}J;{\blf P}_S ; {\blf p} \sigma  ' \tau
|A,S,P\rangle ~
{ \bar u}_\beta({\bf \tilde p},\sigma)~ \Bigr ]~,
\label{corr4}\ee
\ewt
where the equality 
\be
\hat{b}^{\tau\dagger}_{\sigma }({\bf \tilde p''})~
 |{\blf P}'_{S};J' J'_{z} \epsilon', \alpha'; T'_{S} \tau'_{S}
\rangle_{LF}
\nonu= | {\blf  p''} \sigma\tau;
{\blf P}'_{S};J' J'_{z} \epsilon', \alpha'; T'_{S} \tau'_{S}~,
\rangle_{LF}
\label{creaz}\ee
has been used, i.e. a free particle $| {\blf  p''} \sigma\tau\rangle$, with momentum ${\blf  p''}$
in the lab frame, has been created. Moreover, it has been taken into account
that 
the operator $\hat P^-$ acts on $ |{\blf P}_{S};J 
J_{z} \epsilon, \alpha;T_{S}, \tau_{S}\rangle_{LF}$
as follows
\be
\hat P^-~ |{\blf P}_{S};J J_{z} \epsilon, \alpha; T_{S} 
\tau_{S}\rangle_{LF}~=P^-_S~ |{\blf P}_{S};J J_{z} \epsilon, \alpha; T_{S} 
\tau_{S}\rangle_{LF}~
\nonu
=~{M^2_S+|{\bf P}_{S\perp}|^2 \over  P^+_S}~
 |{\blf P}_{S};J J_{z} \epsilon, \alpha; T_{S} 
\tau_{S}\rangle_{LF}~,
\label{Pmeno}
\ee
with $M_S$
 the mass of the interacting $(A-1)$-particle system (for $A-1= 2$, one has $M_S^2 = 4m^2+4m\epsilon$).

By considering that the LF-momentum is conserved (the interaction is contained 
only in the minus component of the momenta) and  the kinematical nature of the LF-boosts, {one has the  exact separation of  the intrinsic 
dof's  from the CM one}
(see Appendix A of \cite{Pace:2020ned}), obtaining 
\be
 |{\blf p}\sigma\tau;{\blf P} _{S}; J J_{z} 
\epsilon, \alpha; T_{S}, \tau_{S}\rangle_{LF}
 = \sqrt{{ E_S \over {\cal{M}}_0[1,(A-1)]}}
 \nonu  \times  ~|{\blf p} + {\blf P} _{S}\rangle_{LF} ~ |{\bma{\blf \kappa}}\sigma\tau; J J_{z} 
\epsilon, \alpha; T_{S} \tau_{S}\rangle_{LF}~  ,
\label{ortimpT1} 
\ee
 where  $|{\blf p} + {\blf P} _{S}\rangle_{LF}$ is the total LF momentum eigenstate of the cluster $[1, (A-1)]$.
The intrinsic state $\ |{\bma{\blf \kappa}} \sigma\tau; J J_{z} 
\epsilon, \alpha; T_{S} \tau_{S}\rangle_{LF}$ is composed by 
a fully-interacting intrinsic state of ~$(A-1)$ constituents and
 a plane wave, describing a constituent that freely
moves in the {\it intrinsic frame of the whole cluster} $[1, (A-1)]$
 {with LF-momentum ${\bma{\blf \kappa}}$ (see Eq. \eqref{kappa}).}


In Eq. (\ref{ortimpT1}) one has $E_S = \sqrt{M_S^2 + |{\bma \kappa}|^2}$, 
and ${\cal{M}}_0[1,(A-1)]$ is defined 
by Eq. (\ref{calM}).
 The factor $\sqrt{{ E_S / {\cal{M}}_0[1,(A-1)]}}$ 
 takes care of the proper normalization
of the momentum eigenstates $|{\blf p} \rangle_{LF}$, $| {\blf P}_{S}\rangle_{LF}$ 
and $|{\blf p} + {\blf P} _{S}\rangle_{LF}$ (see Ref. \cite{Pace:2020ned}).

{Summarizing, the following overlap, present  in Eq. \eqref{corr4}, can be written as follows } 
 \be
\langle P,S,A|{\blf p}\sigma\tau;{\blf P} _{S}; J J_{z} 
\epsilon, \alpha; T_{S} \tau_{S}\rangle_{LF}
 \nonu= 2P^+ (2\pi)^3 ~\sqrt{{ E_S \over {\cal{M}}_0[1,(A-1)]}}~
 ~\delta^3({\blf P} -{\blf P}_{S}-{\blf p}) 
\nonu \times ~\langle  A,S,{int}|{\bma {\blf \kappa}}\sigma\tau; J J_{z} 
\epsilon, \alpha; T_{S} \tau_{S}\rangle_{LF}~,
\label{ort1} 
\ee
where 
the orthogonality of the plane waves is given by  
 $\langle{\bf \tilde P}|{\bf \tilde P'}\rangle=2P^+(2\pi)^3
\delta({\bf \tilde P}-{\bf \tilde P'})$ and $ |int,S,A\rangle \equiv |{\psi}_{\cal{J}\cal{M}};S, T T_z \rangle$ 
 is the intrinsic eigenstate of the system.

The normalization for the intrinsic overlaps  
$_{LF} \langle \tau_{S}T_{S} ; \alpha,\epsilon; J_{z}J ; {\bma {\blf \kappa}} \sigma  \tau
 |{int},S,A \rangle$ (see Eq. (\ref{normFSLF1}))
reads
\be 
\int {d{ \tilde{\bma \kappa}}\over 2\kappa^+ 
(2\pi)^3}  ~\sumint {\rho(\epsilon) d\epsilon} 
 ~ \sum_{\sigma}\sum_{T_{S} \tau_{S} }\sum_{J J_{z} \alpha}
 \nonu \times  _{LF} \langle 
 \tau_{S}T_{S} ; 
\alpha,\epsilon J_{z}J ; {\bma {\blf \kappa}} \sigma  \tau
 |{int},S,A \rangle |^2 =1~.
\label{norm1}
\ee
{Eventually}, with the help of Eq. (\ref{ort1}) 
the  general expression for the valence contribution to the semi-inclusive fermion  correlator 
{becomes}
\bwt\be
\left[\Phi^{\tau}_V(p,P,S)\right]_{\alpha,\beta} =
2\pi~\left({ P^+\over p^+}\right)^2
~\sum_{J J_{z} \alpha} 
  \sum_{T_{S} \tau_{S} }\sumint {  \rho(\epsilon)  d\epsilon~ {\delta(P^-_S +p^--P^-) \over (P^+ - p^+)}
  }  { E_S \over {\cal{M}}_0[1,(A-1)]} 
~ \quad \quad  \quad  \quad  \quad 
\nonu
\times
 ~\sum_{\sigma\sigma '}
 \left [~u_{\alpha}( \tilde {\bf p},\sigma ')
\langle  A,S,{int}|{\bma {\blf \kappa}}\sigma\tau; J J_{z} 
\epsilon, \alpha; T_{S} \tau_{S}\rangle_{LF}
 ~ _{LF} \langle 
 \tau_{S}T_{S} ;
\alpha,\epsilon J_{z}J ; {\bma {\blf \kappa}} \sigma ' \tau
 |{int},S,A \rangle_{LF}  { \bar u}_\beta( \tilde {\bf p},\sigma)~\right ]~
~, 
\label{corr7}
\ee
{where $P^-_S= (M^2_S +|{\bf P}_{S\perp}|^2)/P^+_S$,   (see  Eq. \eqref{Pmeno}). Once the mass $M_S$ is expressed 
in
terms of the intrinsic energy,  the integration on $\epsilon$ can be easily performed, obtaining a relation between the correlator and the spin-dependent LF spectral function defined
in Eq. (\ref{LFspf})}.
\ewt
In the case where $(A-1)=2$ 
and in the reference frame where ${\bf P}_{\perp}=0$, the $\delta$ function in Eq. (\ref{corr7}) implies the equation
\be
{M^2 \over P^+} ~ = ~ p^- ~ + ~ {4 m^2 + 4m\epsilon + |{\bf P}_{S\perp}|^2  \over P^+ - p^+ } ~,
\label{epsi1}
\ee
i.e.
\be
\epsilon = 
{({M^2 \over P^+} ~ - ~ p^-)~ (P^+ - p^+) ~ - |{\bf P}_{S\perp}|^2  \over 4m } - m
\label{epsi}
\ee
with ${\bf P}_{S\perp} = - {\bf p}_{\perp}= - {\bma \kappa}_{\perp} $.
Then for $A=3$ one obtains
\be
\left[\Phi^{\tau}_V(p,P,S)\right]_{\alpha,\beta} ~
=~{2\pi ~ (P^+)^2 \over (p^+ )^2 ~ 4m } ~{ E_S \over {\cal{M}}_0[1,(23)]}
~\sum_{\sigma\sigma '}
\nonu \times~\left \{~u_{\alpha}({\tilde {\bf p}},\sigma')~
{\cal P}^{\tau}_{{\cal M},\sigma'\sigma}({ \tilde {\bma \kappa}},\epsilon,S)~
{ \bar u}_\beta({\tilde {\bf p}},\sigma)~ \right \} ~.
\label{abc}
\ee
{Due to Eq. (\ref{pcc}),} the following equation holds :
\be
{\bar{u}_{}( {\tilde{\bf p}},\sigma')
\Phi^{\tau }_V (p,P,S)u_{}({\tilde {\bf p}},\sigma)}
\nonu
 = 
{2\pi~m ~ E_S \over {\cal{M}}_0[1,(23)]} ~ \left ({P^+ \over p^+} \right )^2
 ~
{\cal P}^{\tau}_{{\cal{M}},\sigma'\sigma}({ \tilde {\bma \kappa}},\epsilon,S) ~.
\label{iden}
\ee

From Eq. (\ref{abc}), using the relation (see Appendix C of Ref. \cite{Lev:2000vm})
\be
\bar{u}_{}( {\tilde {\bf p}}',\sigma')~\gamma^+~u_{}({\tilde {\bf p}},\sigma)=
\delta_{ \sigma' \sigma}~2~\sqrt{{p'}^+ p^+} ~ ,
\label{gammapiu}
\ee
one has
\be
{\Tr}\Bigl[\gamma^+\Phi^{\tau }_V(p,P,S)\Bigr]
= { \pi ~ {(P^+)^2} \over m ~ p^+}  
{E_S \over {\cal{M}}_0[1,(23)]} 
\nonu \times ~
\sum_{\sigma} {\cal P}^{\tau}_{{\cal{M}},\sigma\sigma}({\tilde {\bma \kappa}},\epsilon,S)  ~ .
\ee
Therefore, since (see Ref. \cite{DelDotto:2016vkh})
\be
{\partial x \over \partial \kappa_z}={(1- x) ~ \kappa^+ \over E_S ~ E({\bf \kappa})}
\label{derparz}\ee
and (see Eq. (\ref{epsi1}))
\be
{\partial p^- \over \partial \epsilon}= - {4 m\over(P^+ - p^+)} ~ ,
\label{derparz1}
\ee
 one obtains (recall that $x = p^+/P^+$ and ${\bf p}_\perp = {\bma \kappa}_\perp$) 
\be
\frac{1}{2(2\pi)^4}\int dp^-\int {{dp^+\over 2P^+}}\int d{\bf p}_\perp~Tr\Bigl[\gamma^+\Phi^{\tau }_V (p,P,S)\Bigr] 
 \nonu
= \frac{1}{(2\pi)^4}\int d\epsilon   
 \int d{\bma \kappa}  {  \pi \over   E({\bf \kappa})}  
\sum_{\sigma} {\cal P}^{\tau}_{{\cal M},\sigma\sigma}({\tilde {\bma \kappa}},\epsilon,S) = 1  ,
\ee
because of the spectral function normalization (see Eq. (\ref{normFSLF1})).

Eventually we have the normalization condition for the particle correlator
\be
\int \frac{d^4p}{(2\pi)^4}\frac{1}{2P^+}~ {\Tr}\Bigl[\gamma^+\Phi^{\tau }_V(p,P,S)\Bigr]= \frac{1}{2P^+}
\frac{1}{(2\pi)^4}
\nonu \times \frac{1}{2}\int dp^-\int dp^+\int d{\bf p}_\perp {\Tr}\Bigl[\gamma^+\Phi^{\tau }_V(p,P,S)\Bigr] 
= 1 .
\label{gpb}
\ee

For a generic value of $A$, 
the above equations can be easily generalized.

 \section{{{T-even twist-two}} Transverse Momentum Distributions}
 
 As shown in Appendix \ref{corre},  in valence approximation the leading-twist TMDs are related to the scalar functions
  ${{b}}_{i,{\cal M}}^{}$, that contain the relevant information on the dynamics inside the bound system, by the equations

\be
 f(x, |{\bf p}_{\perp}|^2 ) ~ = ~ {{b}}_{0}^{} \, ,
 \label{Ai1}
\ee
\be
 S_z \, \Delta f +
  \frac1{M} \, {\bf{p}_\perp{\cdot}\bf{S}_\perp } \, g_{1T} \,   
   =  ~ \Bigl[ S_{z}  ~
 {b}_{1,{\cal M}}
 \nonu
 + ~({\bf  S} \cdot \hat {\bf k}_{\perp})~
  {b}_{4,{\cal M}} ~
  + ~
 ({\bf  S} \cdot \hat {z})~ 
 {b}_{5,{\cal M}} 
\Bigr ] \, ,
\label{Ai2}
\ee
\be
    S_x ~ h_{1T} +
  \frac{S_z}{M}\, p_x\, h^\perp_{1L} +
  \frac{{\bf{p}_\perp{\cdot}\bf{S}_\perp}}{M^2} \,  \, p_x \, 
h^\perp_{1T} 
 = 
\Bigl [ {S}_x ~
 {b}_{1,{\cal M}}
 \nonu
 + ~{ k_x \over k_{\perp}}
~({\bf  S} \cdot \hat {\bf k}_{\perp})~
{b}_{2,{\cal M}}  
~ + ~{ k_x \over k_{\perp}} ~({\bf  S} \cdot \hat z)~ 
{b}_{3,{\cal M}} 
\Bigr ] \, ,
\label{Ai3}
\ee
\be
    S_y  ~ h_{1T} +
  \frac{S_z}{M}\, p_y\,  h^\perp_{1L}  +
  \frac{{\bf{p}_\perp{\cdot}\bf{S}_\perp}}{M^2} \,  \, p_y \, 
h^\perp_{1T}  
= 
 \Bigl [ S_{y} ~
 {b}_{1,{\cal M}}
 \nonu
 + ~
{ k_y \over k_{\perp}} ~({\bf  S} \cdot \hat {\bf k}_{\perp})~
{b}_{2,{\cal M}}  
~ + ~{ k_y \over k_{\perp}} ~({\bf  S} \cdot \hat z)~ 
{b}_{3,{\cal M}} 
\Bigr ] ~ ,
\label{Ai3a}
\ee
where  
(see Ref. \cite{Goeke:2005hb})
\be
 h_{1T} = \int \Bigl[d{ p^+} d{ p^-}\Bigr] ~  A^V_3 \, .
\label{a3v}
\ee
{{Let us recall that for a three-body system with total angular momentum ${\cal J} = 1/2$, the dependence of 
${{b}}_{i,{\cal M}}^{}$ ($i=0,...,5$) on ${\bf S}$ is absent}}
{{and that the ${{b}}_{i,{\cal M}}^{}$ are invariant for rotations of ${\bf {k}}_\perp$ around the $z$ axis
(see Appendix \ref{mdistr}).}} Then any dependence on ${\bf S}$ and on the direction of 
${\bf {k}}_\perp$ in the right-hand side of Eqs. \eqref{Ai2}, \eqref{Ai3}, \eqref{Ai3a} is explicitly written down.
 
  {To obtain the explicit  expressions of the TMDs in terms of the scalar functions  
  ${{b}}_{i,{\cal M}}^{}$ from Eqs. (\ref{Ai2}), (\ref{Ai3}) and (\ref{Ai3a})}, we consider specific orientations of the target spin.
%
From Eq. (\ref{Ai2}) one obtains : 
\begin{itemize}
\item
{for ${\bf S} = (0,0,1)$ 
\be
\Delta f  =  
{b}_{1,{\cal M}}^{}
 + ~ {b}_{5,{\cal M}}^{} ~ ,
\label{Sz}
\ee
}
\item
{for ${\bf S} = (1,0,0)$ and  ${\bf S} = (0,1,0)$
\be
g_{1T} = {{M} \over  |{\bf p}_{\perp}|}  ~ 
{b}_{4,{\cal M}}^{} ~ .
\label{Sx}
\ee

}
\end{itemize}
From Eq. (\ref{Ai3}) one obtains  
\begin{itemize}
\item
for ${\bf S} = (0,0,1)$ 
\be
h^\perp_{1L} = {M \over |{\bf p}_{\perp}|}  ~
 {b}_{3,{\cal M}}^{}~,
\label{Sz6}
\ee

\item
for ${\bf S} = (1,0,0)$ 
\be
{{h_{1T}(S_x=1)}} + {|{\bf p}_{\perp}|^2 cos^2 \phi  \over M^2} ~h^\perp_{1T}
\nonu =     
{b}_{1,{\cal M}}^{}  + \left ({ k_x \over k_{\perp}} \right )^2 
{b}_{2,{\cal M}}^{} ~,
\label{Sx6}
\ee

\item
for ${\bf S} = (0,1,0)$ 
\be
h^\perp_{1T} =  { M^2 \over |{\bf p}_{\perp}|^2}      ~
{b}_{2,{\cal M}}^{} ~.
\label{Sy8}
\ee

\end{itemize}

Eventually, from Eq. (\ref{Ai3a}) one has
\begin{itemize}
\item
for ${\bf S} = (0,1,0)$ 
\be
{{{h_{1T}(S_y=1)}} }+ {|{\bf p}_{\perp}|^2 sen^2 \phi  \over M^2} ~h^\perp_{1T}
\nonu =     
{b}_{1,{\cal M}}^{}   + \left ({ k_y \over k_{\perp}} \right )^2 
{b}_{2,{\cal M}}^{} ~,
\label{Sy77}
\ee

\item
for  ${\bf S} = (1,0,0)$ an equation identical to (\ref{Sy8})  is obtained,

\item
for  ${\bf S} = (0,0,1)$ an equation identical to (\ref{Sz6}) is obtained.

\end{itemize}
If Eq. (\ref{Sy8}) is inserted in Eqs. \eqref{Sx6} and  \eqref{Sy77} one obtains
\be
{{{h_{1T}}}} = ~{b}_{1,{\cal M}}^{} ~.  
\label{Sy771}
\ee



The sum of Eqs. (\ref{Sx6}) and  (\ref{Sy77}) gives
\be
\hspace{-5mm} h_{1T} + {|{\bf p}_{\perp}|^2 \over 2 M^2} ~ h^\perp_{1T} = 
 \Delta'_T   f
= { 1  \over 2} ~ 
\Bigl ( ~ 2 ~ {b}_{1,{\cal M}}^{} ~  +  
~
{b}_{2,{\cal M}}^{} ~  \Bigr) ~.
\label{A3Vp1}
\ee
In conclusion, from Eqs. (\ref{Ai2}), (\ref{Ai3}) and (\ref{Ai3a}) and the three possible directions of the polarization vector ${\bf S}$, nine equations are obtained. However only five out of these nine equations are independent and allow one to determine the five TMDs, 
$\Delta f, g_{1T}, h^\perp_{1L}, h^\perp_{1T,}$, and $\Delta '_T f$.
Summarizing our results, {we can write the following expressions for the leading-twist TMDs in valence 
approximation (recall ${\bf p}_{\perp}={\bf k}_{\perp}={\bma \kappa}_{\perp}$)} 

\be
f^\tau(x, |{\bf p}_{\perp}|^2 )
   =  {{b}}_{0}^{\tau}~,
    \label{Tmd1}  
  \\  &&
\Delta f^\tau(x, |{\bf p}_{\perp}|^2 )
   = 
   {b}_{1,{\cal M}}^{\tau} + ~ {b}_{5,{\cal M}}^{\tau} ~,
 \label{Tmd2} 
  \\ &&
 g^\tau_{1T}(x, |{\bf p}_{\perp}|^2 )
= {{M} \over  |{\bf p}_{\perp}|} ~ {b}_{4,{\cal M}}^{\tau}~,
\label{Tmd3}
\\ &&
\Delta'_T   f^\tau(x, |{\bf p}_{\perp}|^2 )
= { 1  \over 2} ~ 
\Bigl( ~ 2 ~ {b}_{1,{\cal M}}^{\tau} ~  +  
~
{b}_{2,{\cal M}}^{\tau} ~  \Bigr)~,
\label{Tmd4}
\\ &&
h^{\perp\tau}_{1L}(x, |{\bf p}_{\perp}|^2 )
= {M \over |{\bf p}_{\perp}|} ~ {b}_{3,{\cal M}}^{\tau} ~ ,
\label{Tmd5}
\\ &&
h^{\perp\tau}_{1T}(x, |{\bf p}_{\perp}|^2 )
   =  { M^2 \over |{\bf p}_{\perp}|^2}   ~
{b}_{2,{\cal M}}^{\tau} ~,
\label{Tmd6}
\ee
where the isospin index $\tau$ has been reintroduced.

Since for a three-body system with total angular momentum ${\cal J} = 1/2$, the dependence of 
${{b}}_{i,{\cal M}}^{}$ 
on ${\bf S}$ is absent
and the ${{b}}_{i,{\cal M}}^{}$ are invariant for rotations of ${\bf {k}}_\perp$ around the $z$ axis,
the transverse momentum distributions $\Delta f(x, |{\bf p}_{\perp}|^2 )$,  $g_{1T}(x, |{\bf p}_{\perp}|^2 )$, 
$ \Delta'_T   f^\tau(x, |{\bf p}_{\perp}|^2 )$, $ h^{\perp}_{1L}(x, |{\bf p}_{\perp}|^2 )$, 
$ h^{\perp}_{1T}(x, |{\bf p}_{\perp}|^2 )$ do not depend on the direction of ${\bf k}_\perp$, as expected.


{{
In Appendix \ref{sdmd},
explicit expressions for
the 
{{functions}} ${b}_{i,{\cal M}}^{\tau}$ 
are obtained in terms
of the three-body wave function. From these expressions,
according to Eqs. (\ref{Tmd1}-\ref{Tmd6}),
the twist-two T-even transverse momentum distributions can be
evaluated, directly
from the wave function,
without a cumbersome
analysis
of the spectral properties
of the system, described
by the spectral function.
}}

\section{Applications}
{In this Section, the above  formalism developed for the valence contribution to the leading-twist TMDs  for a ${\cal J}=1/2$ system is applied to the  proton and neutron inside 
$^3$He. The quantitative analysis allows one to clearly show the impact of the inner dynamics on the evaluation of the TMDs, confirming the
expectation of reaching a detailed 3D picture  of the investigated system
{{in momentum space}}. Moreover, the numerical information we have obtained could be
exploited for motivating further experimental efforts for measuring the $^3$He TMDs.}

\subsection{ TMDs of the $^3$He nucleus}
 
{The T-even TMDs for $^3$He are evaluated by using 
the $^3$He wave function of Ref. \cite{Kievsky:1994mxj,Kievsky:1995uk}
 with the realistic nuclear interaction of
Ref.\cite{Wiringa:1994wb}, but neglecting the small effect of the 
Coulomb repulsion between the protons.
The results
are shown in Figs. \ref{fig1} to \ref{fig7}.}
\begin{figure*}
\begin{center}
\includegraphics[width=8.50cm]{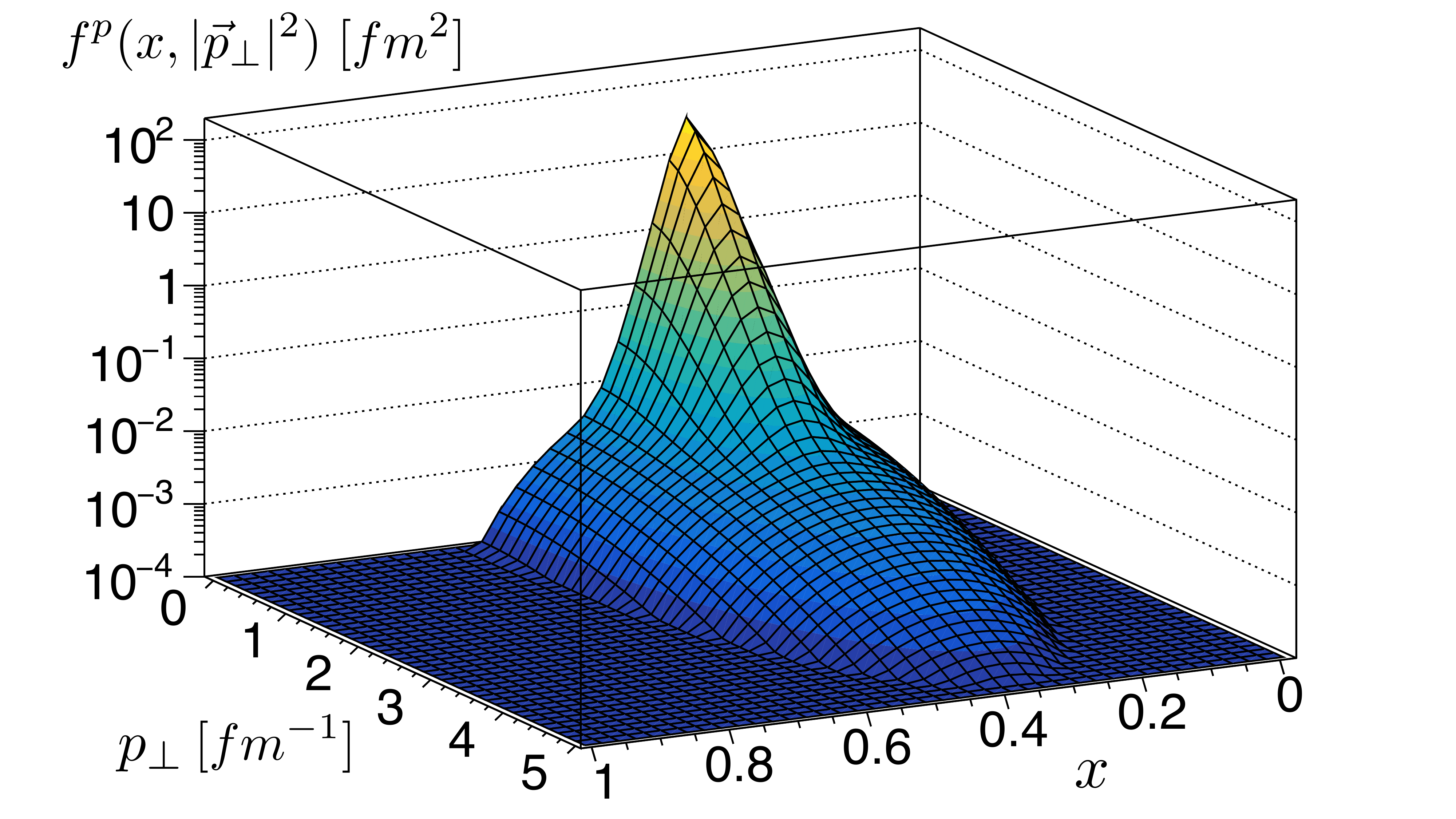}
\includegraphics[width=8.50cm]{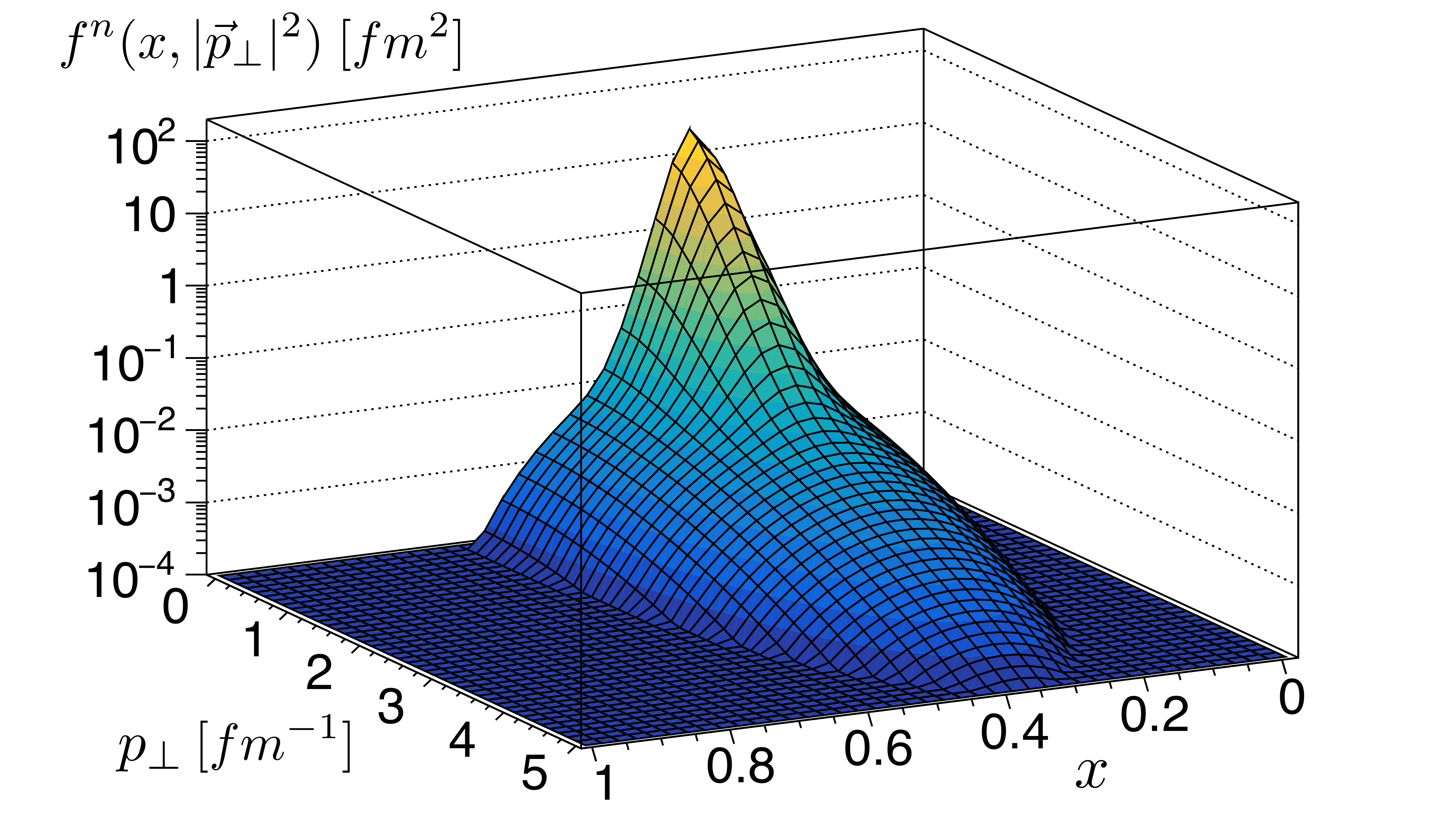}
 \caption{(color online) Nucleon momentum distribution $f^\tau(x, |{\bf p}_{\perp}|^2 )$ in an unpolarized $^3$He for the  proton 
 (left panel) and the neutron (right panel).} \label{fig1}
 \end{center}
\end{figure*}
\begin{figure*}
\begin{center}
\includegraphics[width=8.50cm]{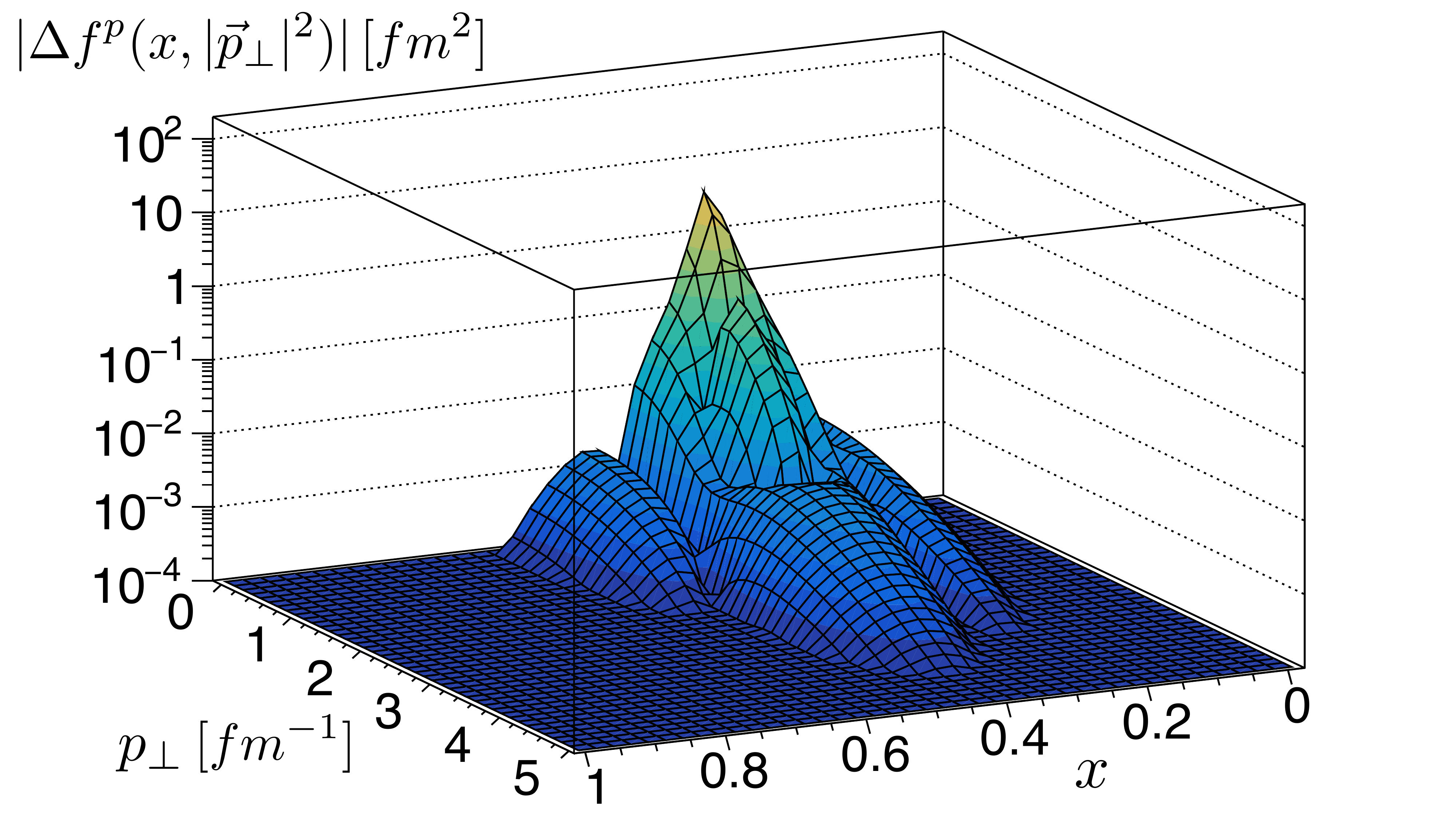}
\includegraphics[width=8.50cm]{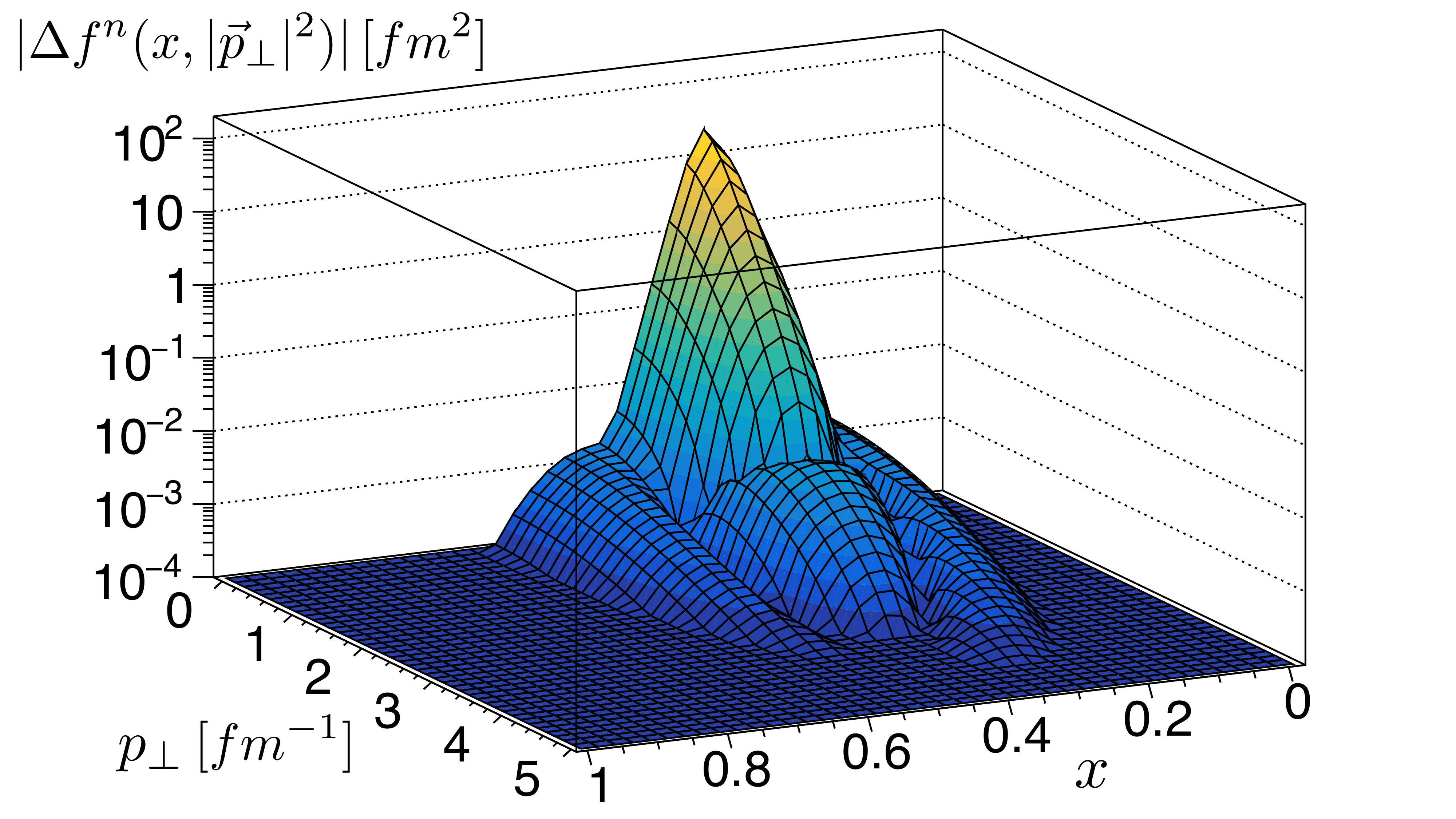}
 \caption{(color online) {Absolute value of the  nucleon longitudinal-polarization distribution}, 
 $\Delta f^\tau(x, |{\bf p}_{\perp}|^2 )$, in a longitudinally polarized $^3$He  for the  proton (left panel) 
 and the neutron (right panel).}
\label{fig2}
\end{center}
\end{figure*}
\begin{figure*}
\begin{center}
\includegraphics[width=8.50cm]{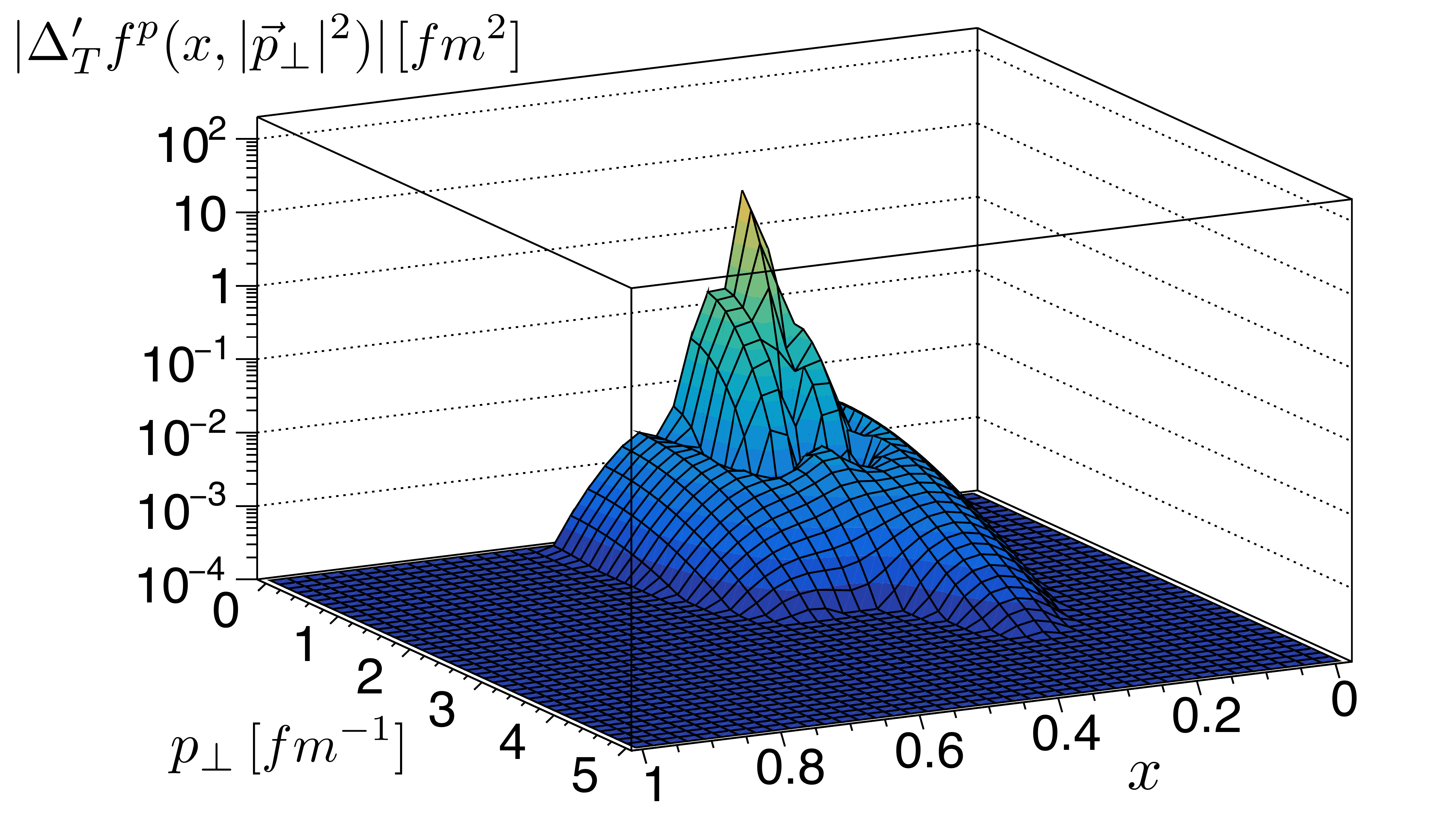}
\includegraphics[width=8.50cm]{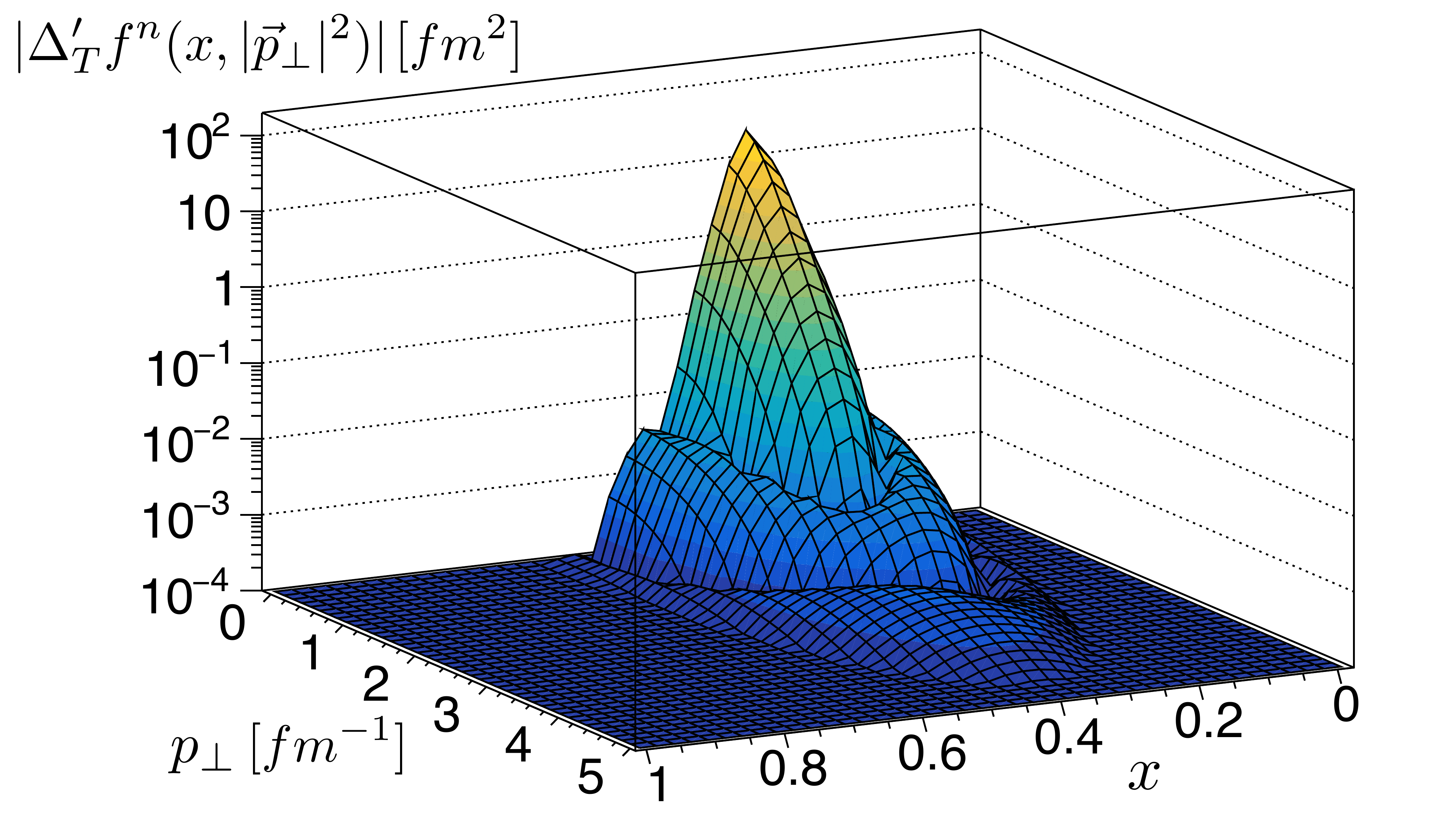}
 \caption{(color online) {Absolute value of the nucleon transverse-polarization  distribution,} 
 $\Delta'_T   f^\tau(x, |{\bf p}_{\perp}|^2 )$, in a $^3$He nucleus transversely polarized  in the same
  direction of the nucleon polarization, for the  proton (left panel) and the neutron (right panel).}
\label{fig3}
\end{center}
\end{figure*}
\begin{figure*}
\begin{center}
\includegraphics[width=8.50cm]{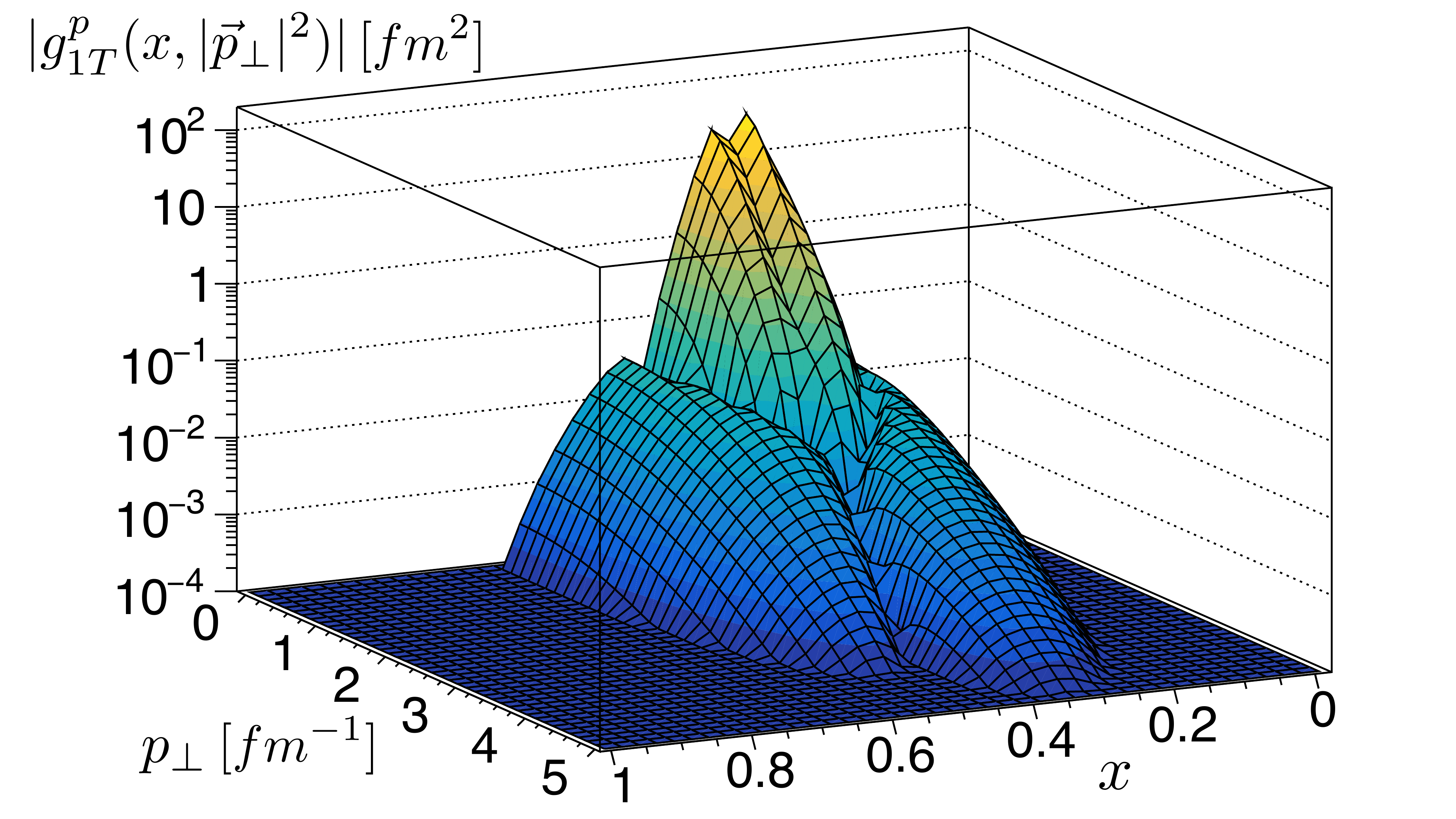}
\includegraphics[width=8.50cm]{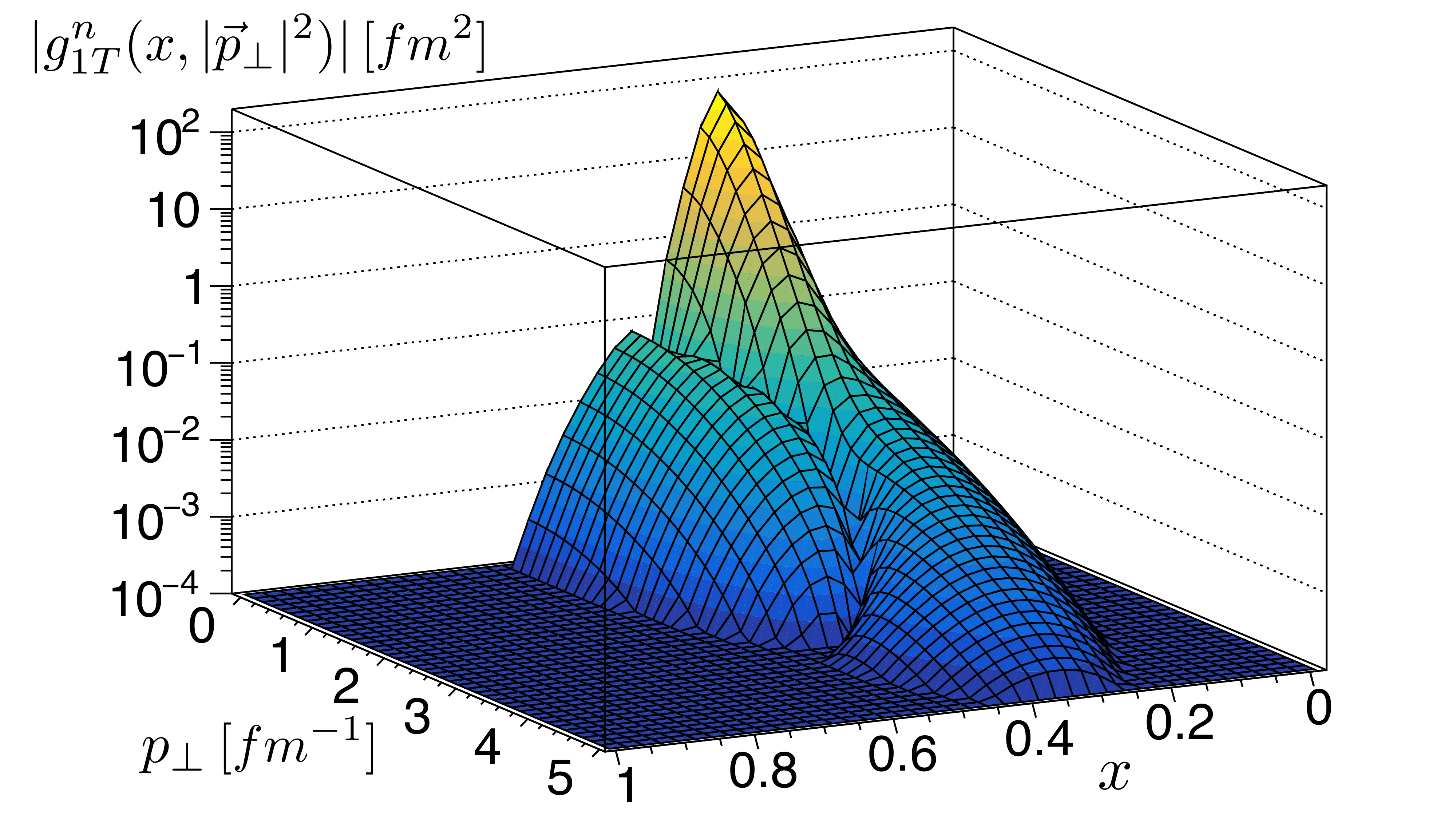}
 \caption{(color online) {Absolute value of the nucleon  longitudinal-polarization distribution,} 
  $ g^\tau_{1T}(x, |{\bf p}_{\perp}|^2 )$, 
 in a transversely polarized $^3$He  for the  proton (left panel) and the neutron (right panel).}
\label{fig5}
\end{center}
\end{figure*}
\begin{figure*}
\begin{center}
\includegraphics[width=8.50cm]{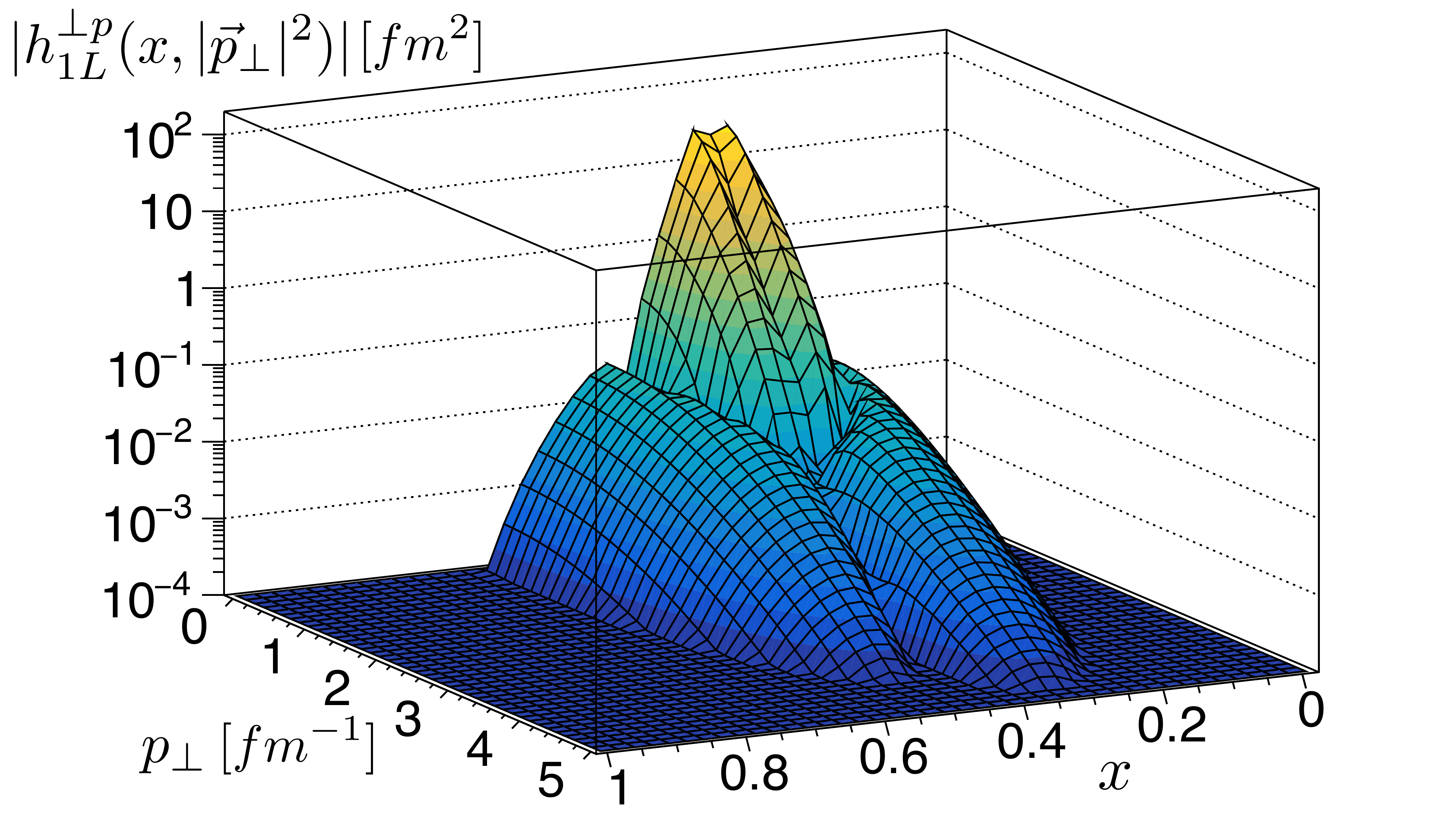}
\includegraphics[width=8.50cm]{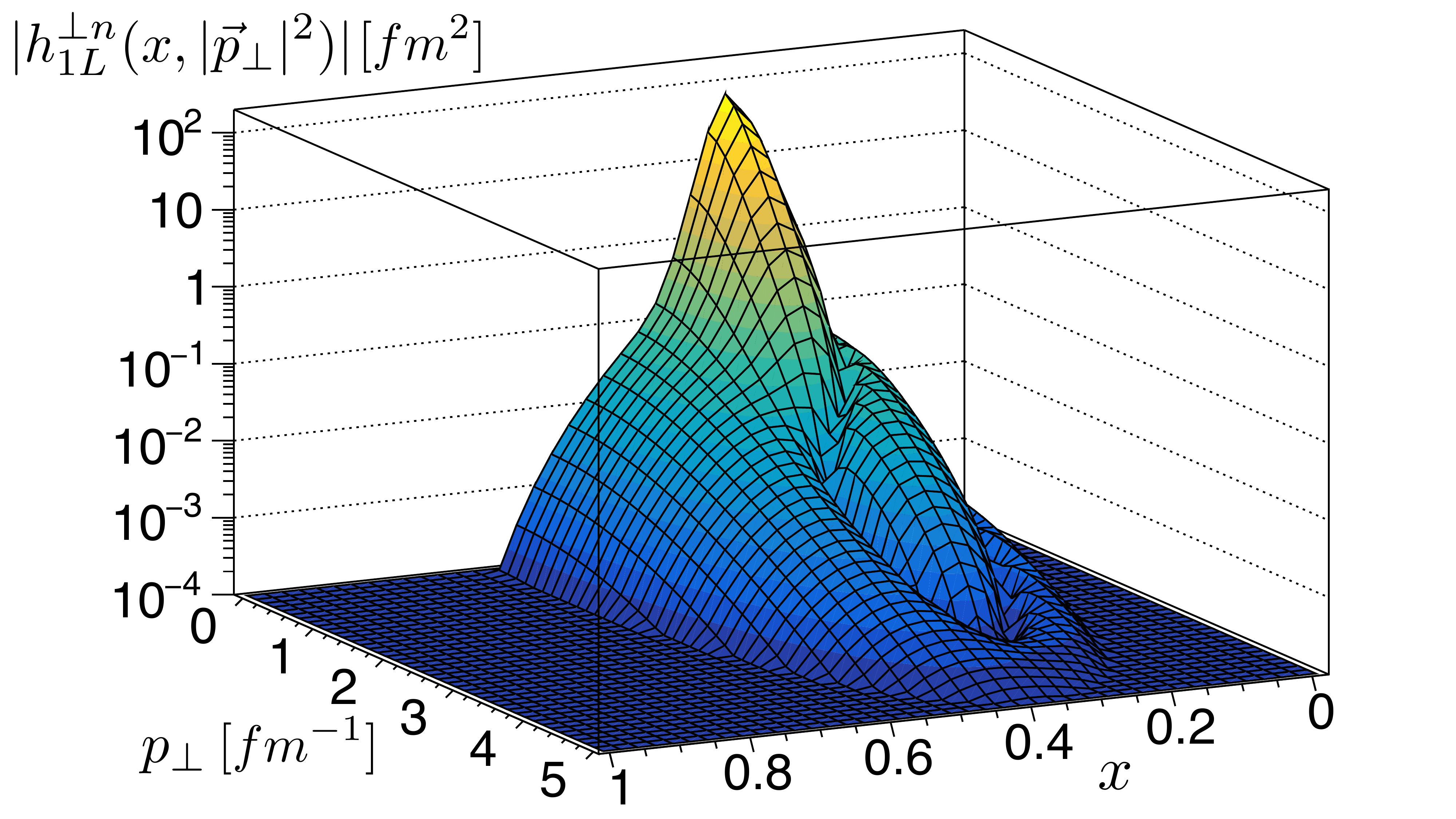}
\caption{(color online) {Absolute value of the nucleon transverse-polarization  distribution,}
$h^{\perp\tau}_{1L}(x, |{\bf p}_{\perp}|^2 )$ in 
a longitudinally polarized $^3$He for the  proton (left panel) and the neutron (right panel).}
\label{fig6}
\end{center}
\end{figure*}
\begin{figure*}
\begin{center}
\includegraphics[width=8.50cm]{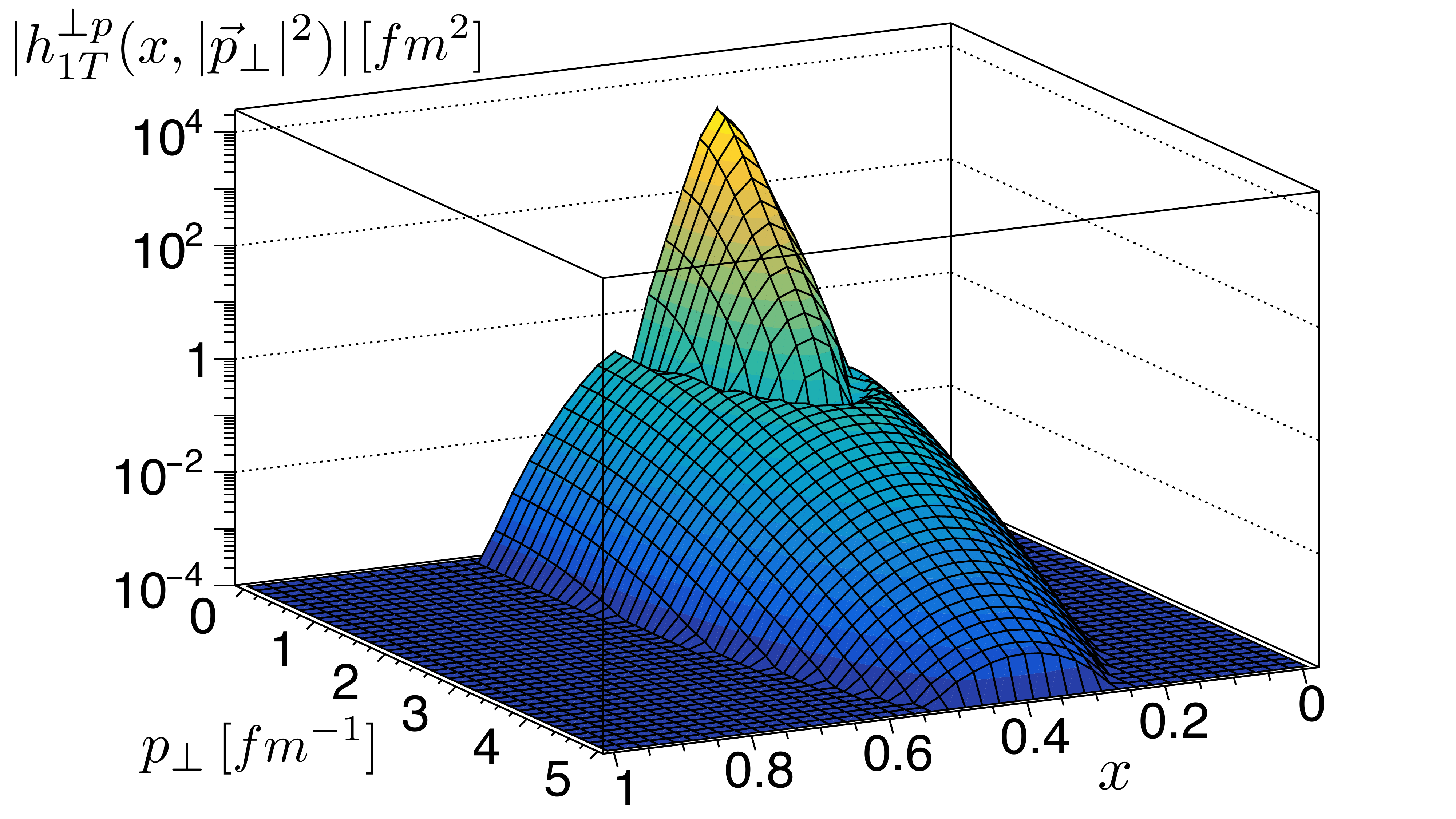}
\includegraphics[width=8.50cm]{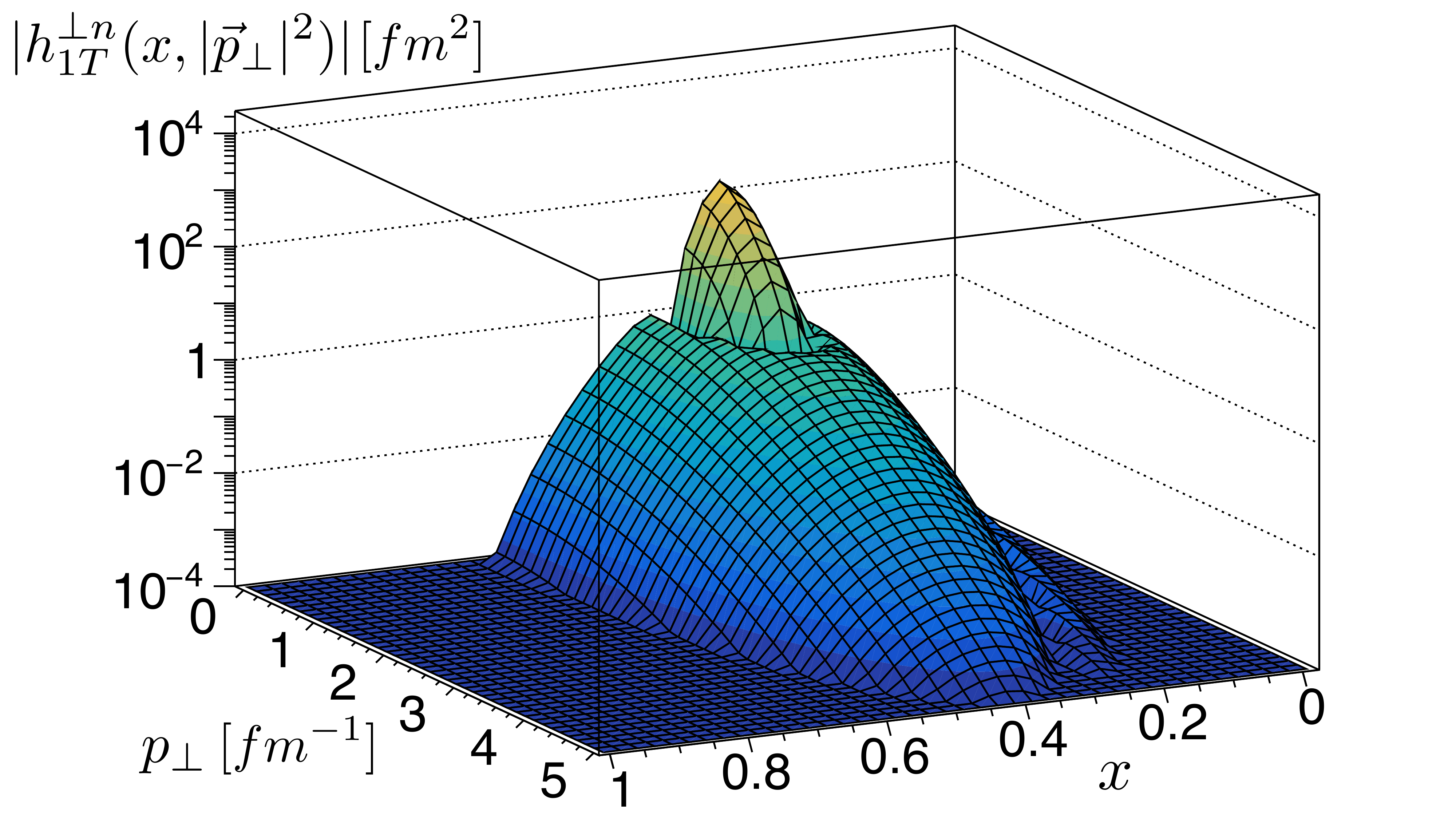}
 \caption{(color online)  {Absolute value of the nucleon transverse-polarization  distribution, }
 $h^{\perp\tau}_{1T}(x, |{\bf p}_{\perp}|^2 )$, 
 in a $^3$He nucleus transversely polarized  in a direction orthogonal to the direction of the nucleon polarization, 
 for the  proton (left panel) and the neutron (right panel).}
\label{fig7}
\end{center}
\end{figure*}

As a first observation one can notice that all of the TMDs are distributed around $x=1/3$, as expected.
The structure of $f^\tau(x, |{\bf p}_{\perp}|^2 )$, {presented in Fig. \ref{fig1}}, is very 
{{smooth}}, while the other five distributions have a rich structure, as a function both of $x$ and 
  $|{\bf p}_{\perp}|$.
{{Another relevant observation arises
from the comparison of the TMDs
$\Delta   f^\tau(x, |{\bf p}_{\perp}|^2 )$,
Eq. \eqref{Tmd2},
{{with}}
$\Delta'_T   f^\tau(x, |{\bf p}_{\perp}|^2 )$, Eq. \eqref{Tmd4}, shown in Figs. \ref{fig2} and
  \ref{fig3}, respectively.
As  is well known, these distributions
should be equal in a non relativistic framework, where boosts and transverse rotations commute 
\cite{Jaffe:1991kp,Jaffe:1996zw,
Barone:1996un,Scopetta:1997qg}. The same does not hold {in
 a relativistic treatment, as
the LFHD one adopted here. Indeed, the  comparison shows  that the two distributions are actually different, 
a signature of 
{remarkable} relativistic corrections, even in a system, the $^3$He nucleus, where the ratio of 
the average nucleon momentum to the nucleon mass is rather small. 

{The  TMDs presented in Figs. \ref{fig5}, \ref{fig6} and  \ref{fig7}, namely $ g^\tau_{1T}(x, |{\bf p}_{\perp}|^2 )$, $h^{\perp\tau}_{1L}(x, |{\bf p}_{\perp}|^2 )$
and $h^{\perp\tau}_{1T}(x, |{\bf p}_{\perp}|^2 )$, respectively, have relevant peaks at low values of 
 $|{\bf p}_\perp|$. 
  }
 Interestingly, $h^{\perp\tau}_{1T}(x, |{\bf p}_{\perp}|^2 )$ shows a
sizable  secondary bump at $|{\bf p}_\perp|\sim 2.5$ ${ 
fm^{-1}
}$, due to the presence of the squared transverse-momentum in 
${b}_{2,{\cal M}}^{\tau}$, differently  from the linear power
 occurring in the other functions ${b}_{i,{\cal M}}^{\tau}$.  It is very important to notice that in valence
 approximation $|g^\tau_{1T}(x, |{\bf p}_{\perp}|^2 )| $, shown in Fig. \ref{fig5}, and $|h^{\perp\tau}_{1L}(x, |{\bf
 p}_{\perp}|^2 )|$, shown in Fig. \ref{fig6}, are
 {very similar}, as expected by inspecting   the two scalar functions, $b^\tau_{3,{\cal M}}$ and 
 $b^\tau_{4,{\cal M}}$  given in  Eqs. \eqref{b30},  \eqref{b32}, \eqref{b40} and   \eqref{b42}, and recalling that the effect of the Melosh
 rotations, parametrized through the angle $\varphi$, is small (see below Eq. \eqref{phidef}).

The shape of the presented distributions  demonstrates that}}}
a comparison {{of these results}} with
the TMDs extracted from future measurements of appropriate
spin asymmetries in 
{{{
$^3{\vec {\rm He}}(\vec e, e'  p)X$
}}}
experiments \cite{exclus} could give very detailed information on the $^3${{He}} wave function,
{{on the validity of the LF description}}
and consequently on the nuclear interaction, once the possible final state interaction is properly taken care of.

\subsection{Effective polarizations}
As a {first} application of our results, let us  evaluate the LF longitudinal, $p^\tau_{||}$, and 
transverse, $p^\tau_{\perp}$,  effective polarizations for the proton and for the neutron, viz.
\be
p^\tau_{||}= \int_0^1 dx ~ \int d {\bf p}_\perp ~ \Delta f^\tau(x, |{\bf p}_{\perp}|^2 ) ~ ,
\label{leffpa}
\ee
\be
p^\tau_{\perp}= \int_0^1 dx ~ \int d {\bf p}_\perp ~ \Delta'_T   f^\tau(x, |{\bf p}_{\perp}|^2 ) ~ .
\label{teffp}
\ee
{They are} used in the extraction of neutron polarized structure functions and of neutron Collins and Sivers 
{{single spin}}
asymmetries, respectively, from the corresponding quantities measured for $^3$He (see, e.g., Refs. \cite{CiofidegliAtti:1993zs,E154:1997xfa}, 
and Refs. \cite{Scopetta:2006ww,DelDotto:2017jub,JeffersonLabHallA:2011ayy} respectively). 
{{This kind of extraction is based on the validity of the Impulse Approximation (IA), i.e. no final state interaction between the struck particle and the interacting spectator system, in the kinematics of the corresponding experiments. 
In the SIDIS case and in the kinematics of both Jefferson Lab
and  the future EIC,
it has been shown {{\cite{DelDotto:2017jub}} that 
effective polarizations,
can be used for the single spin asymmetries, even beyond the IA, {including final state interaction effects}.
}}

From the {{the discussion in the previous section,}} and from the expressions of Eqs. 
(\ref{leffpc1}) and (\ref{teffpc1}) 
it is clear that the longitudinal and the transverse effective polarizations are not anymore equal, as 
{it occurs} in the
non relativistic approximation. The difference between the two effective polarizations is due to the effect of the Melosh rotations for the spin. Indeed, without this effect one has (see Appendix (\ref{effpol}))
\be
p^\tau_{||} = p^\tau_{\perp} = {(-1)^{{\cal M} + 1/2} }~{\sqrt{{3}}}  
 \int d { k}_{23}~k^2_{23} 
 \nonu \times~\int_0^\infty {k}^2 ~ d {k}  ~
 2 ~ {\cal H} ^\tau(0,1,k_{23},k)
 \, ,
\label{teffpc6}
\ee
{where the function ${\cal H} ^\tau(0,1,k_{23},k)$ can be obtained from  Eq. \eqref{distr47}.}
The LF results obtained for the effective polarizations of proton and neutron in $^3$He 
 with the nuclear interaction AV18 of Ref. \cite{Wiringa:1994wb}, 
 {without} the Coulomb repulsion, 
 are shown in Table \ref{table:I}, together with the corresponding normalizations,
 and compared with the non relativistic result. In the first and in the second line the normalizations obtained from 
 the {proton and neutron spectral functions} 
 {through} Eq. (\ref{normFSLF1}) and directly from the wave function 
 are shown, respectively. 
 In the following three lines the LF  
 {calculations} for the longitudinal and the transverse 
 polarizations and for the polarizations without the Melosh rotations are presented. 
 {Eventually}, in the last line,  
 the results for the non relativistic polarizations with the same wave function are shown.
 \begin{table*}
 \begin{center}
\begin{tabular}{ |c|c|c| } 
 \hline
Normalization and effective polarizations & proton & neutron \\ 
  \hline
  normalization from the spectral function  & ~ 0.99915  ~  & ~ 0.99885 ~\\
  normalization from the wave function &  0.99929 &  0.99897 \\
 LF longitudinal polarization & -0.02299 & 0.87261 \\ 
LF  transverse polarization & -0.02446& 0.87314 \\ 
~ LF  polarization without Melosh rotations ~ & -0.02407& 0.87698 \\ 
 non relativistic polarization &  -0.02118  &   0.89337 \\
 \hline
\end{tabular}
  \caption{Normalization and effective longitudinal and transverse polarizations for the proton and the neutron 
  in $^3{{He}}$}
 \label{table:I}
 \end{center}
\end{table*}
{The comparison between the two normalizations allows one to assess  the numerical accuracy, that can be 
estimated of the order of  a few parts}
in $10^4$. 
{Therefore} the difference {of a few parts in $10^3$} between  longitudinal and  transverse polarizations 
is  a meaningful result.
{However, these small difference  indicates} that the effects of the Melosh rotations are tiny (see {{the result in the 
{{last but one}} line and}} the comment below Eq. \eqref{phidef}),
although for the proton it becomes  sizable in percentage. 
{Interestingly}, the difference of the LF polarizations with respect to the non relativistic results 
are larger, up to $2 \%$ in the neutron case
{{and should be ascribed to the  transformations performed between
the symmetric and non symmetric intrinsic coordinate systems, not to the Melosh rotations involving the spins.}}
In any case, the important point is that the relativistic results do not differ 
too much from the non relativistic ones,
{{currently used by the experimental collaborations.}} Therefore,  the 
values of the effective polarizations used
to extract the neutron polarized 
structure functions and the Collins and Sivers  asymmetries from $^3$He data  
{can be
considered reliable also from a  Poincar\'e-covariant point of view},
  although for a more precise determination the new values for
   the effective polarizations should be
   {adopted}. 
{{In closing this subsection, we observe that the longitudinal and transverse polarizations of the nucleons in
 $^3$He are 
 {analogous} to the 
axial and tensor charges of the nucleon in terms of its constituent quarks, in valence approximation. Since the beginning of the transversity studies, their difference has been always considered a signature of the relativistic content of the system
\cite{Jaffe:1991kp,Jaffe:1996zw,Barone:1996un,Scopetta:1997qg,Pasquini:2008ax}.

\subsection{Approximate relations}
In Refs. \cite{JMR,Pasquini:2008ax}, which define the TMDs as in Eqs. (\ref{ftr}), (\ref{str}) and (\ref{tracce}), approximate relations between the TMDs were discussed, i.e.,
 approximate relations between the TMDs were discussed, i.e.,
\be
\Delta f(x, |{\bf p}_{\perp}|^2 )=  \Delta'_T   f(x, |{\bf p}_{\perp}|^2 ) 
  +  {|{\bf p}_{\perp}|^2 \over 2 M^2}~ h^{\perp}_{1T}(x, |{\bf p}_{\perp}|^2 ) \, ,
\nonu\label{app3}
\ee
and 
\be
g_{1T}(x, |{\bf p}_{\perp}|^2 ) = - h^{\perp}_{1L}(x, |{\bf p}_{\perp}|^2 ) \, .
\label{app2} 
\ee
In principle, these relations put clear-cut phenomenological constraints on  the number of independent T-even twist-two TMDs. Therefore,
it is interesting to raise the question to what extent 
 the above relations are valid. To attempt a realistic answer, we tested Eqs. \eqref{app3} and \eqref{app2} in  the 
 case with a refined dynamical content as  the $^3$He nucleus, assuming the nucleons as  constituents.
 The first relation, 
 Eq. (\ref{app3}), is not exactly satisfied, since the equality should hold if 
$b^{\tau}_{2,{\cal M}}=b^{\tau}_{5,{\cal M}}$, as it follows from   Eqs. \eqref{Tmd2}, \eqref{Tmd4} and
\eqref{Tmd6}.  By inspecting  Eqs.   \eqref{b20}, \eqref{b22}, \eqref{b50} and \eqref{b52}, one gets   $b^{\tau(0)}_{2,{\cal
M}}=b^{\tau(0)}_{5,{\cal M}}$,   while  
{$b^{\tau(2)}_{2,{\cal M}}$ and 
$b^{\tau(2)}_{5,{\cal M}}$} have not the same expressions.  {The quantitative difference between the lhs and rhs of Eq. \eqref{app3} is quite small  for the neutron, while is not negligible for the proton} as 
 shown in Fig.  \ref{fig4}. 

{From the evaluation of the effective polarizations we learned that the effects of the Melosh rotations 
are tiny, and  if these effects are neglected in  
${{b}}_{i,{\cal M}}^{}$, i.e. $ sin(\varphi/2) \to 0$ in the expressions presented at the end of   Appendix
\ref{sdmd},  one 
finds that
the second relation,  Eq. (\ref{app2}), holds,
 but with the opposite sign. It is very interesting to analyze such a different sign, that could have far reaching consequences. 
 The functions $b^{\tau}_{4,{\cal M}}$ and
$b^{\tau}_{3,{\cal M}}$ determine  $g_{1T}$  and $ h^{\perp}_{1L}$, respectively, as shown in Eqs. \eqref{Tmd3} and \eqref{Tmd5}. From Eqs.
 \eqref{b30} and \eqref{b40}, where the  
  contribution with $L=0$} is considered, one has  $b^{\tau(0)}_{3,{\cal M}}= - b^{\tau(0)}_{4,{\cal M}} $,
 but  the two functions are of the order $sin(\varphi)$ and therefore very small. The 
 contribution {with $L=2$} leads to $b^{\tau(2)}_{3,{\cal
 M}}\sim b^{\tau(2)}_{4,{\cal M}} $ (see Eqs. \eqref{b32} and \eqref{b42}) and generates the leading term, proportional to {$cos^2(\varphi)$}, 
  of the  two TMDs. Therefore, the sign
in Eq. \eqref{app2}, has to be considered as a signature of the bound-state orbital  
{content}  
that prevails  in the
actual value of $g_{1T}$ and  $h^{\perp}_{1L}$.
{As clear from Appendix \ref{mdistr} (where it is shown that only the values $L=0$ and $L=2$ are possible), if only the contribution from  $L=0$ is present, this implies a vanishing value of the {active fermion} orbital angular-momentum 
$L_{\rho}$
 in the bound-system wave function 
(see Eqs. (\ref{FO6}) - (\ref{distr35})).
{On the other hand,} a non-vanishing value of $L_{\rho}$ implies a non-vanishing value of the contribution from $L=2$ to the momentum distribution.

  For $^3$He, Eq. \eqref{app2} is fulfilled in modulus, as pictorially 
 shown in   Figs.   \ref{fig5}  and  \ref{fig6}.}  
\begin{figure*}
\begin{center}
\includegraphics[width=8.50cm]{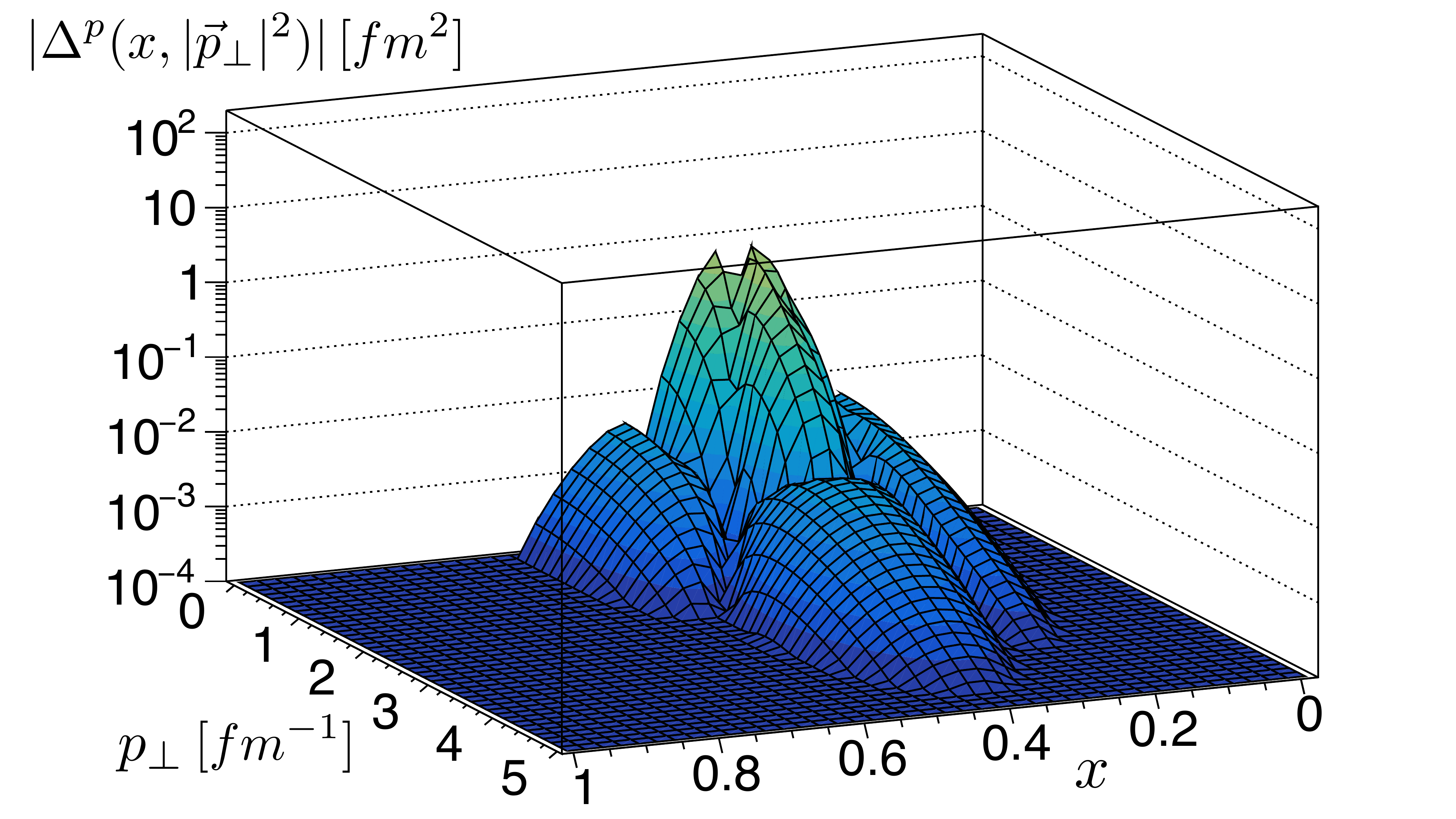}
\includegraphics[width=8.50cm]{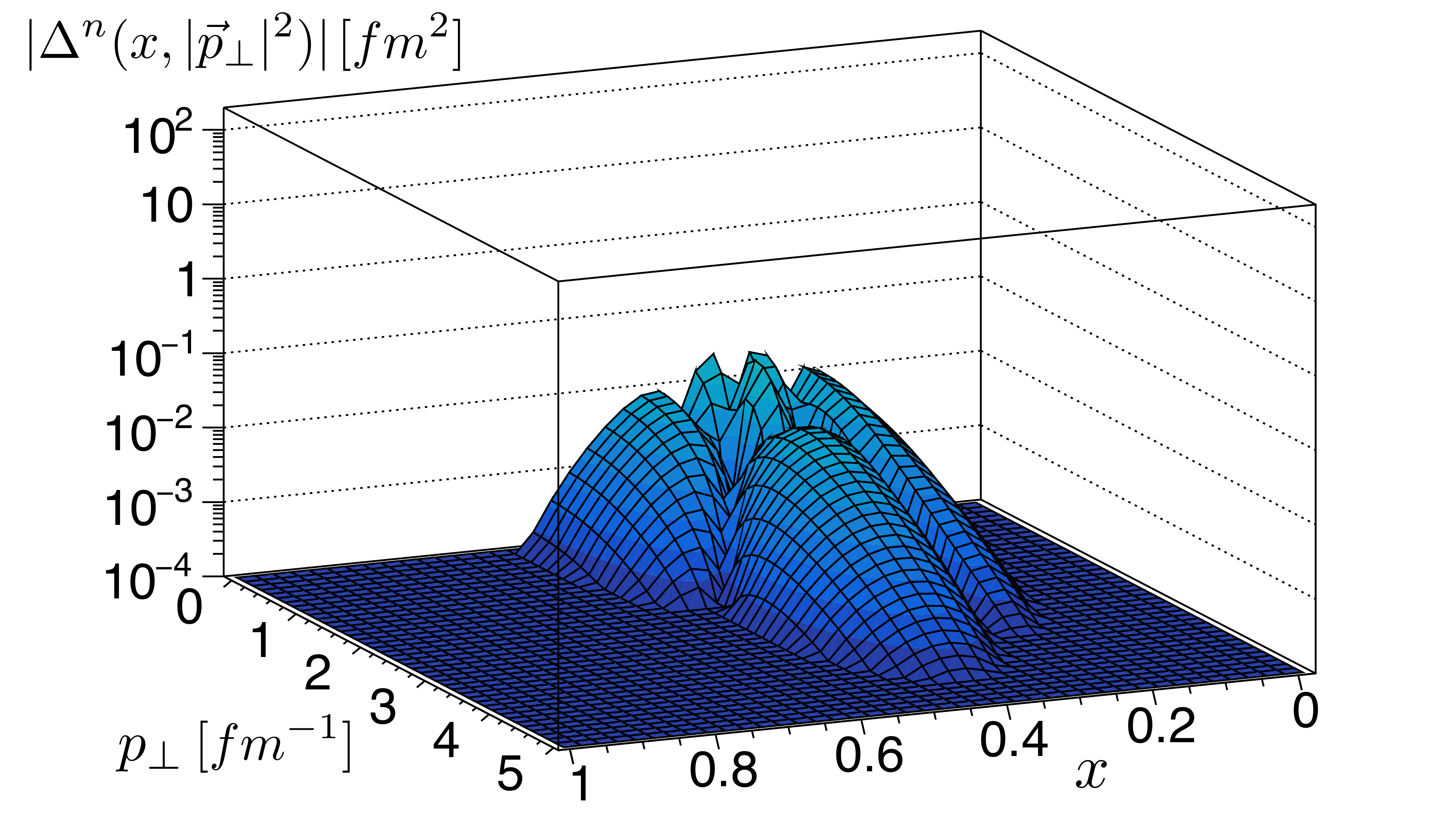}
 \caption{
Absolute value of the difference $ \Delta (x, |{\bf p}_{\perp}|^2 ) = \Delta'_T   f(x, |{\bf p}_{\perp}|^2 ) ~ + ~
 {|{\bf p}_{\perp}|^2 \over 2 M^2}~ h^{\perp}_{1T}(x, |{\bf p}_{\perp}|^2 ) - 
 \Delta f(x, |{\bf p}_{\perp}|^2 )$ in  $^3$He 
 for the  proton (left panel) and the neutron (right panel).}
\label{fig4}
\end{center}
\end{figure*}

For a system with a high average value of the ratio 
{$| {\bf p}_{\perp} |/m$, {as occurs when we shift from the $^3$He case to the nucleon one, where   the constituents are  
the quarks with $m_q < m_N$,}
(see Ref. \cite{Cardarelli:1995dc,Cardarelli:1995ug}), Melosh rotations become  relevant. {If this is the case, from the previous discussion, one expects 
 that the relations
(\ref{app3}) and  (\ref{app2})
hold exactly in valence approximation when the contribution to the transverse momentum distributions from  $L=2$ is very
 small or totally
absent.} 



Therefore, the validity of the equalities (\ref{app3}) and  (\ref{app2}) in a bound-system with a high average value of 
$|{\bf p}_{\perp}|/m$ represent a well-defined constraint  on the allowed dynamics inside the three-body bound system.

Another relation proposed in Refs. \cite{JMR,Pasquini:2008ax} 
i.e.,
\be
\left [ g_{1T}(x, |{\bf p}_{\perp}|^2 ) \right ]^2 = - 2~\Delta'_T   f(x, |{\bf p}_{\perp}|^2 ) ~
h^{\perp}_{1T}(x, |{\bf p}_{\perp}|^2 ) \, ,
\nonu\label{app4}
\ee
does not hold in our approach, even if  the contribution from the angular momentum $L=2$ is absent, 
because of the presence of the integration on $k_{23}$ 
in the expressions of the transverse-momentum distributions (see Eq. (\ref{distr62})). { Namely, a genuine relativistic approach, related to
the implementation of the macroscopic locality, spoils the above relation. Therefore, a signature of such a relativistic effect is given by the 
violation of Eq. \eqref{app4}.}

\section{Conclusions}

 {A Poincar\'e-covariant description of 
 bound systems, with $A$ constituents {of spin $1/2$}, has been developed within 
   the light-front Hamiltonian
dynamics in Ref. \cite{DelDotto:2016vkh} and applied  in the present work to  the T-even twist-two transverse-momentum distributions {of 
${\cal{ J } }= 1/2$ systems}. 
{The explicit expressions for the TMDs are obtained} in terms of LF overlaps, between the 
bound state and states described by  tensor products of a constituent plane wave and an $(A-1)$ fully interacting 
{{system}}. 
{These} LF
overlaps are the basic ingredients of the spin-dependent spectral function \cite{DelDotto:2016vkh}, that is a $2\times 2$ matrix 
 with the main-diagonal terms yielding the distribution probability to find a constituent with given spin and LF momentum, once
the $(A-1)$-spectator system has an assigned mass. Indeed, 
the leading-twist TMDs, in valence approximation,   {is linked to} proper traces of the valence contribution to the semi-inclusive 
correlation
function, that 
{is} linearly related  to the spin-dependent
spectral function.
 The formalism has been applied to the $^3$He nucleus, keeping  a twofold aim in mind: (i)  illustrating  a realistic case, 
 where a theoretical description of the bound state
 takes into account a highly  non trivial dynamics, and consequently assessing the impact {of dynamics} on the actual TMDs; (ii) cumulating quantitative 
 analyses for supporting future experimental efforts dedicated to 
 achieve a detailed 3D picture of the $^3$He.  As a matter of fact, for the $^3$He nucleus there
 exist a reliable LF spin-dependent spectral function, obtained within 
 the Bakamjian-Thomas framework \cite{Bakamjian:1953kh}, suitable  for embedding} the wide knowledge on the nuclear interaction and the successful phenomenology developed  for few-nucleon systems
 in a  Poincar\'e-covariant approach (preliminary calculations of the 
 EMC effect in $^3$He were presented  in Refs. \cite{Pace:2016eiq,Pace:2017dwa,Pace:2020ikl,Pace:2020ned} {and a full calculation will be soon available \cite{Rina}}). 
 
 {Besides the Melosh rotations, the peculiar feature of  our LF approach is the macroscopic locality, implemented through the use of non symmetric intrinsic variables, i.e. intrinsic internal variables for the $A-1$ system and
  the LF momentum of the active fermion in the {intrinsic reference frame of the $[1,(A-1)] $ cluster} \cite{DelDotto:2016vkh}. Their impact on the $^3$He TMDs has been discussed, with a particular attention to the relevance from the
  experimental point of view.} To this end, we have {evaluated}, as a first application, the LF longitudinal and transverse effective 
  polarizations for the proton and  the neutron {in $^3$He.} 
  These two quantities
 are widely used to extract the neutron polarized structure functions and  the neutron Collins and Sivers asymmetries from the corresponding quantities measured  in DIS and SIDIS off $^3$He, respectively.
 {Although this procedure is generally used assuming the validity of the {impulse} approximation, in Ref. \cite{DelDotto:2017jub} it was shown how it can be applied even including the final state interaction.}

We found that, for a system with a small average value of 
$p_\perp /m$, as the three-nucleon 
 system, the effect of the spin Melosh rotation is tiny and  the longitudinal and the transverse  effective polarizations differ very
  little from each other, both for the proton and the neutron. On the contrary, the difference with respect to the non relativistic result 
  is not negligible and this effect,  {ascribed to the use of LF intrinsic variables,} 
  has to be considered for a  more accurate extraction of 
  neutron properties from $^3$He data.  
   
 {A second important result we have discussed  is the validation of the linear relations proposed in \cite{JMR,Lorce:2011zta} between the
 T-even twist-two TMDs. Although those relations were introduced in the context of the nucleon studies in valence approximation,  we have 
 investigated their validity for  
 the $^3$He case,   in order to show  how  a rich dynamics actually impacts the extraction of  important information on the orbital-momentum decomposition 
 of the three-body bound-system.
 Namely, one recovers the above mentioned relations if the state 
 {has a simple S-wave structure}, while in presence of a D-wave and a small effect
 of the Melosh rotations, i.e. a small average value of $p_\perp /m$, the second relation even changes sign. } 
 The other relation proposed  in Refs.
\cite{JMR,Lorce:2011zta},   between the TMDs, a quadratic 
one, does not hold  in any case, since the TMDs are given by integrals over the relative momentum of the {{spectator}} interacting pair, 
{which represent} {an unavoidable feature of a genuinely Poincar\'e-covariant framework.
 Hence, an experimental
 investigation, performed  at high luminosity facilities, could open a window directly on the  orbital-momentum content of the bound state
 and the relativistic regime of the inner dynamics.}

The twist-three TMDs can be also evaluated in our 
scheme and will be the object of another study.

\appendix

\section{Dependence of the spectral function on the polarization vector $\bf S$}
 \label{Sdep}
 The LF spin-dependent spectral function for any direction of the polarization vector $\bf S$ can be written in terms of the Wigner rotation matrixes $D^{\cal J}_{m, \cal M}(\alpha, \beta, \gamma)$ (see Eqs. (\ref{LFspf}) and (\ref{rot}))
   \be
   {\cal P}_{{\cal M},\sigma'\sigma} (\tilde{\bma \kappa},\epsilon,S) 
~   = ~ 
    \sum_{m,m'}  ~ D^{{\cal J}*}_{m, \cal M}(\alpha, \beta, \gamma) ~
D^{\cal J}_{m', \cal M}(\alpha, \beta, \gamma) \nonu \times~
   {\cal P}^{m',m}_{\sigma'\sigma} (\tilde{\bma \kappa},\epsilon,S_z) ~,
 \label{spfS}
   \ee
where $S_z = (0,0,0,1)$ and
    \be
 {\cal P}^{m',m}_{\sigma'\sigma} (\tilde{\bma \kappa},\epsilon,S_z)  =  
 \rho(\epsilon)
 \nonu \times ~
  \sum_{J J_{z}\alpha}\sum_{T_{S}\tau_{S}} 
~_{LF}\langle  \tau_{S}T_{S} ; 
\alpha,\epsilon ;J J_{z}; \tau\sigma',\tilde{\bma \kappa}|{\cal J}m';  \epsilon^A,\Pi;  T T_z\rangle_z
\nonu \times ~ _z\langle T T_z; \Pi, \epsilon^A; 
{\cal J}m |\tilde{\bma \kappa},\sigma\tau; J J_{z}; 
\epsilon, \alpha; T_{S} \tau_{S}\rangle_{LF}  \, .
  \label{zz}
  \ee
with $|{\cal J} m; \epsilon^A,\Pi; T T_z\rangle_z$ the system ground state with energy 
$ \epsilon^A$, parity $\Pi$  and isospin $T, T_z$, polarized along $\hat z$.

For a system with ${\cal J} = 1/2$  Eq. (\ref{spfS}) can be written as follows
  \be
   {\cal P}_{{\cal M},\sigma'\sigma} (\tilde{\bma \kappa},\epsilon,S) 
~   = ~ 
    \sum_{m,m'}  ~ D^{1/2}_{-m, -\cal M}(\alpha, \beta, \gamma) ~ (-1)^{m-{\cal M}}
\nonu \times ~
D^{1/2}_{m', \cal M}(\alpha, \beta, \gamma) ~
   {\cal P}^{m',m}_{\sigma'\sigma} (\tilde{\bma \kappa},\epsilon,S_z) ~ ,
 \label{spfS1}
   \ee

Let us take advantage of 
the following property for the product of two Wigner rotation matrixes with the same arguments (see Eq. (1) at p. 84 of \cite{Varshalovich:1988ye})
\be
D^{j}_{m, k}(\alpha, \beta, \gamma) ~ D^{j'}_{m', k'}(\alpha, \beta, \gamma) = 
\sum_{J=|j-j'|}^{j+j'} \hspace{-.2cm} 
\langle j m, j' m'  |J (m+m') \rangle 
\nonu \times~  \langle j k, j' k'  |J (k+k') \rangle ~
D^{J}_{(m+ m'), (k + k')}(\alpha, \beta, \gamma) ~ .
\label{spfS2}
\ee
\vspace{1mm}
In our case it reads
\be
D^{1/2}_{-m, -{\cal M}}(\alpha, \beta, \gamma) ~ 
D^{1/2}_{m', {\cal M}}(\alpha, \beta, \gamma) 
\nonu
= 
\sum_{J=0}^{1}  
\langle \frac{1}{2} -m, \frac{1}{2}m' |J (m'-m) \rangle ~  
\langle \frac{1}{2} -{\cal M}, \frac{1}{2} {\cal M} |J 0 \rangle 
\nonu \times
D^{J}_{(m'-m),0}(\alpha, \beta, \gamma) 
\nonu
= 
\sum_{J=0}^{1} \hspace{.1cm} 
\langle \frac{1}{2} -m, \frac{1}{2}m' |J (m'-m) \rangle ~ 
 \langle \frac{1}{2} -{\cal M}, \frac{1}{2} {\cal M} |J 0 \rangle 
 \nonu \times~ (-1)^{m'-m} \sqrt{{4 \pi \over 2J+1}} ~
Y_{J (m-m')}(\beta,\alpha) ~.
\label{spfS3}
\ee
Then the product of the two $D^{j}_{m, k}(\alpha, \beta, \gamma)$ matrixes is a sum of two terms. The first one, with $J=0$, is independent of $\beta,\alpha$ (i.e. of $\bf S$). In the second one, with $J=1$, the spherical harmonics $ Y_{1 (m-m')}(\beta,\alpha) $ can be raplaced by the spherical components $S_i$ of  the polarization vector \cite{Varshalovich:1988ye}
\be
Y_{J (m-m')}(\beta,\alpha) = \sqrt {{3 \over 4 ~ \pi }} ~ S_{(m-m')}~.
\label{spfS4}
\ee
Hence the spectral function depends 
linearly on $\bf S$. Since in the expansion (\ref{dv1}) of the pseudovector ${\cal F}^\tau_{\cal M}$ a linear dependence on  $\bf S$ already explicitly appears in the five pseudovectors 
$\bf{S}$, $~\hat {\bf k}_{\perp} ({\bf  S} \cdot \hat {\bf k}_{\perp})$, 
$~\hat {\bf k}_{\perp} ({\bf  S} \cdot \hat z)$, 
 $~\hat z ~({\bf  S} \cdot \hat {\bf k}_{\perp})$, and $~ \hat z ~({\bf  S} \cdot \hat {z})$, as a consequence  for a system with total angular momentum ${\cal J}=1/2$ the quantities ${\cal B}_i$ for $i=1, . . .,5$ can depend only on $ |{\bf k}_{\perp}|,x$, and $\epsilon$.

\section{Semi-inclusive fermion correlator, twist-two transverse momentum distributions and LF spectral function}
\label{corre}
 In this Appendix equations that link in valence approximation the leading-twist TMDs  to the scalar functions
  ${{b}}_{i,{\cal M}}^{}$, that contain the relevant information on the dynamics inside the bound system, are obtained.
  
For the sake of completeness, let us summarize 
the decomposition of the semi-inclusive fermion correlator in terms of the twist-two T-even TMDs 
as presented in Ref. \cite{Barone:2001sp}. {Let us recall that ${\bf p}_\perp=
{\bf k}_\perp={\bma \kappa}_\perp$, and to  simplify the notation 
 the isospin index  $\tau$ is omitted in what follows.}  The 
semi-inclusive correlator
at leading twist is given by 
 \be
 \Phi(p, P, S)  =  {1 \over 2} ~{\slash \! \!\! \!P}~A_1 
 \nonu 
 + {1 \over 2} ~ \gamma_5~{\slash \!\! \! \!P}
  \left[
    A_2 \, S_z \,  +
    \frac1{M} \, \widetilde{A}_1 \, \bf{p}_\perp{\cdot}\bf{S}_\perp\right]
    \nonu
        + {1 \over 2} \gamma_5~{\slash \! \! \!\!P}\left[
    A_3 \,  {\slash \! \! \!\!S}_\perp
    +
    \widetilde{A}_2 \, \frac{S_z}{M}
     \, {\slash \! \! \!p}_\perp
   + 
    \widetilde{A}_3   \, {{\bf{p}_\perp{\cdot}\bf{S}_\perp }
      \, {\slash \! \! \!p}_\perp\over M^2}
  \right] ~,
  \label{corr}
 \ee
 where $M$ is the mass of the system, {and the scalar functions, $A_l$ and $\widetilde A_l$, contain the
 information on the inner dynamics ($l=1,2,3$).} 
 By performing the traces of the {correlator with  suitable  combinations of Dirac matrixes}
 one
 has
\be
  \frac1{2P^+} \, \Tr(\gamma^+ \Phi)
   = 
  A_1 \, ,
\label{ftr} \\ &&
  \frac1{2P^+} \, \Tr(\gamma^+ \gamma_5 \Phi)  = 
 S_z \, A_2 +
  \frac1{M} \, {\bf{p}_\perp{\cdot}\bf{S}_\perp} \, \widetilde{A}_1 \, ,
\label{str} \\&&
  \frac1{2P^+} \, {\Tr}(i \sigma^{j+} \gamma_5 \Phi)  = 
  -\frac1{2P^+} \, {\Tr}( \gamma^j \gamma^+ \gamma_5 \Phi)
  \nonu=
  S_\perp^j \, A_3 +
  \frac{S_z}{M}\, p_\perp^j \, \widetilde{A}_2 +
  \frac1{M^2} \, {\bf{p}_\perp{\cdot}\bf{S}_\perp} \, p_\perp^j \, 
\widetilde{A}_3
  \label{tracce}
\ee
where j=1,2.
Finally, by integrating   {{proper combinations of}} $A_l$, and  $\widetilde{A}_l$ on $p^+$ and $p^-$, one gets
   the TMDs as follows \cite{Barone:2001sp}
\be
 f(x, |{\bf p}_{\perp}|^2 )
   =  \int \Bigl[d{ p^+} d{ p^-}\Bigr] ~A_1 \, , 
 \label{tmd1}  \\ &&
\Delta f(x, |{\bf p}_{\perp}|^2 )
   = \int \Bigl[d{ p^+} d{ p^-}\Bigr]~  A_2 \, , 
 \label{tmd2}  \\ &&
 g_{1T}(x, |{\bf p}_{\perp}|^2 )
= \int \Bigl[d{ p^+} d{ p^-}\Bigr]~ \widetilde{A}_1 \, , 
\label{tmd3}\\ &&
\Delta'_T   f(x, |{\bf p}_{\perp}|^2 )
= \int \Bigl[d{ p^+} d{ p^-}\Bigr]
\left ({A}_3 + {|{\bf p}_{\perp}|^2 \over 2M^2} \widetilde{A}_3\right )  , 
\label{tmd4}\\ &&
h^{\perp}_{1L}(x, |{\bf p}_{\perp}|^2 )
= \int \Bigl[d{ p^+} d{ p^-}\Bigr] ~\widetilde{A}_2 \, , 
\label{tmd5}\\ &&
h^{\perp}_{1T}(x, |{\bf p}_{\perp}|^2 )
   = \int \Bigl[d{ p^+} d{ p^-}\Bigr]~  \widetilde{A}_3 \, . 
\label{tmd6}
\ee
with
\be \int \Bigl[d{ p^+} d{ p^-}\Bigr] =\frac1{2} \int \frac{d{ p^+} d{ p^-}}{(2\pi)^4} \delta[p^+ - x P^+] P^+
\ee
Notice that in the right hand side of the {above equations, a factor of 2 is missing with respect 
to the expressions in } Ref. \cite{Barone:2001sp}, because of the different definitions of the variables $v^{\pm}$.
If only the {valence contribution} to the 
{{correlator}}
 is retained,
the full 
$\Phi(p,P,S)$ is approximated by $\Phi_V(p,P,S)$, 
and in turn $\Phi_V(p,P,S)$ is expanded in analogy with Eq. (\ref{corr}) in terms of 
$A^V_l$ and $\widetilde{A}^V_l$. Hence, instead of 
Eqs. \eqref{ftr}, \eqref{str} and \eqref{tracce}
one can write
\be
 \frac1{2P^+} \, \Tr(\gamma^+ \Phi) ~ \sim ~ \frac1{2P^+} \, \Tr(\gamma^+ \Phi_V)
   = 
  A^V_1 \, ,
\label{ftrV}
\\ &&
\frac1{2P^+} \, {\Tr}(\gamma^+ \gamma_5 \Phi) ~ \sim ~ \frac1{2P^+} \, {\Tr}(\gamma^+ \gamma_5 \Phi_V)  
\nonu = 
  S_z \, A^V_2 +
  \frac1{M} \, {\bf{p}_\perp{\cdot}\bf{S}_\perp } \, \widetilde{A}^V_1 \, ,
\label{strV} 
\\&&
 \frac1{2P^+} \, {\Tr}(i \sigma^{j+} \gamma_5 \Phi) 
  ~ \sim  
  - ~ \frac1{2P^+} \, {\Tr}( \gamma^j \gamma^+ \gamma_5 \Phi_V) =
  \nonu=
  S_\perp^j \, A^V_3 +
  \frac{S_z}{M}\, p_\perp^j \, \widetilde{A}^V_2 +
  \frac{{\bf{p}_\perp{\cdot}\bf{S}_\perp}}{M^2} \,  \, p_\perp^j  
\widetilde{A}^V_3
  \, ,
  \label{tracceV}
\ee
where $A^V_l$ and $\widetilde{A}^V_l  $ are the valence approximations for 
$A_l$ and  $\widetilde{A}_l$, respectively.
Because of  Eqs.  (\ref{abc})   and  (\ref{iden}), the above traces of $\Phi_V$ 
can be expressed by means of traces of the spectral function,
as shown in Appendix \ref{traces} (see also Eq. (\ref{gammapiu})):
\be
 \frac1{2P^+} \, {\Tr}(\gamma^+ \Phi_V) = 
{P^+ \over p^+}  { 2\pi \over 4  m}  {E_S \over  {\cal{M}}_0[1,(23)]}
\nonu \times ~
 {\Tr}\left[ {\bma {\widehat {\cal P}}}^{}_{\cal M}(\tilde {\bma \kappa},\epsilon,S)
\right]~,
\label{trpiu} \\ &&
\frac1{2P^+} \, {\Tr}(\gamma^+ \gamma_5 ~ \Phi_V) 
= {P^+ \over p^+} ~ { 2\pi \over 4  m} ~ {E_S \over  {\cal{M}}_0[1,(23)]}
\nonu \times ~
{\Tr}\left[ \sigma_z ~ {\bma {\widehat {\cal P}}}^{}_{\cal M}(\tilde{\bma \kappa},\epsilon,S) \right]\, ,
\label{trpiu5}\\ &&
- ~ \frac1{2P^+} \, {\Tr}( \gamma^j ~\gamma^+ ~\gamma_5 ~ \Phi_V) 
= {P^+ \over p^+} ~ { 2\pi \over 4  m} ~ {E_S \over {\cal{M}}_0[1,(23)]}
\nonu \times ~
{\Tr}\left[ \sigma^j
~ {\bma {\widehat {\cal P}}}^{}_{\cal M}
(\tilde{\bma \kappa},\epsilon,S) \right] \, .
\label{trperpiu5}
\ee
As noted in Section II, the LF spectral function can be written in terms of
 six  scalar quantities, ${\cal B}_{i,{\cal M}}$.
  
Therefore the following equations hold :
\be
  {\Tr}( {\bma {\widehat {\cal P}}_{\cal M}} I) = ~ {\cal {B}}_{0,{\cal M}}^{}
  \label{tri} \, ,
\\
&&
  {\Tr}(  {\bma {\widehat {\cal P}}_{\cal M}} \sigma_x ) =
  {S}_x ~
 {\cal B}_{1,{\cal M}}
 ~
 + ~{ k_x \over k_{\perp}}
~({\bf  S} \cdot \hat {\bf k}_{\perp})~
{\cal B}_{2,{\cal M}}  
\nonu + ~{ k_x \over k_{\perp}} ~({\bf  S} \cdot \hat z)~ 
{\cal B}_{3,{\cal M}} \, ,
\label{trz}
\\
&&
{\Tr}(  {\bma {\widehat {\cal P}}_{\cal M}} \sigma_y ) =
S_{y} ~
 {\cal B}_{1,{\cal M}}
 ~
 + ~
{ k_y \over k_{\perp}} ~({\bf  S} \cdot \hat {\bf k}_{\perp})~
{\cal B}_{2,{\cal M}}  
\nonu
 + ~{ k_y \over k_{\perp}} ~({\bf  S} \cdot \hat z)~ 
{\cal B}_{3,{\cal M}} \, ,
\label{trpery}
 \\
&&
{\Tr}(  {\bma {\widehat {\cal P}}_{\cal M}} \sigma_z )=
S_{z}  ~
 {\cal B}_{1,{\cal M}}
 ~
 + ~({\bf  S} \cdot \hat {\bf k}_{\perp})~
  {\cal B}_{4,{\cal M}} ~
 \nonu  + ~
 ({\bf  S} \cdot \hat {z})~ 
 {\cal B}_{5,{\cal M}}
~.
\label{trper}
\ee
From Eqs. \eqref{ftrV}  
 to \eqref{trper} one obtains

\be
A^V_1 ~ =  {\pi \over 2 m} {E_S \over \kappa^+} ~ {\cal {B}}_{0,{\cal M}}^{}
 \label{A1}
\ee
\be
\hspace{-3mm} S_z \, A^V_2 +
  \frac1{M} \, {\bf{p}_\perp{\cdot}\bf{S}_\perp } \, \widetilde{A}^V_1 \,   
   =  {\pi \over 2 m} {E_S \over \kappa^+} 
   \nonu
    \hspace{-3mm}\times~ \Bigl[ S_{z}  ~
 {\cal B}_{1,{\cal M}}
 ~
 + ~({\bf  S} \cdot \hat {\bf k}_{\perp})~
  {\cal B}_{4,{\cal M}} ~
  + ~
 ({\bf  S} \cdot \hat {z})~ 
 {\cal B}_{5,{\cal M}} 
\Bigr ]
\label{A2}
\ee
\be
    S_x  A^V_3 +
  \frac{S_z}{M}\, p_x\, \widetilde{A}^V_2 +
  \frac{{\bf{p}_\perp{\cdot}\bf{S}_\perp}}{M^2} \,  \, p_x \, 
\widetilde{A}^V_3 = 
 {\pi \over 2 m} {E_S \over \kappa^+} 
 \Bigl [ {S}_x ~
 {\cal B}_{1,{\cal M}}
 \nonu + ~{ k_x \over k_{\perp}}
~({\bf  S} \cdot \hat {\bf k}_{\perp})~
{\cal B}_{2,{\cal M}}  
~ + ~{ k_x \over k_{\perp}} ~({\bf  S} \cdot \hat z)~ 
{\cal B}_{3,{\cal M}} 
\Bigr ]~,
\label{A3}
\ee
\be
    S_y  A^V_3 +
  \frac{S_z}{M}\, p_y\, \widetilde{A}^V_2 +
  \frac{{\bf{p}_\perp{\cdot}\bf{S}_\perp}}{M^2} \,  \, p_y \, 
\widetilde{A}^V_3 = 
 {\pi \over 2 m} {E_S \over \kappa^+}
 ~ \Bigl [ S_{y} ~
 {\cal B}_{1,{\cal M}}
\nonu + ~
{ k_y \over k_{\perp}} ~({\bf  S} \cdot \hat {\bf k}_{\perp})~
{\cal B}_{2,{\cal M}}  
~ + ~{ k_y \over k_{\perp}} ~({\bf  S} \cdot \hat z)~ 
{\cal B}_{3,{\cal M}} 
\Bigr ] ~.
\label{A3a}
\ee

Let us integrate Eqs. \eqref{A1}, \eqref{A2}, \eqref{A3} and \eqref{A3a} on $p^+$ and $p^-$ as in Eqs. 
\eqref{tmd1} - \eqref{tmd6}. Then in valence approximation one has
\be
 f(x, |{\bf p}_{\perp}|^2 ) ~ = ~ {{b}}_{0}^{} \, ,
 \label{Api1}
\ee
\be
 S_z \, \Delta f +
  \frac1{M} \, {\bf{p}_\perp{\cdot}\bf{S}_\perp } \, g_{1T} \,   
   =  ~ \Bigl[ S_{z}  ~
 {b}_{1,{\cal M}}
 \nonu
 + ~({\bf  S} \cdot \hat {\bf k}_{\perp})~
  {b}_{4,{\cal M}} ~
  + ~
 ({\bf  S} \cdot \hat {z})~ 
 {b}_{5,{\cal M}} 
\Bigr ] \, ,
\label{Api2}
\ee
\be
    S_x ~ h_{1T} +
  \frac{S_z}{M}\, p_x\, h^\perp_{1L} +
  \frac{{\bf{p}_\perp{\cdot}\bf{S}_\perp}}{M^2} \,  \, p_x \, 
h^\perp_{1T} 
 = 
\Bigl [ {S}_x ~
 {b}_{1,{\cal M}}
 \nonu
 + ~{ k_x \over k_{\perp}}
~({\bf  S} \cdot \hat {\bf k}_{\perp})~
{b}_{2,{\cal M}}  
~ + ~{ k_x \over k_{\perp}} ~({\bf  S} \cdot \hat z)~ 
{b}_{3,{\cal M}} 
\Bigr ] \, ,
\label{Api3}
\ee
\be
    S_y  ~ h_{1T} +
  \frac{S_z}{M}\, p_y\,  h^\perp_{1L}  +
  \frac{{\bf{p}_\perp{\cdot}\bf{S}_\perp}}{M^2} \,  \, p_y \, 
h^\perp_{1T}  
= 
 \Bigl [ S_{y} ~
 {b}_{1,{\cal M}}
 \nonu
 + ~
{ k_y \over k_{\perp}} ~({\bf  S} \cdot \hat {\bf k}_{\perp})~
{b}_{2,{\cal M}}  
~ + ~{ k_y \over k_{\perp}} ~({\bf  S} \cdot \hat z)~ 
{b}_{3,{\cal M}} 
\Bigr ] ~ ,
\label{Api3a}
\ee
where  
(see Ref. \cite{Goeke:2005hb})
\be
 h_{1T} = \int \Bigl[d{ p^+} d{ p^-}\Bigr] ~  A^V_3 \, .
\label{ap3v}
\ee

     \section{The spin-dependent momentum distribution and the three-body wave function}
  \label{mdistr}   
 One can obtain the LF momentum distribution dependent upon the spin directions, $ \left [ {\bma {\cal N}}_{\cal M}^\tau(x,{\bf k}_{\perp};\bf S) \right ]_{\sigma  \sigma '} $, {Eq. \eqref{N}},  for any direction of the polarization vector $\bf S$ of the system,
 using {Eq. \eqref{spfS} and Eq. \eqref{zz}
\bwt
\be
\left [ {\bma {\cal N}}_{\cal M}^\tau(x,{\bf k}_{\perp};\bf S) \right ]_{\sigma  \sigma '} ~ =  ~
 \sumint {d\epsilon ~} 
{1 \over 2 ~ (2\pi)^3}~
{ E_S \over (1- x)~ \kappa^+}~
\rho(\epsilon)
\sum_{J J_{z}\alpha}\sum_{T_{S}\tau_{S} }\sum_{m}D^{\cal J}_{m, {\cal M}}
(\alpha, \beta, \gamma)
\nonu \times _{LF} \langle  \tau_{S}T_{S} ; 
\alpha,\epsilon ;J J_{z}; \tau\sigma,\tilde{\bma \kappa}|
{\cal J} j_z=m;
\epsilon^3,\Pi; T T_z\rangle
 \sum_{m'}[D^{\cal J}_{m', {\cal M}}(\alpha, \beta, \gamma)]^* 
 \langle T T_z; \Pi,\epsilon^3; {\cal J}j_z=m'; 
|\tilde{\bma \kappa},\sigma'\tau; J J_{z}; 
\epsilon, \alpha; T_{S} \tau_{S}\rangle_{LF} ~ ,
\nonu\label{distr71}
\ee
\ewt
where $|{\cal J}j_z=m; \epsilon^3,\Pi; T T_z\rangle$ is the three-body ground state, 
polarized along $\hat z$.
Using the explicit expression for the overlaps given by Eq. (62) of Ref. \cite{DelDotto:2016vkh}, the two-body completeness 
\be
\sum_{J,J_z\alpha} \sum_{T T_z}\sumint d\epsilon ~\rho(\epsilon)  
\langle{\bf k}' |JJ_z; \epsilon,\alpha; T_{S} \tau_{S} \rangle 
\nonu \times ~
\langle \tau_{S}  T_{S};\alpha,\epsilon;J_zJ|{\bf k}\rangle=
\delta^3({\bf k}'-{\bf k}) ~ ,
\label{nrcompl1}
\ee
 and the unitarity of the $D$ matrixes, we have
\be
 \left [ {\bma {\cal N}}_{\cal M}^\tau(x,{\bf k}_{\perp};\bf S) \right ]_{\sigma  \sigma '}  ~ =  ~
   \sum_m  ~ D^{{\cal J}}_{m, {\cal M}}(\alpha, \beta, \gamma) 
   \nonu \times~
 \sum_{m'}  ~  [D^{{\cal J}}_{m', {\cal M}}(\alpha, \beta, \gamma)]^* ~
 {\cal F}^{\tau m m'}_{\sigma\sigma'} (x,{\bf k}_{\perp})~,
 \label{distr4}
 \ee
 where
\bwt \be
{\cal F}^{\tau m m'}_{\sigma\sigma'} (x,{\bf k}_{\perp}) = {1 \over  (1- x)  } ~
 \sum_{\tau_2\tau_3}  ~
\int d {\bf k}_{23} ~ 
E({\bf k}^{})~ {E_{23}\over k^{+}}
~\sum_{\sigma_1 { \sigma'_1}}~
D^{{1 \over 2}} [{\cal R}_M ({\blf k})]_{\sigma\sigma_1}~  
D^{{1 \over 2}*} [{\cal R}_M ({\blf k} )]_{\sigma'{\sigma}'_1}~
\nonu \times 
\sum_{\sigma'_2,\sigma'_3}~  
\langle \sigma'_3, \sigma'_2,\sigma_1; \tau_3,\tau_2,\tau; {\bf k}_{23},{\bf k}|j,j_z = m;
\epsilon^3_{int},\Pi; TT_z \rangle ~
\langle  {\sigma}'_3, {\sigma}'_2,{\sigma}'_1; {\tau}_3,{\tau}_2,{\tau}; 
{\bf k}_{23},{\bf k}|j,j_z = m'; \epsilon^3_{int},\Pi; T T_z \rangle^*    ~ ,
\label{mdexspin1}
\ee
\ewt
with ${\cal R}_M $  the Melosh rotation (see Appendix \ref{Melosh})
and $\langle \sigma'_3, \sigma'_2,\sigma'_1; \tau_3,\tau_2,\tau; {\bf k}_{23},{\bf k}|j,j_z = m;
\epsilon^3_{int},\Pi; T T_z \rangle$  the momentum-space instant-form wave function. 

The quantities ${{\bf k}}$, $E({\bf k})$, and $E_{23}$ in Eq. (\ref{mdexspin1}) are easily determined  from the
variables $x$, $\bf {k_{\perp}} $, ${{\bf k}_{23}}$ \cite{DelDotto:2016vkh}
\be
k^+=x~ M_0 (1,2,3)
\label{FO13}
\ee
with
\be
M^2_0 (1,2,3)={{m^2+{\bf k}^2_{\perp}}\over x}+{{M^2_{23}+{\bf k}^2_{\perp}}\over (1-x)}`,
\nonu
M_{23}=2~\sqrt{{m^2+|\bf k}_{23}|^2} 
\ee
and eventually
\be
k_{z}={1\over 2}(k^+ -k^-)= {1\over 2}\left ( k^+-{{m^2+{\bf k}^2_{\perp}}\over k^+} \right )
~,
\label{FO18}
\ee
while
\be
E({\bf k})= \sqrt{{m^2+|{\bf k}|^2} } ~~ , \quad 
E_{23}=\sqrt{M^2_{23}+{{\bf k}}^2 } ~.
\label{FO16}
\ee
Let us recall that in the system rest frame one has $\bma P_\perp$ = 0 . \\

  \subsection{Evaluation of ${\cal F}^{\tau m m'}_{\sigma\sigma'}$}
  \label{Fmm}
 Let us take the instant-form three-body wave function as the $^3$He non-relativistic wave function \cite{DelDotto:2016vkh}.
The $^3$He wave function in momentum space can be written as follows from the wave function in coordinate space \cite{Kievsky:1994mxj,Kievsky:1995uk}
\bwt\be
 \langle \sigma_1,\sigma_2,\sigma_3; 
\tau_1, \tau_{2}, \tau_3; 
{\bf k}_{23},{\bf k}|^3{\rm He};
{1 \over 2} m; 
{1 \over 2} T_z\rangle 
=\sum_{l_{23} \mu_{23} }\sum_{L_{\rho} M_{\rho}}{ Y}_{ l_{23} \mu_{23} } (\hat {\bf k}_{23}) ~
{ Y}_{ L_{\rho} M_{\rho}} (\hat {\bf k})
\sum_{T_{23},\tau_{23}}\langle  {1 \over 2} \tau_2  {1 \over 2} \tau_3 | T_{23} \tau_{23} \rangle 
\langle T_{23} \tau_{23} {1 \over 2} \tau_1 |{1 \over 2} T_z \rangle
\nonu\times
\hspace{-1mm}\sum_{X M_{X}} \sum_{j_{23} m_{23}} \hspace{-1mm}\langle XM_X L_\rho M_\rho|{ 1 \over 2}m \rangle  \langle j_{23}\ m_{23}{1 \over 2}\sigma_1|XM_X \rangle
\sum_{s_{23}\sigma_{{23}}} \langle {1 \over 2}\sigma_2 {1 \over 2}
\sigma_3| s_{23}\sigma_{{23}} \rangle  \langle l_{23} \mu_{23} s_{23} \sigma_{{23}} 
|j_{23} \ m_{23} \rangle {\bma {\cal G}}_{{L_{\rho}}X}^{j_{23} l_{23}s_{23}}(k_{23},k)
\label{FO6}
\ee 
\ewt
with
\be
 {\bma {\cal G}}_{{L_{\rho}}X}^{j_{23} l_{23}s_{23}}(k_{23},k)={2(-1)^{{l_{23} + L_{\rho}}\over 2}\over \pi} 
 \int r^2 d{r}~j_{l_{23}}( k_{23} r)
 \nonu \times~\int {\rho}^2~ d{\rho}~j_{L_{\rho}}(k \rho)~
 \phi^{j_{23}l_{23} s_{23}}_{L_\rho X} ( |{\bf r}|,|{\bma \rho}|) ~.
 \label{FO5}
\ee
The antisymmetrization of the wave function requires
$l_{23}+s_{23}+T_{23}$, where $T_{23}$ is the isospin
of the pair $23$, to be odd. In addition, 
$l_{23}+L_\rho$ has to be even, due to the parity
of $^3$He.

Then, inserting Eq. (\ref{FO6}) in Eq. (\ref{mdexspin1}),
the function $ {\cal F}^{\tau m m'}_{\sigma\sigma'}(x,{\bf k}_{\perp})$  becomes
\bwt
\be
 {\cal F}^{\tau m m'}_{\sigma\sigma'}(x,{\bf k}_{\perp})={1\over (1- x)}~ \sum_{\tau_2\tau_3}  ~\sum_{\sigma'_2\sigma'_3}
 \int d {\bf k}_{23} ~{E_{23}~E({\bf k})\over k^+_1} ~\sum_{\sigma_1}D^{{1 \over 2}} [{\cal R}_M ({\blf k} )]_{\sigma\sigma_1}
 \sum_{l_{23} \mu_{23} }\sum_{L_{\rho} M_{\rho}}{ Y}_{ l_{23} \mu_{23} } (\hat {\bf k}_{23}) ~{ Y}_{ L_{\rho} M_{\rho}} (\hat {\bf k})
\nonu\times~\sum_{T_{23},\tau_{23}}\langle  {1 \over 2} \tau_2  {1 \over 2} \tau_3 | T_{23} \tau_{23} \rangle 
~\langle T_{23} \tau_{23} {1 \over 2} \tau |{1 \over 2} T_z \rangle
\sum_{X M_{X}} \sum_{j_{23} m_{23}}\langle XM_X L_\rho M_\rho|{ 1 \over 2}m \rangle ~ \langle j_{23}\ m_{23}{1 \over 2}\sigma_1|XM_X \rangle
\sum_{s_{23}\sigma_{{23}}} ~\langle {1 \over 2}\sigma'_2 {1 \over 2}
\sigma'_3| s_{23}\sigma_{{23}} \rangle 
\nonu\times
~ \langle l_{23} \mu_{23} s_{23} \sigma_{{23}} 
|j_{23} \ m_{23} \rangle ~{\bma {\cal G}}_{{L_{\rho}}X}^{j_{23} l_{23}s_{23}}(k_{23},k)
\sum_{{\sigma}'_1}D^{{1 \over 2}*} [{\cal R}_M ({\blf k} ]_{\sigma'{\sigma}'_1}\sum_{l'_{23} \mu'_{23} }\sum_{L'_{\rho} M'_{\rho}}{ Y}^*_{ l'_{23} \mu'_{23} } (\hat {\bf k}_{23}) ~{ Y}^*_{ L'_{\rho} M'_{\rho}} (\hat {\bf k})
\nonu\times
\sum_{T'_{23},\tau'_{23}}\langle  {1 \over 2} \tau_2  {1 \over 2} \tau_3 | T'_{23} \tau'_{23} \rangle ~\langle T'_{23} \tau'_{23} {1 \over 2} \tau |{1 \over 2} T_z \rangle
\sum_{X' M'_{X}} \sum_{j'_{23} m'_{23}}\langle X'M'_X L'_\rho M'_\rho|{ 1 \over 2}m' \rangle ~ \langle j'_{23}\ m'_{23}{1 \over 2}\sigma'_1|X'M'_X \rangle
\nonu\times
\sum_{s'_{23}\sigma'_{{23}}} ~\langle {1 \over 2}\sigma'_2 {1 \over 2}
\sigma'_3| s'_{23}\sigma'_{{23}} \rangle ~ \langle l'_{23} \mu'_{23} s'_{23} \sigma'_{{23}} 
|j'_{23} \ m'_{23} \rangle ~{\bma {\cal G^*}}_{{L'_{\rho}}X'}^{j'_{23} l'_{23}s'_{23}}(k_{23},k) ~.
\label{FO8}
\ee
\ewt

Since ${{\bf k}}$ is only function of  $|{\bf k}_{23}|$, then we are allowed to integrate the spherical harmonics in Eq. (\ref{FO8}) over $ d {\Omega_{\hat {\bf k}_{23}}}$.
Therefore, taking care of the orthogonality of the Clebsch-Gordan coefficients, we can write:
\bwt
\be
 {\cal F}^{\tau m m'}_{\sigma\sigma'}= {1\over (1- x)}~ \int_0^\infty d { k}_{23}~k^2_{23} ~{E_{23}~E({\bf k})\over k^+}~
\sum_{\sigma_1}D^{{1 \over 2}} [{\cal R}_M ({\blf k})]_{\sigma\sigma_1} 
\sum_{{\sigma}'_1}D^{{1 \over 2}*} [{\cal R}_M ({\blf k} ]_{\sigma'{\sigma}'_1}
 ~ \sum_{s_{23}}  \sum_{j_{23}}  \sum_{l_{23} }  \sum_{T_{23},\tau_{23}} 
 \nonu \times\langle T_{23} \tau_{23} {1 \over 2} \tau |{1 \over 2} T_z \rangle 
 \langle T_{23} \tau_{23} {1 \over 2} \tau |{1 \over 2} T_z \rangle
 \sum_{L_{\rho}} \sum_{L'_{\rho}} \sum_{X} \sum_{X'} 
{\cal R}^{j_{23}, mm', \sigma_1 {{\sigma}'_1}}_{L_{\rho} L'_{\rho} X X'}(k_{23},{\bf k})
{\bma {\cal G}}_{{L_{\rho}}X}^{j_{23} l_{23}s_{23}}(k_{23},k)
{\bma {\cal G^*}}_{{L'_{\rho}}X'}^{j_{23} l_{23}s_{23}}(k_{23},k) ~,
\label{FO9}
\ee 
where
\be
{\cal R}^{j_{23}, mm', \sigma_1 {{\sigma}'_1}}
_{L_{\rho} L'_{\rho} X X'}(k_{23},{\bf k}) =
 \sum_{M_{\rho}} ~{ Y}_{ L_{\rho} M_{\rho}} (\hat {\bf k}) \sum_{M_{X}} \sum_{m_{23}}
 ~
\langle XM_X L_\rho M_\rho|{ 1 \over 2}m \rangle ~ \langle j_{23}\ m_{23}{1 \over 2}\sigma_1|XM_X \rangle
\nonu
\times \sum_{M'_{\rho}} ~(-1)^{-M'_{\rho}}{ Y}_{ L'_{\rho} -M'_{\rho}} (\hat {\bf k})
\sum_{M'_{X}}~\langle X'M'_X L'_\rho M'_\rho|{ 1 \over 2}m' \rangle 
~
 \langle j_{23}\ m_{23}{1 \over 2}
\sigma'_1|X'M'_X \rangle~.
\label{R1}
\ee
\ewt
The quantity
${\cal R}^{j_{23}, mm', \sigma_1 {{\sigma}'_1}}_{L_{\rho} L'_{\rho} X X'}(k_{23},{\bf k})$ is invariant for parity, since $L_{\rho}$ and $L'_{\rho}$ have the same parity.

\subsection{Sum of products of five 3j symbols}
\label{prod3j}
By using 
the properties of the product of two spherical harmonics with the same argument \cite{Varshalovich:1988ye} and 
3j symbols, Eq. (\ref{R1}) becomes
\bwt\be
{\cal R}^{j_{23}, mm', \sigma_1 {{\sigma}'_1}}_{L_{\rho} L'_{\rho} X X'}(k_{23},{\bf k}) =
\sum_{L M} \sqrt{{(2L_{\rho}+1)(2L'_{\rho}+1)}\over {4\pi (2L+1)}}
 ~
\langle L_{\rho}~ 0~ L'_{\rho}~ 0|L~ 0 \rangle ~ Y_{L M}(\theta,\phi)~(-1)^{(L_{\rho}-L'_{\rho}+M)}
\nonu
\times \sum_{M_{\rho}M'_{\rho}}~(-1)^{-M'_{\rho}}~\sqrt{2L+1}
\begin{pmatrix}
L_{\rho}&L'_{\rho}&L \\
M_{\rho}&-M'_{\rho}&-M
\end{pmatrix}
\sum_{M_{X} M'_{X}}~\sum_{m_{23}}~(-1)^{(X-L_{\rho}+m)}~(-1)^{(j_{23}-{1\over 2}+M_X)}
\nonu\times
(-1)^{(X'-L'_{\rho}+m')}~(-1)^{(j_{23}-{1\over 2}+M'_X)}~\sqrt{2}\sqrt{2}~\sqrt{2X+1}~\sqrt{2X'+1}~
\nonu\times
\begin{pmatrix}
X&L_{\rho}&{1\over 2} \\
M_X&M_{\rho}&-m
\end{pmatrix}
\begin{pmatrix}
j_{23}&{1\over 2}&X \\
m_{23}&\sigma_1 &-M_X
\end{pmatrix}
\begin{pmatrix}
X'&L'_{\rho}&{1\over 2} \\
M'_X& M'_{\rho}&-m'
\end{pmatrix}
\begin{pmatrix}
j_{23}&{1\over 2}&X' \\
m_{23}&{\sigma'_1} &-M'_X
\end{pmatrix}~,
\label{distr6}
\ee
\ewt
where the angles $\theta$ and $\phi$ define the direction of $\hat {\bf k}$. Only even values of $L$ are allowed to satisfy the parity invariance of ${\cal R}^{j_{23}, mm', \sigma_1 {{\sigma}'_1}}_{L_{\rho} L'_{\rho} X X'}(k_{23},{\bf k})$.

Through  permutations of the columns in the 3j symbols, to have the indices $m$ and $m'$ in the middle, and changing the sign of the third momentum components in the 3j symbols where $X$ appears, we obtain (see Eq. (16) at p. 457 of Ref. \cite{Varshalovich:1988ye} )
\bwt
\be
 {\cal R}^{j_{23}, mm', \sigma_1 {{\sigma}'_1}}_{L_{\rho} L'_{\rho} X X'}(k_{23},{\bf k})
=-2~(-1)^{\sigma_1'} ~ (-1)^{m} ~ (-1)^{(X+m)}~(-1)^{(X'+m')}~ (-1)^{j_{23}} ~\sqrt{{(2L_{\rho}+1)(2L'_{\rho}+1)}\over {4\pi}}~
\nonu
\nonu
\hspace{-.0cm} \times
~ \langle L_{\rho}~ 0~ L'_{\rho}~ 0|L~ 0 \rangle ~\sqrt{2X+1}~\sqrt{2X'+1}
~\sum_{L M }(-1)^{(L+M)}~ Y_{L M}(\theta,\phi) ~
 (-1)^{(X' - L'_{\rho} -1/2 -1/2 -L - 1/2)}
\nonu
\nonu
 \times ~\sum_{a \alpha y \eta}(-1)^{(a-\alpha+y-\eta)}\Pi^2_{ay}
 \begin{pmatrix}
{1\over 2}&{1\over 2}&a\\ \\
{\sigma'_1}&-\sigma_1&\alpha
\end{pmatrix}
\begin{pmatrix}
a&L&y \\ \\
-\alpha&-M &-\eta
\end{pmatrix}
\begin{pmatrix}
y&{1\over 2}&{1\over 2} \\\\
\eta &m &-m'
\end{pmatrix}
\begin{Bmatrix}
{1\over 2}& {1\over 2}&a \\ \\
X &X' &j_{23}
\end{Bmatrix}
\begin{Bmatrix}
{1\over 2}&{1\over 2}&y \\ \\
X&X' &a \\ \\
L_{\rho} &L'_{\rho} & L
\end{Bmatrix}~.
\label{distr9}
\ee
\subsection{Sums involving the $D$ matrixes for the system polarization and the Melosh factors}
\label{melofac}
Let us consider the following combination of Wigner functions and Melosh rotations
\be
 D^{j{\cal M}, a \, y}_{\sigma  \sigma '}({\bf S},{\blf k}) ~ =   - ~ 2 ~ 
~\sum_{L M }
(-1)^{M}
~~Y_{L M}(\theta,\phi) \sum_m  ~ 
D^{j}_{m, {\cal M}}(\alpha, \beta, \gamma)  ~ \sum_{m'}  ~ [D^{j}_{m', {\cal M}}(\alpha, \beta, \gamma)]^* ~ 
~\sum_{\sigma_1}~D^{{1 \over 2}} [{\cal R}_M ({\blf k} )]_{\sigma\sigma_1}
\nonu
 \times 
 \sum_{{ \sigma'_1}}
D^{{1 \over 2}*} [{\cal R}_M ({\blf k} )]_{\sigma'{ \sigma}'_1}~ 
\sum_{\mu,  \eta}~ (-1)^{-\mu - \eta}
\begin{pmatrix}
1\over 2&{1\over 2}&a\\
{\sigma'_1}&-\sigma_1 &\mu
\end{pmatrix} 
 \begin{pmatrix}
a&L&y \\
-\mu&-M&-\eta
\end{pmatrix}
\begin{pmatrix}
y&{1\over 2}&{1\over 2} \\
\eta& m&-m'
\end{pmatrix} 
 ~(-1)^{\sigma_1} ~ (-1)^{2m + m' }
~.
 \label{distr5}
\ee
If one applies Eq. (\ref{spfS2}) to the product
$D^{j}_{m, {\cal M}}(\alpha, \beta, \gamma)~ D^{j}_{-m', -{\cal M}}(\alpha, \beta, \gamma) $ ,
then Eq. (\ref{distr5}) becomes
\be
 D^{j{\cal M}, a \, y}_{\sigma  \sigma '}({\bf S},{\blf k}) ~ 
=  ~ \sqrt {2}~ {(-1)^{a  + 1} \over \sqrt {2a + 1}}  ~\sum_{L M }~ (-1)^{M}
~Y_{L M}(\theta,\phi)~ \sum_{J=0}^{2j}  ~
\langle j  {\cal M}, j  -{\cal M} |J 0 \rangle ~ \sum_{ \mu,  \eta}~ (-1)^{ - \eta} ~ \sum_{m   m'}
(-1)^{m' - {\cal M}} ~
\nonu
 \times ~ 
\hspace{-.0cm} 
\langle j m, j -m'  |J (m-m') \rangle ~   ~
D^{J}_{(m - m'), 0}(\alpha, \beta, \gamma)~ ~
\sum_{\sigma_1}~D^{{1 \over 2}} [{\cal R}_M ({\blf k} )]_{\sigma\sigma_1}~
 \sum_{{ \sigma'_1}}
D^{{1 \over 2}*} [{\cal R}_M ({\blf k} )]_{\sigma'{ \sigma}'_1}~
 (-1)^{{\sigma_1}} ~(-1)^{({1} - a)}
\nonu
 \times
~\langle {1 \over 2}  \sigma_1 , {1 \over 2} - {\sigma'_1}~ | ~a \mu \rangle ~
 \begin{pmatrix}
a&L&y \\
-\mu&-M&-\eta
\end{pmatrix} 
 ~ (-1)^{1/2  + 2y}
~ (-1)^{1/2 -m} \sqrt{{2 \over 2y +1}} ~
\langle {1 \over 2} m , {1 \over 2} ~ -m' ~ | ~y - \eta \rangle ~.
\label{distr30}
\ee
One has to recall that $j = 1/2$ and that $(m-m')$ has to be equal to $-\eta$.
 Then we obtain
\be
 D^{1/2{\cal M}, a \, y}_{\sigma  \sigma '}({\bf S},{\blf k}) ~ 
=
-
~ {2}~ {1 \over \sqrt {2a + 1}}  ~\sum_{L M }~ (-1)^{M}~~Y_{L M}(\theta,\phi)~    ~
\langle {1 \over 2}  {\cal M}, {1 \over 2}  -{\cal M} |y 0 \rangle ~ 
\sum_{ \mu,  \eta}~  
(-1)^{ - {\cal M}} ~ Y^*_{y-\eta}(\beta,\alpha) ~\sqrt{ {4 \pi \over 2y+1}}
\nonu
\nonu
 \times  
    ~ 
~\sum_{\sigma_1}~D^{{1 \over 2}} [{\cal R}_M ({\blf k} )]_{\sigma\sigma_1}~
 \sum_{{ \sigma'_1}}
D^{{1 \over 2}*} [{\cal R}_M ({\blf k} )]_{\sigma'{\sigma}'_1}~
\langle {1 \over 2}  \sigma_1 , {1 \over 2} - {\sigma'_1}~ | ~a \mu \rangle  ~
 (-1)^{{\sigma_1}}
~
 \begin{pmatrix}
a&L&y \\
-\mu&-M&-\eta
\end{pmatrix} 
 ~
~
 {1 \over \sqrt{2y +1}} ~.
\label{distr31}
\ee

\subsection{Spin-dependent momentum distribution}
\label{sdmd}
Making use of Eq. (\ref{distr31}) to express the quantity 
$D^{j{\cal M}, a \, y}_{\sigma  \sigma '}({\bf S},{\blf k})$,
we can now summarize our results  for the spin-dependent momentum distribution, {Eq. \eqref{N}, as follows }

 \be
 {\left [ {\bma {\cal N}}_{\cal M}^\tau(x,{\bf k}_{\perp};\bf S) \right ]_{\sigma  \sigma '} }   =  
 ~ { (-1)^{ {\cal M}} ~ 2 \over (1- x)}~ \int _0^\infty d { k}_{23}~k^2_{23} ~{E_{23}~E({\bf k})\over k^+}~
 \sum_{s_{23}}  \sum_{j_{23}}  \sum_{l_{23} }  \sum_{T_{23},\tau_{23}} \langle T_{23} \tau_{23} {1 \over 2} \tau |{1 \over 2} T_z \rangle \langle T_{23} \tau_{23} {1 \over 2} \tau |{1 \over 2} T_z \rangle
\nonu
 \times ~  (-1)^{j_{23}} ~ \sum_{L_{\rho}} \sum_{L'_{\rho}} \sum_{X} \sum_{X'} ~ 
\sqrt{{(2L_{\rho}+1)(2L'_{\rho}+1)}}~ 
\langle L_{\rho}~ 0~ L'_{\rho}~ 0|L~ 0 \rangle ~\sqrt{2X+1}~\sqrt{2X'+1}~
 (-1)^{(- L'_{\rho} -1/2)}
\nonu
\times
~{\bma {\cal G}}_{{L_{\rho}}X}^{j_{23} l_{23}s_{23}}(k_{23},k)
~{\bma {\cal G^*}}_{{L'_{\rho}}X'}^{j_{23} l_{23}s_{23}}(k_{23},k) 
~(-1)^{X}~\sum_{a ~ y} (-1)^{a} ~ \Pi^2_{ay}
~\sum_{L M }  ~ Y_{L M}(\theta,\phi) ~ 
 {1 \over \sqrt {2a + 1}}  ~    ~
\langle {1 \over 2}  {\cal M}, {1 \over 2}  -{\cal M} |y 0 \rangle 
\nonu
\times ~
\sum_{ \mu,  \eta}~  
  Y^*_{y-\eta}(\beta,\alpha) ~{1 \over \sqrt{2y+1}} 
    ~ 
~\sum_{\sigma_1}~D^{{1 \over 2}} [{\cal R}_M ({\blf k})]_{\sigma\sigma_1}~
 \sum_{{ \sigma'_1}}
D^{{1 \over 2}*} [{\cal R}_M ({\blf k})]_{\sigma'{\sigma}'_1}~
\langle {1 \over 2}  \sigma_1 , {1 \over 2} - {\sigma'_1}~ | ~a \mu \rangle  ~
 (-1)^{{\sigma_1}}
 \nonu
   \times ~ \langle a \mu L M | ~ y -\eta \rangle  ~ (-1)^{\mu} 
~
 {1 \over {2y +1}}
 \begin{Bmatrix}
{1\over 2}& {1\over 2}&a \\ \\
X &X' &j_{23}
\end{Bmatrix}
\begin{Bmatrix}
{1\over 2}&{1\over 2}&y \\ \\
X&X' &a \\ \\
L_{\rho} &L'_{\rho} & L
\end{Bmatrix}~.
 \label{distr35}
 \ee
 It has to be noticed that $(a + y)$ is an even number, as can be easily shown exchanging $X$ with  $X'$ and $L_\rho$ with $L'_\rho$. Furthermore both $a$ and $y$ can be only $0$ or $1$.

Then, if $L=0$ one has $a=y=0$, or $a=y=1$. If $L=2$,  only  $a=y=1$ is possible.

Let us define the quantity
\be
{\cal Z}^\tau _{\sigma \sigma'}(k_{23},{\bf k}, {\bf S}, a, y, L) =  \Pi^2_{ay} 
~ {1 \over \sqrt {2a + 1}}  
~{1 \over {(2y +1)}^{3/2}}
   ~
\langle {1 \over 2}  {\cal M}, {1 \over 2}  -{\cal M} |y 0 \rangle ~ {\cal H}^\tau  (L,a,k_{23},k)
 \sum_{M }  ~ Y_{L M}(\theta,\phi) 
\nonu
\times ~ 
 \sum_{\sigma_1}~D^{{1 \over 2}} [{\cal R}_M ({\blf k} )]_{\sigma\sigma_1}~
 \sum_{{ \sigma'_1}}
D^{{1 \over 2}*} [{\cal R}_M ({\blf k} )]_{\sigma'{ \sigma}'_1}
~ 
  \sum_{\mu \eta}   
\hspace{-.0cm} 
 ~  Y^*_{y -\eta}(\beta,\alpha)  
 \hspace{-.0cm}   
 ~\langle a \mu, LM | y -\eta \rangle ~ (-1)^{\mu} 
 \nonu
\times ~  \langle {1 \over 2}  \sigma_1 , {1 \over 2} - {\sigma'_1}~ | ~a \mu \rangle  ~
 (-1)^{{\sigma_1} - 1/2}~
 \label{distr41}
\ee

where
\be
 {\cal H}^\tau (L,a,k_{23},k) = (-1)^{a} ~
 \sum_{j_{23}}  \sum_{l_{23} } \sum_{s_{23}}  \sum_{T_{23},\tau_{23}} 
 \langle T_{23} \tau_{23} {1 \over 2} \tau |{1 \over 2} T_z \rangle
 \langle T_{23} \tau_{23} {1 \over 2} \tau |{1 \over 2} T_z \rangle
 ~  (-1)^{j_{23}} ~ 
 \nonu
 \times ~\sum_{L_{\rho}} \sum_{L'_{\rho}}  ~ \sum_{X} \sum_{X'} ~ (-1)^{(X+1/2)}~
\sqrt{{(2L_{\rho}+1)(2L'_{\rho}+1)}}~ 
  ~\langle L_{\rho}~ 0~ L'_{\rho}~ 0|L~ 0 \rangle ~\sqrt{2X+1}~\sqrt{2X'+1}
  \nonu
 \times
~(-1)^{X+  X' }~
 \begin{Bmatrix}
{1\over 2}& {1\over 2}&a \\ \\
X &X' &j_{23}
\end{Bmatrix}
\begin{Bmatrix}
L_{\rho}&X&{1\over 2} \\ \\
L'_{\rho}&X' &{1\over 2} \\ \\
L & a & y
\end{Bmatrix} ~(-1)^{l_{23}}
 ~{\bma {\cal G}}_{{L_{\rho}}X}^{j_{23} l_{23}s_{23}}(k_{23},k)
~{\bma {\cal G^*}}_{{L'_{\rho}}X'}^{j_{23} l_{23}s_{23}}(k_{23},k) ~.
 \label{distr47}
\ee
Then the momentum distribution can be written as a function of the three independent quantities
${\cal Z}^\tau _{\sigma \sigma'}(k_{23},{\bf k}, {\bf S}, 0, 0, 0) $, ${\cal Z}^\tau _{\sigma \sigma'}(k_{23},{\bf k}, {\bf S}, 1, 1, 0) $, and ${\cal Z}^\tau _{\sigma \sigma'}(k_{23},{\bf k}, {\bf S}, 1, 1, 2) $ that notably depend on
$ {\cal H}^\tau (0,0,k_{23},k)$, $ {\cal H}^\tau (0,1,k_{23},k)$, and $ {\cal H}^\tau (2,1,k_{23},k)$,
viz.
\be
 {\left [ {\bma {\cal N}}_{\cal M}^\tau(x,{\bf k}_{\perp};\bf S) \right ]_{\sigma  \sigma '} }  ~  
=  
(-1)^{{\cal M} + 1/2} 
~ {2 \over (1- x)}~
 \int_0^\infty d { k}_{23}~k^2_{23} ~{E_{23}~E({\bf k})\over k^+}~
\left \{ ~ {\cal Z}^\tau _{\sigma \sigma'}(k_{23},{\bf k}, {\bf S}, 0, 0, 0) ~ \right .
\nonu
 \left . + ~{\cal Z}^\tau _{\sigma \sigma'}(k_{23},{\bf k}, {\bf S}, 1, 1, 0) ~ + ~
{\cal Z}^\tau _{\sigma \sigma'}(k_{23},{\bf k},{\bf S}, 1, 1, 2) ~ \right \}
~.
\label{distr36}
\ee

We evaluate separately ${\cal Z}^\tau _{\sigma \sigma'}(k_{23},{\bf k}, {\bf S}, a, y, L)$ for the three possible cases of the variables $L, a, y$.
The first two quantities to evaluate are

\be
{\cal Z}^\tau _{\sigma \sigma'}(k_{23},{\bf k}, {\bf S}, 0, 0, 0) =  \Pi^2_{00} ~  
   ~
\langle {1 \over 2}  {\cal M}, {1 \over 2}  -{\cal M} | 0 0 \rangle ~ {\cal H}^\tau  (0,0,k_{23},k)
 ~  \sum_{\sigma_1}~D^{{1 \over 2}} [{\cal R}_M ({\blf k} )]_{\sigma\sigma_1}
\nonu
\times ~ 
~
 \sum_{{ \sigma'_1}}
D^{{1 \over 2}*} [{\cal R}_M ({\blf k} )]_{\sigma'{\sigma}'_1}
~  Y_{00}(\theta,\phi)   
 ~  Y^*_{00}(\beta,\alpha)   
 ~\langle 00, 00 | 00 \rangle 
 ~  \langle {1 \over 2}  \sigma_1 , {1 \over 2} - {\sigma_1'}~ | ~0 0 \rangle  ~
 (-1)^{{\sigma_1}  - 1/2}
 \nonu
= ~\delta_{\sigma \sigma'} ~ (-1)^{1/2 - {\cal M}}   
 ~ {\cal H}^\tau  (0,0,k_{23},k)
~  {1\over 8 \pi}~
 \label{distr39}
 \ee
 and
 \be
{\cal Z}^\tau _{\sigma \sigma'}(k_{23},{\bf k}, {\bf S}, 1, 1, 0) =
 \Pi^2_{1,1} 
~ {1 \over \sqrt {3}}  ~
~{1 \over {(3)}^{3/2}}
   ~
\langle {1 \over 2}  {\cal M}, {1 \over 2}  -{\cal M} |1 0 \rangle ~ {\cal H}^\tau  (0,1,k_{23},k)
~ \sum_{\sigma_1}~D^{{1 \over 2}} [{\cal R}_M ({\blf k} )]_{\sigma\sigma_1}
\nonu
\times ~
 \sum_{{ \sigma'_1}}
D^{{1 \over 2}*} [{\cal R}_M ({\blf k} )]_{\sigma'{ \sigma}'_1}
~ 
 \sum_{\mu} ~ 
  ~ Y_{00}(\theta,\phi)  
  \sum_{\eta}~    
 ~  Y^*_{1 -\eta}(\beta,\alpha) ~ 
 \hspace{-.0cm}   
 ~\langle 1 \mu, 00 | 1 -\eta \rangle 
 ~ (-1)^{\mu} 
~  \langle {1 \over 2}  \sigma_1 , {1 \over 2} - {\sigma'_1}~ | ~1 \mu \rangle  ~
 (-1)^{{\sigma_1}  - 1/2}
 \nonu
 = ~ ~{1 \over \sqrt{2}} ~ {1 \over \sqrt{4\pi}}   ~ {\cal H}^\tau  (0,1,k_{23},k) ~
\sum_{\sigma_1}~D^{{1 \over 2}} [{\cal R}_M ({\blf k} )]_{\sigma\sigma_1}~
 \sum_{{\sigma'_1}}
D^{{1 \over 2}*} [{\cal R}_M ({\blf k} )]_{\sigma'{ \sigma}'_1}
 ~ {1 \over 2} ~ 
 \sqrt{{3 \over 2 \pi}} ~\left [ {\bma  \sigma} \cdot {\hat {\bf S}} \right ]_{\sigma _1 { \sigma}'_1}~,
 \label{distr38}
 \ee
 where the identity (see 
 Ref. \cite{Varshalovich:1988ye})
 \be
  \sum_{\mu}    
 ~  Y_{1 -\mu}(\beta,\alpha) ~     
 ~
 (-1)^{\mu} 
~  \langle {1 \over 2}  \sigma_1 , {1 \over 2} - {\sigma'_1}~ | ~1 \mu \rangle  ~
 (-1)^{{{\sigma}_1'}}
 = ~{1 \over 2} ~ (-1)^{1/2} \sqrt{{3 \over 2 \pi}} ~
 \left [ {\bma  \sigma} \cdot {\hat {\bf S}} \right ]_{\sigma_1 {\sigma}'_1}~,
 \label{distr50}
 \ee
 has been used.

 Through the actual expressions of the Melosh rotations  (see Appendix \ref{Melosh} and in particular Eqs. (\ref{54}) and (\ref{A58})), Eq. (\ref{distr38}) becomes
 \be
 {\cal Z}^\tau _{\sigma \sigma'}(k_{23},{\bf k}, {\bf S}, 1, 1, 0) =
{1 \over 2} ~ 
  {1 \over \sqrt{8\pi}}   ~ \sqrt{{3 \over 2 \pi}} ~ {\cal H}^\tau  (0,1,k_{23},k) 
 \left [ \cos {\varphi \over 2} ~ + ~i ~ \sin {\varphi \over 2} ~ \hat {\bf n} \cdot  {\bma  \sigma}  \right ]_{\sigma\sigma_1}
 \left [ {\bma  \sigma} \cdot {\hat {\bf S}} \right ]_{\sigma _1 {\sigma}'_1}
\nonu
 \nonu
 \times ~  \left [ \cos {\varphi \over 2} ~ - ~i ~\sin {\varphi \over 2} ~ \hat {\bf n} \cdot  {\bma  \sigma}\right ]_{{\sigma}'_1\sigma'}
 \nonu
=  {\sqrt{{3}} \over {8\pi}}    ~ {\cal H}^\tau  (0,1,k_{23},k) 
  ~
\left \{  \left [ {\bma  \sigma} \cdot {{\bf S}} \right ]_{{\sigma}\sigma'} ~
 - ~2 ~ \sin {\varphi \over 2} ~\cos {\varphi \over 2} ~  \left \{ \left [  \left  ( {\bf S} \cdot \hat{\bf k}_\perp \right )   {\hat z }   -  ({\bf  S} \cdot \hat z)~
 { \hat{\bf k}_{\perp}  }
\right  ] \cdot  {\bma  \sigma}
 \right \}_{{\sigma}\sigma'}  \right .
 \nonu
 \nonu
\left . - ~2 ~ \sin^2 {\varphi \over 2} ~  \left \{ \left [ \left ( {\bf  S} \cdot \hat z  \right ) \hat z - 
 \left  ( {\bf S} \cdot \hat{\bf k}_\perp \right ) \hat{\bf k}_\perp~
   \right  ] \cdot \sigma \right \}_{{\sigma}\sigma'}    \right \}~,
\label{55}
\ee
where $ \sin {\varphi / 2} $ and $ \cos {\varphi / 2} $ are defined in Eq. (\ref{52}).

  The last quantity,
  ${\cal Z}^\tau _{\sigma \sigma'}(k_{23},{\bf k}, {\bf S}, 1, 1, 2)$, is
 \be
 {\cal Z}^\tau _{\sigma \sigma'}(k_{23},{\bf k}, {\bf S}, 1, 1, 2)   
= ~ \Pi^2_{1, 1}  
~ {1 \over \sqrt {3}}  ~
~{1 \over {(3)}^{3/2}}
   ~
\langle {1 \over 2}  {\cal M}, {1 \over 2}  -{\cal M} |1 0 \rangle ~ {\cal H}^\tau  (2,1,k_{23},k)
 \sum_{\sigma_1}~D^{{1 \over 2}} [{\cal R}_M ({\blf k} )]_{\sigma\sigma_1}~
\nonu
 \times  
~  \sum_{{\sigma'_1}}
D^{{1 \over 2}*} [{\cal R}_M ({\blf k} )]_{\sigma'{\sigma}'_1}
~ \sum_{M }  ~ Y_{2 M}(\theta,\phi)  
  \sum_{\mu \eta}    
 ~  Y^*_{1 -\eta}(\beta,\alpha) ~    
 ~\langle 1 \mu, 2M | 1 -\eta \rangle  ~
 (-1)^{\mu} 
 ~  \langle {1 \over 2}  \sigma_1 , {1 \over 2} - {\sigma'_1}~ | ~1 \mu \rangle  ~
 (-1)^{{\sigma_1} - 1/2}
 \nonu
=    ~ - {1 \over {2}}
 ~ {\cal H}^\tau  (2,1,k_{23},k) ~ \sum_{\sigma_1}~D^{{1 \over 2}} [{\cal R}_M ({\blf k} )]_{\sigma\sigma_1}~
 \sum_{{ \sigma'_1}}
D^{{1 \over 2}*} [{\cal R}_M ({\blf k} )]_{\sigma'{ \sigma}'_1}
~  \sum_\mu  ~(-1)^{\mu} 
 ~ \langle {1 \over 2}  \sigma_1 , {1 \over 2} - {\sigma'_1}~ | ~1 \mu \rangle  ~
 (-1)^{{\sigma_1}- 1/2}
 \nonu
 \times ~ 
\left [ Y^*_{1 \mu}(\theta,\phi) ~{\sqrt 3} ~Y_{1, 0} ({\hat {\bf k}} \cdot \hat {\bf  S}) 
- Y^*_{1 \mu}(\hat {\bf  S}) ~ {1 \over \sqrt {4 \pi}} \right ]~.
 \label{distr37}
\ee
In the last step the first of the bipolar harmonics in Eq. (A5) of Ref. \cite{CiofidegliAtti:1994cm} has been used. 
Therefore, using again Eq. (\ref{distr50}) 
and the results of Appendix \ref{Melosh} (see Eq. (\ref{54})) we have
 \be
 {\cal Z}^\tau _{\sigma \sigma'}(k_{23},{\bf k}, {\bf S}, 1, 1, 2)   
  = ~  - {1 \over  {4}} ~ \sqrt{{3 \over 8}} ~ {1 \over \pi}
 ~ {\cal H} ^\tau (2,1,k_{23},k) \left [ ~ {3} ~{\hat {\bf k}^{}} \cdot \hat {\bf  S}  ~
\sum_{\sigma_1 { \sigma'_1}}~D^{{1 \over 2}} [{\cal R}_M ({k^{+}, {\bf k}_{\perp}})]_{\sigma\sigma_1}~  
\left [ {\bma  \sigma} \cdot {\hat {\bf k}} \right ]_{\sigma_1 {\sigma}'_1}
\right .
\nonu
\left . \times ~
D^{{1 \over 2}} [{\cal R}_M ({{k^{+}, -{\bf k}_{\perp}}})]_{{\sigma}'_1 \sigma'} 
~ -  ~\sum_{\sigma_1  {\sigma'_1}}~
D^{{1 \over 2}} [{\cal R}_M ({k^{+}, {\bf k}_{\perp}})]_{\sigma\sigma_1}~
 \left [ {\bma  \sigma} \cdot {\hat {\bf S}} \right ]_{\sigma_1 { \sigma}'_1}
 ~
D^{{1 \over 2}} [{\cal R}_M ({k^{+}_1, -{\bf k}_{\perp}} )]_{{ \sigma}'_1 \sigma'}
  \right ] ~ .
 \label{distr57}
\ee

Eventually from Eqs. (\ref{A58}, \ref{A59}) we have

\be
{\cal Z}^\tau _{\sigma \sigma'}(k_{23},{\bf k}, {\bf S}, 1, 1, 2)   
=   ~  - {1 \over  {4}} ~ \sqrt{{3 \over 8}} ~{1 \over \pi}
 ~ {\cal H}^\tau  (2,1,k_{23},k) 
 ~ \left \{ ~ {{3} \over k}~{\hat {\bf k}^{}} \cdot \hat {\bf  S}  ~ \left [ ~
\left (  \cos^2 {\varphi \over 2} - \sin^2 {\varphi \over 2} \right )~ \left [ {\bma  \sigma} \cdot {{\bf k}} \right ]_{{\sigma}\sigma'} \right .  \right .
\nonu
\nonu
\left .  \left . 
 - ~2 ~ \sin {\varphi \over 2} ~\cos {\varphi \over 2} ~ \left [  \left (   {k}_{\perp}  ~{\hat z }  
  - {k}_{z} ~ {\hat {\bf k}}_{\perp}
 \right ) \cdot  {\bma  \sigma}
 \right ]_{{\sigma}\sigma'}  \right ]
 ~ - ~ \left \{   ~
 \left [ {\bma  \sigma} \cdot {{\bf S}} \right ]_{{\sigma}\sigma'} 
 - ~2 ~ \sin {\varphi \over 2} ~\cos {\varphi \over 2} ~  \left [ \left  (  ({\bf  S} \cdot \hat {\bf k}_{\perp}) ~ {\hat z }   - ({\bf  S} \cdot \hat z)~
 { \hat{\bf k}_{\perp}  }
 \right ) \cdot  {\bma  \sigma}
 \right ]_{{\sigma}\sigma'}  
 \right .
  \right .
 \nonu
  \nonu
\left . 
  \left .
 - ~2 ~ \sin^2 {\varphi \over 2} ~  
 \left [   \left (   ({\bf  S} \cdot \hat z)  ~ {\hat z } + ({\bf  S} \cdot \hat {\bf k}_{\perp} ) ~ { \hat{\bf k}_{\perp}  } \right )
 \cdot  {\bma  \sigma} ~  \right ]_{{\sigma}\sigma'}~  
 \right \}  \right \} ~.
 \label{distr60}
\ee

Summarizing our results, the spin-dependent momentum distribution 
 in terms of the pseudovectors ~
  $\bf{S}$, $~\hat {\bf k}_{\perp} ({\bf  S} \cdot \hat {\bf k}_{\perp})$, 
$~\hat {\bf k}_{\perp} ({\bf  S} \cdot \hat z)$, 
 $~\hat z ~({\bf  S} \cdot \hat {\bf k}_{\perp})$, and $~ \hat z ~({\bf  S} \cdot \hat {z})$
 ~ is
\be
 {\left [ {\bma {\cal N}}_{\cal M}^\tau(x,{\bf k}_{\perp};\bf S) \right ]_{\sigma  \sigma '} }  ~ =  ~
~ {(-1)^{{\cal M} + 1/2}  \over 4 \pi ~(1- x)}~
 \int_0^\infty d { k}_{23}~k^2_{23} ~{E_{23}~E({\bf k})\over k^+}~
\left \{   ~\delta_{\sigma \sigma'} ~ (-1)^{1/2 - {\cal M}}   
 ~ {\cal H}^\tau  (0,0,k_{23},k) ~ 
\right .
\nonu
+ ~ {\sqrt{{3}}}    ~ \left [ ~  {\cal H}^\tau  (0,1,k_{23},k) 
+ ~ \sqrt{{1 \over 2 }}
 ~ {\cal H}^\tau  (2,1,k_{23},k) ~ 
\right ] \left [ ~ \left [ {\bma  \sigma} \cdot {{\bf S}} \right ]_{{\sigma}\sigma'} 
 - ~2 ~ \sin {\varphi \over 2} ~\cos {\varphi \over 2} ~  \left [ \left ( ({\bf S} \cdot {\hat {\bf k}}_{\perp}  )~ {\hat z }   -  \left ({\bf S} \cdot   {\hat z } \right )~
 {\hat {\bf k}}_{\perp}  
\right  ) \cdot  {\bma  \sigma}
 \right ]_{{\sigma}\sigma'}  \right .
 \nonu
 \nonu
\left . - ~2 ~ \sin^2 {\varphi \over 2} ~  \left [   \left (   \left ({\bf S} \cdot   {\hat z } \right )  ~ {\hat z } + 
({\bf S} \cdot {\hat {\bf k}}_{\perp}  ) ~ { \hat{\bf k}_{\perp} 
 } \right ) 
 \cdot  {\bma  \sigma} ~  \right ]_{{\sigma}\sigma'} \right ]  -   \sqrt{{3 \over 2}}
 ~ {\cal H}^\tau  (2,1,k_{23},k) ~{3 \over k^2} 
  \left [ 
 {{\bf k}}_{\perp} \cdot  {\bf  S}    \left [
\left (  \cos^2 {\varphi \over 2} - \sin^2 {\varphi \over 2} \right )
 \left ( {\bf k}_{\perp} \cdot {\bma  \sigma}  \right )_{{\sigma}\sigma'} \right . \right .
 \nonu 
 \nonu
 \left . \left . +
 \left (  \cos^2 {\varphi \over 2} - \sin^2 {\varphi \over 2} \right )
  {k}_{z}  
(  {\hat z }  \cdot  {\bma  \sigma} )_{{\sigma}\sigma'} 
 - ~ 2 \sin {\varphi \over 2} \cos {\varphi \over 2}  \left [  \left ( k_\perp ~{\hat z }  ~
  - ~ {k}_{z} ~ {\hat {\bf k}}_{\perp}
 \right ) \cdot  {\bma  \sigma} 
 \right ]_{{\sigma}\sigma'}  + ~ {k}_{z} ~  \left (  {\hat z }  \cdot  {\bf  S}  \right ) 
  \left [
\left (  \cos^2 {\varphi \over 2} - \sin^2 {\varphi \over 2} \right )
  \right . \right . \right .
 \nonu
 \nonu
 \left . \left . \left .
\times ~\left ( {\bf k}_{\perp} \cdot {\bma  \sigma}  \right )_{{\sigma}\sigma'} +
 \left (  \cos^2 {\varphi \over 2} - \sin^2 {\varphi \over 2} \right )
{k}_{z}  
(  {\hat z }  \cdot  {\bma  \sigma} )_{{\sigma}\sigma'}  - 2 \sin {\varphi \over 2} \cos {\varphi \over 2}  \left [  \left ( k_\perp ~{\hat z }  
  - {k}_{z} ~ {\hat {\bf k}}_{\perp}
 \right ) \cdot  {\bma  \sigma} 
 \right ]_{{\sigma}\sigma'} \right ] 
   \right ]
 \right \}~.
  \label{distr62}
\ee

From a comparison of Eqs. (\ref{distr1}), (\ref{dv2}) and (\ref{distr62}) one can immediately obtain explicit expressions for the functions $b^\tau_{i,{\cal M}}\left [ |{\bf k}_{\perp}|,x,({\bf  S}\cdot \hat {\bf k}_{\perp})^2,({\bf  S} \cdot {\hat z})^2,{{  (  \hat {\bf k}_{\perp} \times {\hat z} ) \cdot  {\bf S} }} \right] $ ($i= 0,1, . . .,5)$ and verify that actually they do not depend on ${\bf  S}$ {and then they do not depend on the direction of ${\bf k}_{\perp}$}.
For $i=0$ one has
\be
 b^\tau _{0,{\cal M}}\left [x, |{\bf k}_{\perp}|
 \right]
= ~ {(-1)  \over 2 \pi ~(1- x)}~
 \int_0^\infty d { k}_{23}~k^2_{23} ~{E_{23}~E({\bf k})\over k^+}   
 ~ {\cal H}^\tau  (0,0,k_{23},k) ~ . 
\label{b0}
\ee

It can be useful to decompose the functions $b^\tau _{i,{\cal M}}$ ($i = 1, . . . ,5$) according to the values $0$ or $2$ of the momentum $L$ 
\be
b^\tau _{i,{\cal M}} = b^{\tau(L=0) }_{i,{\cal M}} + b^{\tau(L=2)}_{i,{\cal M}}~,
\label{decomp}
\ee
and one obtains 
\be
b^{\tau (0)}_{1,{\cal M}}\left(x, |{\bf k}_{\perp}| \right)
=
~ {(-1)^{{\cal M} + 1/2}  \over 2 \pi ~(1- x)}~{\sqrt{{3}}}
 \int_0^\infty d { k}_{23}~k^2_{23} ~{E_{23}~E({\bf k})\over k^+}    ~
   {\cal H} ^\tau(0,1,k_{23},k) 
 ~, 
 \label{b10}
\ee
\be
b^{\tau(2)}_{1,{\cal M}}\left(x, |{\bf k}_{\perp}| \right)
\label{b1}
=
~ {(-1)^{{\cal M} + 1/2}  \over 2 \pi ~(1- x)}~{\sqrt{{3}}}
 \int_0^\infty d { k}_{23}~k^2_{23} ~{E_{23}~E({\bf k})\over k^+}    ~
  ~ \sqrt{{1 \over 2 }}
 ~ {\cal H}^\tau(2,1,k_{23},k) ~ ,
 \label{b12}
\ee
\be
b^{\tau(0)}_{2,{\cal M}}\left(x, |{\bf k}_{\perp}| \right)
=
~ - ~{(-1)^{{\cal M} + 1/2}  \over 2 \pi ~ (1- x)}~\sqrt{{3}}
 \int_0^\infty d { k}_{23}~k^2_{23} ~{E_{23}~E({\bf k})\over k^+}~
  {\cal H} ^\tau (0,1,k_{23},k)  ~ 
 ~2 ~ \sin^2 {\varphi \over 2}~,
\label{b20}
\ee
\be
b^{\tau(2)}_{2,{\cal M}}\left(x, |{\bf k}_{\perp}| \right)
=
~ - ~{(-1)^{{\cal M} + 1/2}  \over 2 \pi ~ (1- x)}~\sqrt{{3}}
 \int_0^\infty d { k}_{23}~k^2_{23} ~{E_{23}~E({\bf k})\over k^+}~
  \sqrt{{1 \over 2 }}  ~ {\cal H} ^\tau (2,1,k_{23},k) 
  \nonu
 \left \{    ~2 ~ \sin^2 {\varphi \over 2} ~
 + ~ 3  ~{1 \over k^2}
 \left [ \left (  \cos^2 {\varphi \over 2}  -   \sin^2 {\varphi \over 2}  \right )
 ~ k_{\perp}^2  + 2  ~ \sin {\varphi \over 2} ~\cos {\varphi \over 2} ~  k_{\perp} ~ k_{z}  \right ]
 \right \}~,
\label{b22}
\ee
\be
b^{\tau(0)}_{3,{\cal M}}\left(x, |{\bf k}_{\perp}| \right)
= ~ {(-1)^{{\cal M} + 1/2}  \over 2 \pi ~ (1- x)}~{\sqrt{{3}}}   ~
 \int_0^\infty d { k}_{23}~k^2_{23} ~{E_{23}~E({\bf k})\over k^+}
 ~  2 ~\sin {\varphi \over 2} ~\cos {\varphi \over 2}  ~ 
   {\cal H} ^\tau (0,1,k_{23},k) ~,  
   \label{b30}
\ee
\be
 b^{\tau(2)}_{3,{\cal M}}\left(x, |{\bf k}_{\perp}| \right)
= ~ {(-1)^{{\cal M} + 1/2}  \over 2 \pi ~ (1- x)}~{\sqrt{{3}}}   ~
 \int_0^\infty d { k}_{23}~k^2_{23} ~{E_{23}~E({\bf k})\over k^+} ~ \sqrt{{1 \over 2 }} ~ 
 {\cal H} ^\tau (2,1,k_{23},k)
\nonu
 \hspace{-.0cm}\times ~ \left \{  ~  2 ~\sin {\varphi \over 2} ~\cos {\varphi \over 2}  ~ 
   -  ~ {3}
 ~{1 \over k^2} ~\left [ k_{\perp} ~ k_{z} 
  \left (  \cos^2 {\varphi \over 2}  -   \sin^2 {\varphi \over 2}  \right ) ~+ ~k_{z}^2 ~2 ~ \sin {\varphi \over 2} 
  ~\cos {\varphi \over 2} \right ] \right \}~,
\label{b32}
\ee
\be
 b^{\tau(0)}_{4,{\cal M}}\left(x, |{\bf k}_{\perp}| \right)
=  - b^{\tau(0)}_{3,{\cal M}}\left(x, |{\bf k}_{\perp}| \right)~,
  \label{b40}
\ee
\be
 b^{\tau(2)}_{4,{\cal M}}\left(x, |{\bf k}_{\perp}| \right)
= ~ - ~ {(-1)^{{\cal M} + 1/2}  \over 2 \pi ~ (1- x)}~{\sqrt{{3}}}  
 \int_0^\infty d { k}_{23}~k^2_{23} ~{E_{23}~E({\bf k})\over k^+} ~
  \sqrt{{1 \over 2 }} ~ {\cal H} ^\tau (2,1,k_{23},k)
  \nonu
  \times
~ \left \{   ~2 ~  \sin {\varphi \over 2} ~\cos {\varphi \over 2}  ~
+  ~ {3}
 ~{1 \over k^2} ~\left [ k_{\perp} ~ k_{z} 
  \left (  \cos^2 {\varphi \over 2}  -   \sin^2 {\varphi \over 2}  \right ) 
  - k^2_{\perp} ~2 ~  \sin {\varphi \over 2} ~\cos {\varphi \over 2} \right ]\right \}~,
\label{b42}
\ee
\be
b^{\tau(0)}_{5,{\cal M}}\left(x, |{\bf k}_{\perp}| \right)
=  b^{\tau(0)}_{2,{\cal M}}\left(x, |{\bf k}_{\perp}| \right)~,
\label{b50}
\ee
\be
 b^{\tau(2)}_{5,{\cal M}}\left(x, |{\bf k}_{\perp}| \right)
= ~ -
~ {(-1)^{{\cal M} + 1/2}  \over 2 \pi ~ (1- x)}~\sqrt{{3}}
 \int_0^\infty d { k}_{23}~k^2_{23} ~{E_{23}~E({\bf k})\over k^+}
  ~ \sqrt{{1 \over 2 }}
 ~ {\cal H} ^\tau (2,1,k_{23},k)  
  \nonu
 \times \left \{ 
 ~2 ~ \sin^2 {\varphi \over 2}  ~ + ~ 3 ~ 
 {1 \over k^2}
  \left [  \left (  \cos^2 {\varphi \over 2}  -  \sin^2 {\varphi \over 2}  \right ) ~
 k_{z}^2   - 2  ~ \sin {\varphi \over 2} ~\cos {\varphi \over 2} ~ k_{\perp} ~ k_{z}  \right ] \right \}~,
\label{b52}
\ee
{where the dependence upon $\varphi$ is generated by the Melosh rotations (see Appendix \ref{Melosh}). It has to be pointed out that, in the case of a three-nucleon bound system, $\varphi$
is small, as one can deduce from its definition in Eq. \eqref{52}, and therefore  $sin(\varphi/2)/cos(\varphi/2) << 1$. }

One can immediately recognize that the quantities $b^\tau _{i,{\cal M}}$ actually 
{{ are invariant for rotations around the $z$ axis}}, while they do depend on $|{\bf k}_{\perp}|$ and $x$. 
The quantity $b^\tau _{0}$ is independent of $\cal M$. For $i = 1, . . . ,5$, the dependence on $\cal M$ is through the 
factor $(-1)^{{\cal M} +1/2}$.


\ewt
\subsection{Effective polarizations}
\label{effpol}
Using the equations (see Appendix B of \cite{DelDotto:2016vkh})
\be
{dx \over d k^+} =  { 1 - x  \over E_{23} } ~,
\nonu {d k^+ \over d k_{z}} =
{ k^+ \over E({\bf k})}~,
\label{norma1}
\ee
the longitudinal and transverse effective polarizations of Eqs. (\ref{leffpa}) and (\ref{teffp}), respectively,  are given by
\be
p^\tau_{||} = {(-1)^{{\cal M} + 1/2} }~{\sqrt{{3}}}  
 \int_0^\infty d { k}_{23}~k^2_{23} ~\int_0^\infty {k}^2 ~ d {k} 
 \nonu \times ~\int_{-1}^1 d \cos \theta ~ f_{||}({\bf k},k_{23}) ~, 
\label{leffpc1}
\ee
and
\be
p^\tau_{\perp} = {(-1)^{{\cal M} + 1/2} }~{\sqrt{{3}}}  
 \int_0^\infty d { k}_{23}~k^2_{23} ~\int_0^\infty {k}^2 ~ d {k} 
 \nonu \times  ~\int_{-1}^1 d \cos \theta ~ f_{\perp}({\bf k},k_{23}) 
\label{teffpc1}
\ee
where
\be
 f_{||}({\bf k},k_{23})  
 =  {\cal H} ^\tau(0,1,k_{23},k) \left ( 1 - 2 ~ \sin^2 {\varphi \over 2} \right )
+
  \sqrt{{1 \over 2 }}
  \nonu\times~
  {\cal H}^\tau(2,1,k_{23},k) 
 \left \{ 1 -
 ~2 ~ \sin^2 {\varphi \over 2}  
 ~- ~ 3 ~ 
  \left [  \left (  \cos^2 {\varphi \over 2} 
 \right .\right. \right .\nonu \left. \left. \left.-  \sin^2 {\varphi \over 2}  \right )   {\cos^2 \theta}~
   - 2  \sin {\varphi \over 2} \cos {\varphi \over 2} {\sin \theta}  {\cos \theta} \right ] \right \} ~
\label{teffpc3}
\ee
and
\be
f_{\perp}({\bf k},k_{23}) =  
  {\cal H} ^\tau(0,1,k_{23},k) \left ( 1 - \sin^2 {\varphi \over 2}  \right ) 
 + ~ \sqrt{{1 \over 2 }} 
 \nonu \times ~ {\cal H} ^\tau (2,1,k_{23},k) 
 \left \{ 1 -   \sin^2 {\varphi \over 2} ~
 - ~ \frac{3}{2} ~
 \left [ \left (  \cos^2 {\varphi \over 2}  
 \right. \right.  \right.  \nonu \left.\left. \left.
 -   \sin^2 {\varphi \over 2}  \right )  \sin^2 \theta
   +  2   \sin {\varphi \over 2} \cos {\varphi \over 2}  \sin \theta \cos \theta  \right ] \right \}~.
\label{teffpc4}
\ee

In Eqs. (\ref{teffpc3}) and (\ref{teffpc4}) we have defined ${\cos \theta} =  k_{z}  / k$ and  $\sin \theta = k_{\perp}/k$.

Let us emphasize that without the effect of the Melosh rotations one has $\sin {\varphi \over 2}=0$ and $ \cos {\varphi \over 2}=1$. Then the two polarizations become equal, viz
\be
p^\tau_{||} = p^\tau_{\perp} = {(-1)^{{\cal M} + 1/2} }~2 ~{\sqrt{{3}}}  
 \nonu 
 \times ~ \int_0^\infty d { k}_{23}~k^2_{23} ~\int_0^\infty {k}^2 ~ d {k}  ~
   {\cal H} ^\tau(0,1,k_{23},k)  
\label{teffpc6b}
\ee
\section{Products of Melosh  and Pauli matrixes}
\label{Melosh}
 The Melosh matrix 
 \be
D^{{1 \over 2}} [{\cal R}_M ({\blf k})]_{\sigma\sigma'}=
 \chi^\dagger_{\sigma}{m+k^+-\imath 
{\bma \sigma} \cdot (\hat z \times
{\bf k}_{\perp}) \over \sqrt{\left ( m +k^+ \right )^2 +|{\bf
k}_{\perp}|^2}}
\chi_{\sigma'}
\label{50}
\ee
can be rewritten as follows
\be
D^{{1 \over 2}} [{\cal R}_M ({\blf k})]_{\sigma\sigma'}
= \left [ \cos {\varphi \over 2} ~ + ~i ~ \sin {\varphi \over 2} ~ \hat {\bf n} \cdot  {\bma  \sigma} 
 \right ]_{\sigma\sigma'}~,
\label{51}
\ee
where 
\be
\varphi = 2 ~ arctg  ~{ |{\bf k}_\perp| \over k^+ + m}~,
\nonu \cos {\varphi \over 2} =
 {k^+ + m \over \sqrt{ (k^+ + m)^2 + {\bf k}^2_\perp} }~,
 \nonu  \sin {\varphi \over 2} =
  { |{\bf k}_\perp| \over \sqrt{ (k^+ + m)^2 + {\bf k}^2_\perp} }~,
\label{52}
\ee
and
\be
\hat {\bf n} = - {\hat z \times {\bf k}_{\perp}  \over   { |{\bf k}_\perp|  }}~.
\label{53}
\ee
Then the equality
\be
D^{{1 \over 2}*} [{\cal R}_M ({k^+, {\bf k}_{\perp}})]_{\sigma\sigma'} = 
D^{{1 \over 2}} [{\cal R}_M ({k^+, -{\bf k}_{\perp}})]_{\sigma'\sigma} 
\nonu =
\left [ \cos {\varphi \over 2} ~ - ~i ~\sin {\varphi \over 2} ~ \hat {\bf n} \cdot  {\bma  \sigma}\right ]_{\sigma'\sigma}
\label{54}
\ee
holds.

Let us now evaluate the sandwich of $\left [ {\bma  \sigma} \cdot \hat{\bf e} \right ]$ between two Melosh matrixes (see Eqs. (\ref{distr38}),(\ref{55}), (\ref{distr57})), with $\hat{\bf e}$  a unit vector. One gets
\be
 {\cal D}_{\sigma\sigma'} ({k^{+}, {\bf k}_{\perp}}, \hat{\bf e}) = \sum_{\sigma'_1 {\tilde \sigma'_1}}~
D^{{1 \over 2}} [{\cal R}_M ({k^{+}, {\bf k}_{\perp}})]_{\sigma\sigma'_1}~
~\left [ {\bma  \sigma} \cdot {\hat{\bf e}} \right ]_{\sigma' _1 {\tilde \sigma}'_1}
\nonu \times~
D^{{1 \over 2}} [{\cal R}_M ({k^{+}, -{\bf k}_{\perp}})]_{{\tilde \sigma}'_1 \sigma'} 
 =   
\left [  \cos^2 {\varphi \over 2} - \sin^2 {\varphi \over 2} \right ]~ 
\left [ {\bma  \sigma} \cdot {\hat{\bf e}} \right ]_{{\sigma}\sigma'} 
\nonu  - ~2 ~ \sin {\varphi \over 2} ~\cos {\varphi \over 2} ~ 
\left \{ [ (\hat{\bf e}\cdot   \hat{\bf k}_{\perp}) ~ {\hat z }   - (\hat{\bf e} \cdot  {\hat z }  ) ~ \hat{\bf k}_{\perp}
 ] \cdot  {\bma  \sigma}
 \right \}_{{\sigma}\sigma'}  
 \nonu
 +  ~2 ~ \sin^2 {\varphi \over 2} ~  \left \{ \left (   \hat {\bf n} \cdot  \hat{\bf e} \right ) ~(\hat {\bf n} \cdot  {\bma  \sigma}) ~
   \right \}_{{\sigma}\sigma'} ~. 
\label{A57}
\ee
For $\hat{\bf e}= {\bf S}$ one has
\be
 {\cal D}_{\sigma\sigma'} ({k^{+}, {\bf k}_{\perp}}, {\bf S}) ~ =  
~
 \left [ {\bma  \sigma} \cdot {{\bf S}} \right ]_{{\sigma}\sigma'} ~
\nonu
 - ~2 ~ \sin {\varphi \over 2} ~\cos {\varphi \over 2} ~ \left \{ \left [ \left  ( {\bf S} \cdot \hat{\bf k}_\perp \right )  ~  {\hat z }   - \left  ( {\bf S} \cdot {\hat z} \right )~
 { \hat{\bf k}_{\perp} }
 \right ] \cdot  {\bma  \sigma}
 \right \}_{{\sigma}\sigma'}  
 \nonu
 \nonu
 - ~2 ~ \sin^2 {\varphi \over 2} ~  
 \left \{   \left [   \left  ( {\bf S} \cdot {\hat z} \right ) ~ {\hat z } + \left  ( {\bf S} \cdot \hat{\bf k}_\perp \right ) ~ { \hat{\bf k}_{\perp} } \right ]
 \cdot  {\bma  \sigma} ~  \right \}_{{\sigma}\sigma'}~,
\label{A58}
\ee
For ${\hat{\bf e}} = \hat{\bf k}$, one obtains
\be
{\cal D}_{\sigma\sigma'} ({k^{+}, {\bf k}_{\perp}}, \hat{\bf k}) 
   =   ~
\left [  \cos^2 {\varphi \over 2} - \sin^2 {\varphi \over 2} \right ]~ \left [ {\bma  \sigma} \cdot {\hat{\bf k}} \right ]_{{\sigma}\sigma'} ~
\nonu
 - ~2 ~ \sin {\varphi \over 2} ~\cos {\varphi \over 2} ~{1 \over k} \left [ [   {k}_{\perp}  ~{\hat z }  
  - {k}_z ~ {\hat {\bf k}}_{\perp}
 ] \cdot  {\bma  \sigma}
 \right ]_{{\sigma}\sigma'} ~, 
\label{A59}
\ee
where $ ~{{k}}_{\perp} = \sqrt{k_x^2 + k_y^2} ~$ and 
$~k_z = {\bf k} \cdot  {\hat z } ~={1\over 2}(k^+ - k^-)= {1\over 2}(k^+ -{{m^2 + {\bf k}^2_{\perp}}\over k^+})$.
\\

\section{Traces of the Valence Correlator and of the light-front Spectral Function}
\label{traces}

Let us  denote by $\Gamma$ a generic $4 \times 4$ matrix. The traces of $  \left [  \Gamma ~ \Phi_V  \right ] $ can be expressed
through  traces of the spectral function or of the spectral function times $\bma \sigma$ matrixes. 
 Indeed with the help of  Eqs.  (\ref{abc})   and  (\ref{iden}) one has
 \be
 {1 \over 2 P^+} ~ Tr \left [ \Gamma ~ \Phi_V \right ] = 
  {1 \over 2 }~
{1 \over 2 ~ m}~\sum_\sigma \sum_{\sigma'} ~ \bar{u}_{}( {\tilde {\bf p}},\sigma)~
\Gamma~u_{}({\tilde {\bf p}},\sigma') ~
\nonu \times~{1 \over p^+}   ~ 
{  \pi  ~ E_S \over  \xi ~{\cal{M}}_0[1,(23)]} ~ 
{ {\cal P}}^{}_{\cal M, \sigma' \sigma}(\tilde{\bma \kappa},\epsilon,S) ~.
 \label{GF}
 \ee
 As in Appendix \ref{corre}, to simplify the notation the isospin index $\tau$ is understood.

The traces of $\Phi_V$ 
in Eqs. (\ref{ftrV},\ref{strV},\ref{tracceV}), i. e. the traces needed when the correlator is expanded at twist-two level considering only the T-even terms, can be expressed by traces of the spectral function with the help of  Eq.  (\ref{GF})   and
 of  the following equalities for the matrix elements of 
$\gamma$ matrixes between LF spinors (see  Ref. \cite{Lev:2000vm}) 
\be
 1) ~ \bar{u}_{}( {\tilde {\bf p}}',\sigma')~\gamma^+~u_{}({\tilde {\bf p}},\sigma)
=
\delta_{ \sigma' \sigma}~2~\sqrt{{p'}^+ p^+}
\label{gp}
\\&&
2) ~ \bar{u}_{}( {\tilde {\bf p}}',\sigma')~\gamma^+~\gamma_5~u_{}({\tilde {\bf p}},\sigma)=
{2 
\sqrt{{p'}^+ p^+}}~\chi^{\dagger}_{\sigma'}\sigma_z\chi_{\sigma}
\label{gp5}\\&&
3) ~ \bar{u}_{}( {\tilde {\bf{p}}}',\sigma')~\gamma^+~\gamma_5~\gamma_x~u_{}({\tilde {\bf p}},\sigma)
\nonu=
~-2~\sqrt{{p'}^+ p^+}~\chi^{\dagger}_{\sigma'}~\sigma_x~\chi_{\sigma} ~,
\label{gp5x}
\ee
where $\chi_{\sigma}$ is the spin eigenfunction. Then one obtains 
\be
 1) ~  \frac1{2P^+} \, \Tr(\gamma^+ \Phi_V)  ~
 = ~ c ~{\Tr}\left[{\bma {\cal P}}^{}_{\cal M}(\tilde{\bma \kappa},\epsilon,S)
\right]
\label{trpiuA} 
\\ &&
\nonu
2) ~  \frac1{2P^+} \, {\Tr}(\gamma^+ \gamma_5 ~ \Phi_V)  = 
 ~ c~
{\Tr}\left[ \sigma_z ~{\bma {\cal P}}^{}_{\cal M}(\tilde{\bma \kappa},\epsilon,S) \right]
\label{trpiu5A}\\ &&
\nonu
3) ~ -  \frac1{2P^+} \, \Tr(\gamma^i~\gamma^+ ~\gamma_5 ~ \Phi_V) 
\nonu    = c ~
{\Tr}\left[ \sigma^i
~{\bma {\cal P}}^{}_{\cal M}(\tilde{\bma \kappa},\epsilon,S) 
\right]~
\label{trperpiu5A}
\ee    
where $~ c ~ = {\pi ~ E_S /( 2 m ~ \kappa^+)} $ and $i=1,2$.


%


\end{document}